\documentclass[11pt,aps,showpacs,notitlepage,nofootinbib]{revtex4-1}
\usepackage{amssymb,amsmath}
\usepackage{bm}
\usepackage[pdftex]{graphicx}
\usepackage{subfig}
\usepackage{color}
\usepackage{amsthm}
\usepackage[update,prepend]{epstopdf}
\usepackage{hyperref}
\usepackage{cleveref}

\usepackage{footnote}

\newcommand{\beq}{\begin{equation}}
\newcommand{\beqn}{\begin{eqnarray}}
\newcommand{\eeq}{\end{equation}}
\newcommand{\eeqn}{\end{eqnarray}}

\begin{document}

\title{Multi-field inflation and preheating in asymmetric $\alpha$-attractors}
\author{
Oksana~Iarygina$^{1}$, Evangelos~I.~Sfakianakis$^{1,2}$, Dong-Gang~Wang$^{1,3}$, Ana~Ach\'ucarro$^{1,4}$
}
\email{\baselineskip 11pt Email addresses: iarygina@lorentz.leidenuniv.nl; e.sfakianakis@nikhef.nl; wdgang@strw.leidenuniv.nl; achucar@lorentz.leidenuniv.nl}
\affiliation{
$^1$Institute Lorentz of Theoretical Physics, Leiden University, 2333 CA Leiden, The Netherlands
\\
$^2$ Nikhef, Science Park 105, 1098 XG Amsterdam, The Netherlands
\\
$^3$ Leiden Observatory, Leiden University, 2300 RA Leiden, The Netherlands
\\
$^4$ Department of Theoretical Physics, University of the Basque Country, UPV-EHU 48080 Bilbao, Spain
}
\date{\today}
\begin{abstract}
We analyze and compare the multi-field dynamics during inflation and preheating in symmetric and asymmetric models of $\alpha$-attractors, characterized by a hyperbolic field-space manifold. We show that the generalized (asymmetric) E- and (symmetric) T-models exhibit identical two-field dynamics during inflation for a wide range of initial conditions. The resulting motion can be decomposed in two approximately single-field segments connected by a sharp turn in field-space. The details of preheating can nevertheless be different. 
For the T-model one main mass-scale dominates the evolution of fluctuations of the spectator field, whereas for the E-model, a  competing mass-scale emerges due to the steepness of the potential away from the inflationary plateau, leading to different contributions to parametric resonance for small and large wave-numbers. 
Our linear multi-field analysis of fluctuations indicates that for highly curved manifolds, both the E- and T-models preheat almost instantaneously. 
For massless fields this is always due to efficient tachyonic amplification of the spectator field, making single-field results inaccurate. Interestingly,  there is  a parameter window corresponding to $r={\cal O}(10^{-5})$ and massive fields, where the preheating behavior is qualitatively and quantitatively different for symmetric and asymmetric potentials. In that case, the E-model can completely preheat due to self-resonance for values of the curvature where preheating in the T-model is inefficient. This provides a first distinguishing feature between models that otherwise behave identically, both at the single-field and multi-field level.  Finally, we discuss how one can describe multi-field preheating on a hyperbolic manifold by identifying the relevant mass-scales that control the growth of inflaton and spectator fluctuations, which can be applied to any $\alpha$-attractor model and beyond.

\end{abstract}

\maketitle

\tableofcontents

\section{Introduction}
\label{sec:Intro}

Our understanding of the early universe is largely based  on two  observationally constrained phases: inflation and big-bang nucleosynthesis (BBN). Inflation remains the leading framework for physics of the very early universe because it provides an elegant solution for the horizon and flatness problems \cite{Guth:1980zm, Linde:1981mu} as well as a mechanism to seed quantum fluctuations which is in excellent agreement with the latest observational tests \cite{Akrami:2018odb} for a wide range of models. At the same time, BBN is based on the detailed information about nuclear reactions and provides predictions for the light-element abundances \cite{Steigman:2007xt}. So far the theoretical predictions of BBN match observations to very high accuracy. On the other hand, the reheating process that provides an exit from inflation and transition to the thermal state of the universe, which is required for BBN, is far less explored or constrained.
The duration of reheating determines the moment of transition to the radiation dominated era, hence it can affect BBN and shift the time at which CMB-relevant scales left the horizon during inflation, thereby altering inflationary predictions.
Therefore, a detailed knowledge of the reheating physics is crucial in the era of precision cosmology, in order to reduce theoretical uncertainties and provide a smooth link between the theory and present (or future) observational data.

Since the energy scale of the early universe is expected to be very high (over or close to $E^{1/4}\sim 10^{16}$Gev), the universe may be populated with multiple scalar fields which could  participate in inflation and affect the relevant dynamics.
Therefore, despite the simplicity of single-field models, there is strong motivation to study multi-field effects and their predictions.
Recent work has revealed an abundance of models with
 strong turns in the inflationary trajectory \cite{Achucarro:2019pux, Christodoulidis:2019jsx, Christodoulidis:2019mkj, Christodoulidis:2018qdw, Bjorkmo:2019fls, Bjorkmo:2019aev, Brown:2017osf, Mizuno:2017idt, Garcia-Saenz:2018ifx, Achucarro:2017ing, Bravo:2019xdo}.
Multi field models of this sort have been shown to possess strong dynamical single-field attractors, which are of a different nature compared to usual single-field inflation. In fact these novel attractors lead to large turn-rate, possibly seeding large non-Gaussianity.
Given the theoretical motivation and the multi-field ``surprises" that have been revealed to occur during inflation in some cases, it is essential to extend inflationary models to include multiple fields. In particular, focusing on two-fields can provide a breadth of novel phenomena, while allowing us to build intuition and easily visualize the dynamics.

Due to the huge number of inflationary models, it is hardly possible to state universal (model independent) physical predictions for the various  observables. In the last few years a broad class of inflationary theories have been discovered, that can be grouped under the name of  ``cosmological attractors".
This includes conformal attractors \cite{Kallosh:2013hoa, Kallosh:2013daa}, universal attractors with non-minimal coupling to gravity \cite{Kallosh:2013tua, KS} and $\alpha$-attractors \cite{Ferrara:2013rsa, Kallosh:2013yoa, Cecotti:2014ipa,  Kallosh:2015lwa, Linde:2018hmx}. This class of models brings together a lot of well-known inflationary models such as the Starobinsky model \cite{Starobinsky:1980te}, the GL
model \cite{Goncharov:1985yu,Linde:2014hfa}, and Higgs inflation \cite{Salopek:1988qh, Bezrukov:2007ep}. All of the models have different setups, yet give very close cosmological predictions for the important
observables.
It is thus important that we clarify the twofold meaning of the term ``attractor'' in the context of inflation. For most multi-field models, the term attractor is used to describe a specific trajectory in field space, toward which  the inflationary evolution will flow, regardless of the initial conditions within a certain basin of attraction. 
For the ``cosmological attractors''  \cite{Kallosh:2013hoa, Kallosh:2013daa, Kallosh:2013tua, KS, Ferrara:2013rsa, Kallosh:2013yoa, Cecotti:2014ipa,  Kallosh:2015lwa, Starobinsky:1980te, Goncharov:1985yu,Linde:2014hfa, Salopek:1988qh, Bezrukov:2007ep}, the term is not mainly used to describe a dynamical attractor in field space, but denotes the fact that in some parameter regime, the observables will ``flow'' to a specific value, which is then largely insensitive to the exact parameter values. 
In particular for the scalar spectral index $n_s$ and the tensor-to-scalar ratio $r$, $\alpha$-attractors and related models give
\begin{equation}
n_s=1-\frac{2}{N_*}, \quad r=\frac{12\alpha}{N_*^2},
 \label{eq:ns,r}
\end{equation}
where $N_*$ is the time in e-folds before the end of inflation, where modes first exit the horizon during inflation and $\alpha$ is a dimensionless parameter that --in some models-- encodes the field-space curvature\footnote{It is interesting to note, that two-field $\alpha$-attractors with  $\alpha={\cal O}(1)$ can lead to the predictions of Eq.~\eqref{eq:ns,r}  without possessing a dynamical single-field attractor \cite{Achucarro:2017ing}.}.
For $N_*\gtrsim 55$, the cosmological attractor predictions lead to very good agreement with the observational data.
 These models can be further used to link inflation to the dark energy (cosmological constant) problem \cite{Akrami:2017cir} and aspects of supersymmetry breaking \cite{AlphaBeyondInflation}.
Frequent use of the term ``$\alpha$-attractors'' is made to describe single-field systems with plateau potentials, usually of the form $V\propto \left | 1-e^{-\phi/\Lambda}\right |^{2n}$ or $V\propto \left | \tanh(\phi/\Lambda) \right |^{2n}$, leading to the predictions of Eq.~\eqref{eq:ns,r}. However the flattening of the potential is merely a by-product of a more general feature of $\alpha$-attractors:  hyperbolic field-space manifolds. As we further demonstrate in the present work,  the presence of a second field is crucial for the full dynamics of $\alpha$-attractors during preheating and must be considered to properly extract the predictions of these models, making the single field analysis generally insufficient.

It is worth mentioning that, despite $\alpha$-attractor models being in a great agreement with the Planck 2018 data, there is still the strong inverse dependence on $N_*$ in Eq.~\eqref{eq:ns,r}. Therefore, the uncertainties from the duration of reheating are becoming increasingly important as more  data are being gathered. In particular, the latest {\it Planck} release \cite{Akrami:2018odb} has shown a slight tension (depending on the exact data sets that are being combined) between the measured value of $n_s$ and the $\alpha$-attractor predictions for $N_*\simeq 50$.

In this paper we focus specifically on $\alpha$-attractor models, which are characterized by a hyperbolic field-space
geometry with the constant negative curvature determined by the parameter $\alpha$.
There have been several constructions of $\alpha$-attractor models, but two of the earliest ones, which are still considered the prototypical workhorses, are T- and E-models. In the single-field limit, they represent potentials  that are respectively symmetric and antisymmetric around the minimum. By construction, $\alpha$-attractors are two-field models, since they are constructed by specific choices of the superpotential and K\"ahler potential
in ${\cal N}=1$ supergravity models of
 a complex scalar field, corresponding to an axion-dilaton system (see Appendix A).
The effects of two-field dynamics in T-model preheating has received attention recently using both numerical \cite{Krajewski:2018moi} and semi-analytical techniques \cite{Iarygina:2018kee}. Here, we complement our analysis of the symmetric two-field T-model, by examining a class of generalized E-model potentials \cite{Carrasco:2015rva}, in which the inflaton potential is asymmetric with respect to the origin, which is also the global minimum of the potential.
We explore differences and similarities in the inflationary dynamics, the duration and the underlying mechanism of preheating for symmetric and asymmetric potentials.

We find interesting two-field dynamics during inflation, leading to a single-field attractor in which the second field (spectator) is stabilized at its minimum. Interestingly, the similarities of the T- and E-model during inflation go beyond the existence of a strong single-field attractor with a large basin of attraction. In fact, we show that the full two-field dynamics of the two models is identical, up to slow-roll corrections. Given the existence of a strong single field attractor, we have analyzed the excitation of fluctuations in the inflaton and spectator field, the latter driven by a tachyonic instability due to the negatively curved field-space manifold.
By analyzing the preheating  efficiency of the E-model, we find qualitative differences with similar studies of the related T-model. In particular,  the parametric resonance of inflaton fluctuations is significantly more enhanced in the E-model, as compared to the T-model. Furthermore, for $10^{-4}\lesssim\alpha \lesssim 10^{-3}$, preheating is efficient for the E-model, but not the T-model. This presents the first example of a difference between these two $\alpha$-attractor models and can lead to different predictions for CMB observables.

This work is organized as follows. In Section \ref{sec:Model} we introduce a generalization of the E-model, with an inflaton $\phi$ and spectator field $\chi$,  and study the background motion with a detailed comparison to the T-model.
We find that during inflation, the approach to the single-field attractor of the E- and T-models is identical, up to slow-roll corrections.
In order to assess the strength of the single-field attractor and treat the two fields on the same footing, regardless of the intricacies of the specific parametrization on the curved field-space manifold, we evaluated the background evolution with initial conditions chosen to lie on several iso-potential surfaces. The resulting motion can be viewed as  approximately single-field trajectories joined by a sharp tun in field-space, followed by a brief period of transient oscillations.
 Section \ref{sec:Fluctuations} provides an overview of the fluctuation analysis for the case of multiple fields. We focus on the parametric excitation of $\chi$ fluctuations --since  the corresponding parametric resonance for $\phi$ fluctuations has been studied in the literature and is  weaker for most parameters of interest--  and extensively study separate contributions to the effective frequency that affect particle production. In Section \ref{sec:massscales} we use  Floquet theory to study particle production and  invoke the various mass-scales  to explain the differences between the T- and E-model results.
We  numerically compute the transfer of energy to radiative degrees of freedom in the linear approximation, neglecting mode-mode coupling and backreaction.
   We focus on the $n=1$ case, where the system close to the minimum is described as consisting of interacting massive particles, and compare the preheating efficency of the T- and E-model.
In Section \ref{sec:Conclusion} we conclude and provide an outlook for further studies.


\section{Model and inflationary dynamics}
\label{sec:Model}

Having studied the preheating behaviour of the generalized two-field T-model in Ref.~\cite{Iarygina:2018kee}, we move to the corresponding generalization of the E-model.
The T- and E- model can be viewed as the prototypical examples of symmetric and asymmetric $\alpha$-attractors. Analyzing them can help us build the toolbox and intuition needed to analyze any current or future $\alpha$-attractor scenario that possesses a late-time single-field attractor.
We consider a model consisting of two interacting scalar fields on a hyperbolic manifold of constant negative curvature.
The specific supergravity construction can be found in Appendix A, leading to  the two field  Lagrangian
\beq
{\cal L} = -{1\over 2} \left ( \partial_\mu \chi \partial^\mu \chi + e^{2b(\chi)} \partial_\mu \phi \partial^\mu \phi \right ) - V(\phi,\chi)  \, ,
\label{eq:Lphichi}
\eeq
where $ b(\chi) = \log\left(\cosh(\beta \chi)\right) $.  The corresponding two-field potential is
\beq
V(\phi,\chi) =  \alpha \mu^2 \left (
1-  {2e^{-\beta \phi} \over \cosh \left (\beta \chi \right ) } + e^{-2\beta\phi}
 \right)^n \left( \cosh(\beta \chi) \right )^{2/\beta^2} \, ,
 \label{eq:potential}
\eeq
where $\beta = \sqrt{2/3\alpha}$. For $\chi=0$ the potential becomes
\begin{equation}
\label{eq:singlefieldE}
V(\phi,0) = \alpha \mu^2 \left [ \left( 1-e^{-\beta \phi(t)}\right)^{2}\right ]^n \, ,
\end{equation}
which is a simple one-parameter family of the single-field E-model described in Ref.~\cite{Carrasco:2015rva}. We follow the conventions used in Ref.~\cite{Iarygina:2018kee} for simplicity.

The background equations of motion for $\phi(t)$ at $\chi(t)=0$ are
\begin{subequations}
\label{eq:phieom}
\beqn
&&\ddot \phi + 3 H \dot \phi +2 \sqrt{\frac{2}{3}}  \frac{  \sqrt{\alpha } n \left[\left(e^{-\beta  \phi }-1\right)^2\right]^n}{e^{\beta  \phi }-1}=0
\\
&&3H ^2 = {1\over 2} \left ( {d\phi \over dt}\right)^2 +\alpha  \left [ \left (1-e^{-\beta\phi}\right)^{2}  \right ]^n=0
\eeqn
\end{subequations}
where we rescaled the field $\phi$ by $M_{\rm Pl}$, time $t$ by $\mu$ and the curvature parameter $\alpha$ by $M_{\rm Pl}^2$, as  in Ref.~\cite{Iarygina:2018kee}. Hence with these conventions the Hubble scale is measured in units of $\mu$. The same is true for the comoving wavenumbers, as we will see in Section~\ref{sec:Fluctuations}.

\subsection{Single-field background motion}\label{subsec: SingleFieldbg}

Eqs.~\eqref{eq:phieom} can be simplified during slow-roll inflation, for $\phi \gg \sqrt\alpha$
\beq
3H\dot \phi + {2\sqrt{2}\over \sqrt{3}} \sqrt{\alpha} \, n e^{-\beta\phi} \simeq 0 \, , \quad 3H^2 \simeq {\alpha\over M_{\rm Pl}^2} \mu^2
\eeq
where we explicitly wrote the dimensions of the various quantities in the equation of the Hubble scale. These equations are very similar to the ones that govern the inflationary behaviour of the T-model \cite{Iarygina:2018kee, Carrasco:2015rva}, and can be solved analogously
\beq
\dot \phi \simeq -{2\sqrt{2} n\over 3} e^{-\beta\phi} \, , \quad N = {3\alpha \over 4n} e^{\beta\phi }
\eeq
leading to the slow-roll quantities
\beq
\epsilon \equiv -{\dot H\over H^2} \simeq {3\alpha\over 4N^2} \, , \quad \eta \equiv {\dot \epsilon \over \epsilon H} \simeq {2\over N}
\eeq
and in turn to the tensor-to-scalar ratio
\beq
r = 16\epsilon = {12\alpha\over N^2} \, .
\eeq

As expected, the results for the slow-roll parameters and Hubble scale during single-field inflation are identical for the generalized T- and E-models. Following the breakdown of the slow-roll analysis close to $\epsilon=1$, inflation can be shown to end at $\phi_{\rm end} = {\cal O}(1) \sqrt{\alpha}$ and the corresponding Hubble scale to be $H_{\rm end}^2 \sim {\cal O}(1) \alpha\, \mu^2$. From the amplitude of the scalar power spectrum
\beq
A_s = {H^2 \over 8\pi M_{\rm Pl}^2 \epsilon} \simeq 2\times 10^{-9}
\eeq
we extract the mass-scale $\mu\simeq 6\times 10^{-6} M_{\rm Pl}$ for $N=55$ $e$-folds. Note that this scale is the same for the $T$ model \cite{Iarygina:2018kee}.

\begin{figure}
\centering
\includegraphics[width=0.45\textwidth]{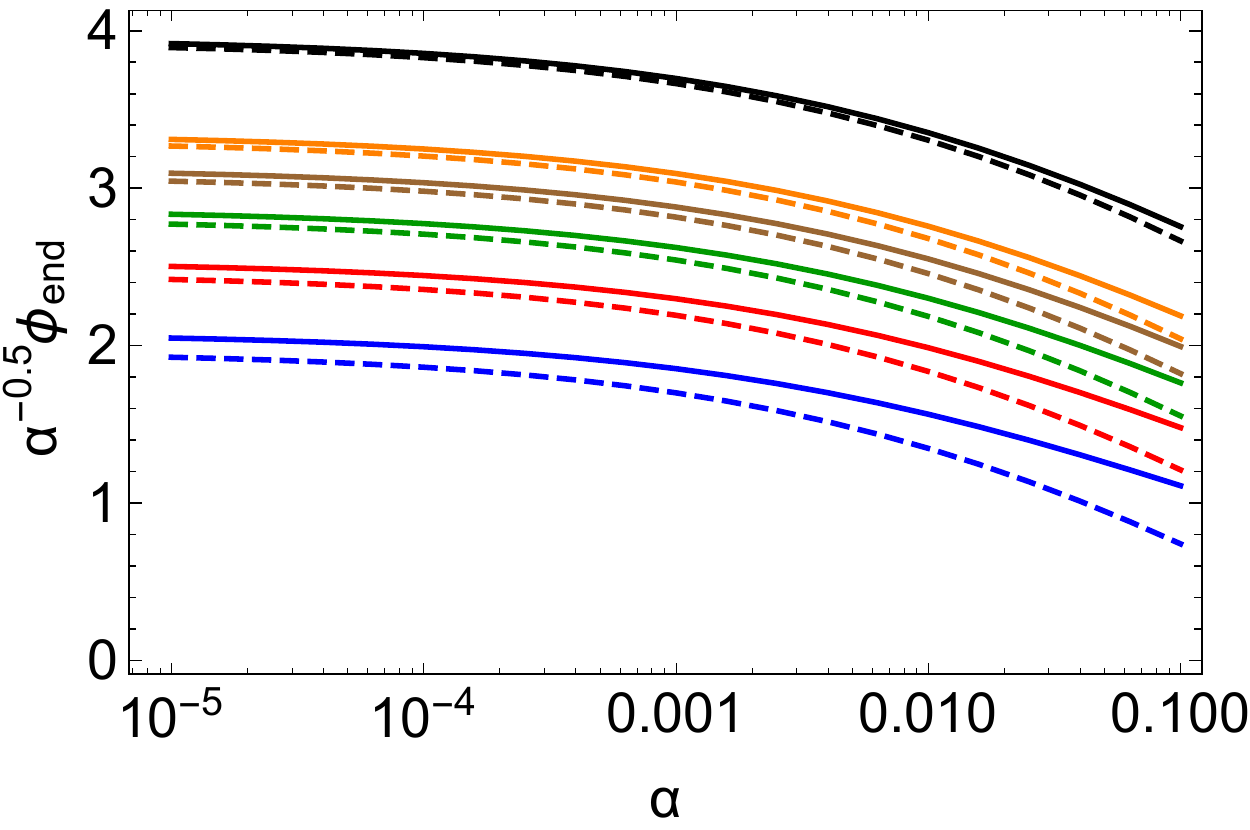} \quad
\includegraphics[width=0.48\textwidth]{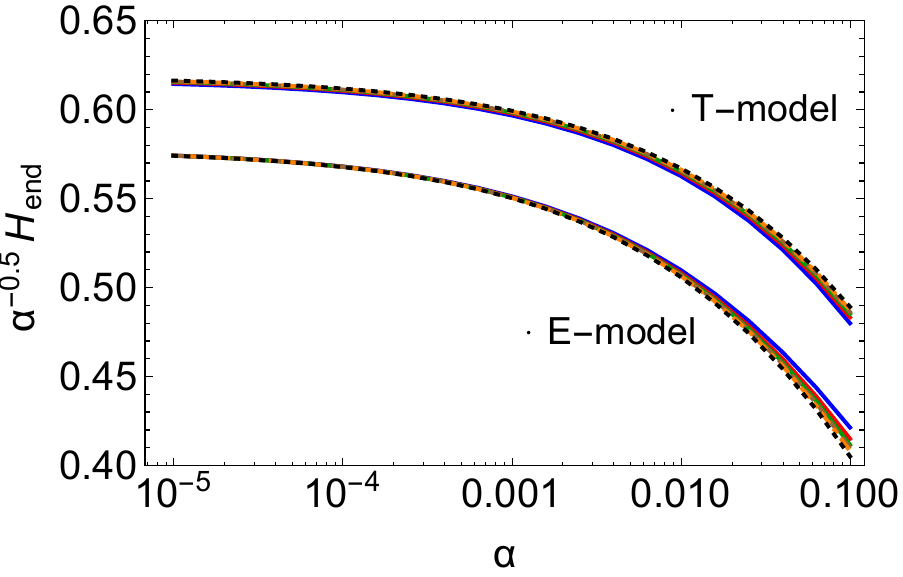}
\caption{
{\it Left:}
The rescaled value of the inflaton field at the end of inflation $\phi_{\rm end}/\sqrt{\alpha}$ as a function of $\alpha$ for $n=1, 1.5, 2, 2.5, 3, 5$ (blue, red, green, brown, orange and black respectively). The solid curves correspond to the  E-model  potential of Eq.~\eqref{eq:singlefieldE}, while the dotted ones correspond to $\phi_{\rm end}$ for the  T-model  potential of Eq.~\eqref{eq:singlefieldT} shifted vertically by $\log(2)/\beta$ according to Eq.~\eqref{eq:phiTvsE}.
We see that the values of $\phi_{\rm end}$ are similar for the T- and E-models and scale as $\phi_{\rm end} \propto \sqrt{\alpha}$ for small $\alpha$.
{\it Right:}
The rescaled Hubble scale at the end of inflation $H_{\rm end}/\sqrt{\alpha}$ for the same parameters and color-coding. The upper / lower curves correspond to the T- and E-model respectively.
The parameter $\alpha$ is measured in units of $M_{\rm Pl}^2$, while $\phi$ is measured in units of $M_{\rm Pl}$ and the Hubble scale is measured in units of $\mu$.
}
 \label{fig:phiendvsa}
\end{figure}

It is worth noting that the generalized T-model, introduced in Ref.~\cite{Carrasco:2015rva} and studied recently in Refs.~\cite{Krajewski:2018moi, Iarygina:2018kee} can be described by a similar field-space metric and a potential (assuming again $\chi=0$)
\beq
V_T(\phi,\chi=0) = \alpha \mu^2 \left [\tanh^2 \left ({\beta\phi\over 2}\right ) \right ]^n
\label{eq:singlefieldT}
\eeq
leading to the same functions of $\epsilon(N)$ and $\eta(N)$, but a slightly different function of $\phi(N)$. In particular
\beq
\phi_T (N) \simeq \phi(N) + {\log (2)\over \beta}
\label{eq:phiTvsE}
\eeq
where $\phi_T$ and $\phi$ correspond to the slow-roll expressions for the T- and E-model respectively, for the same parameters $\alpha$ and $n$. Fig.~\ref{fig:phiendvsa} shows that Eq.~\eqref{eq:phiTvsE} holds very well, even for relating $\phi_{\rm end}$ between the T- and E-models. Furthermore, the Hubble scale at the end of inflation scales as $
H_{\rm end} \sim 0.5 \sqrt{\alpha}$ in units of $\mu$. The scaling is similar for the E- and T-models, with slightly different pre-factors, as shown in Fig.~\ref{fig:phiendvsa}.

\begin{figure}
\centering
\includegraphics[width=0.45\textwidth]{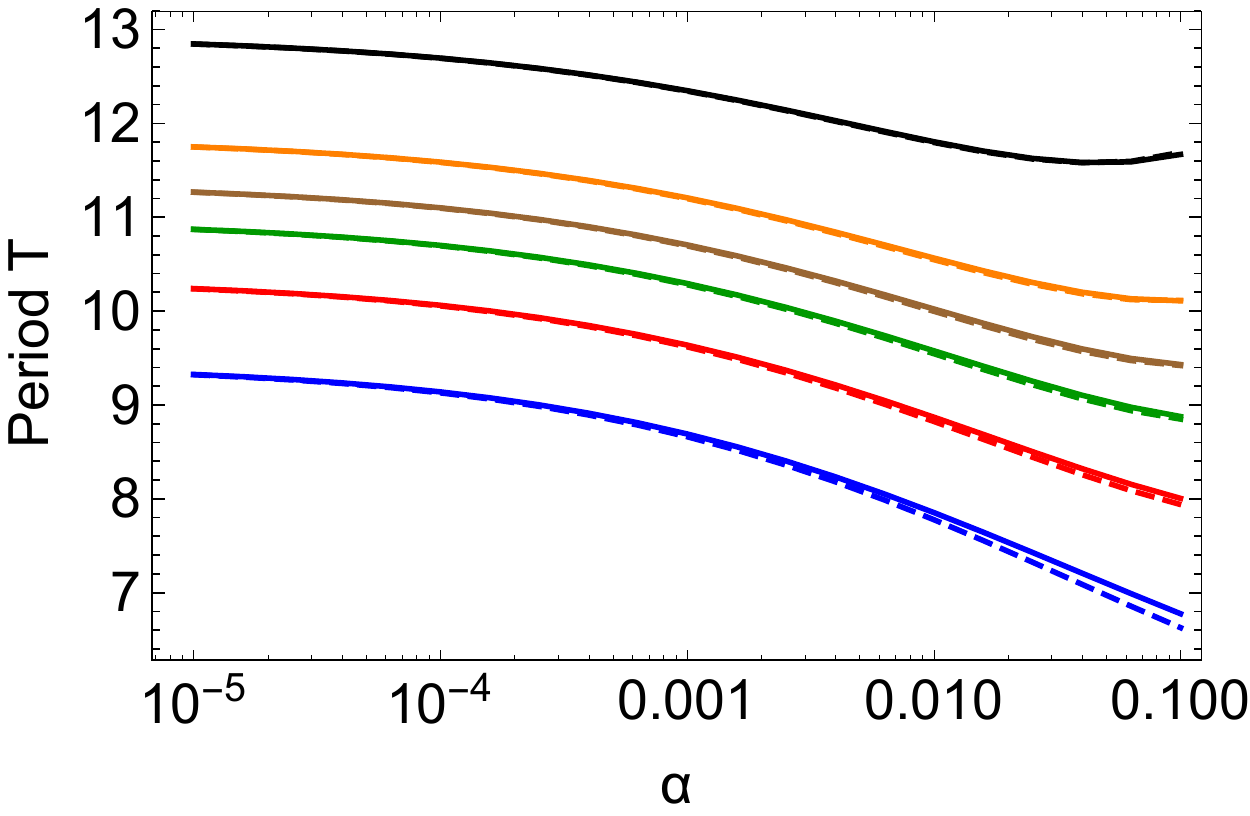} \quad
\includegraphics[width=0.46\textwidth]{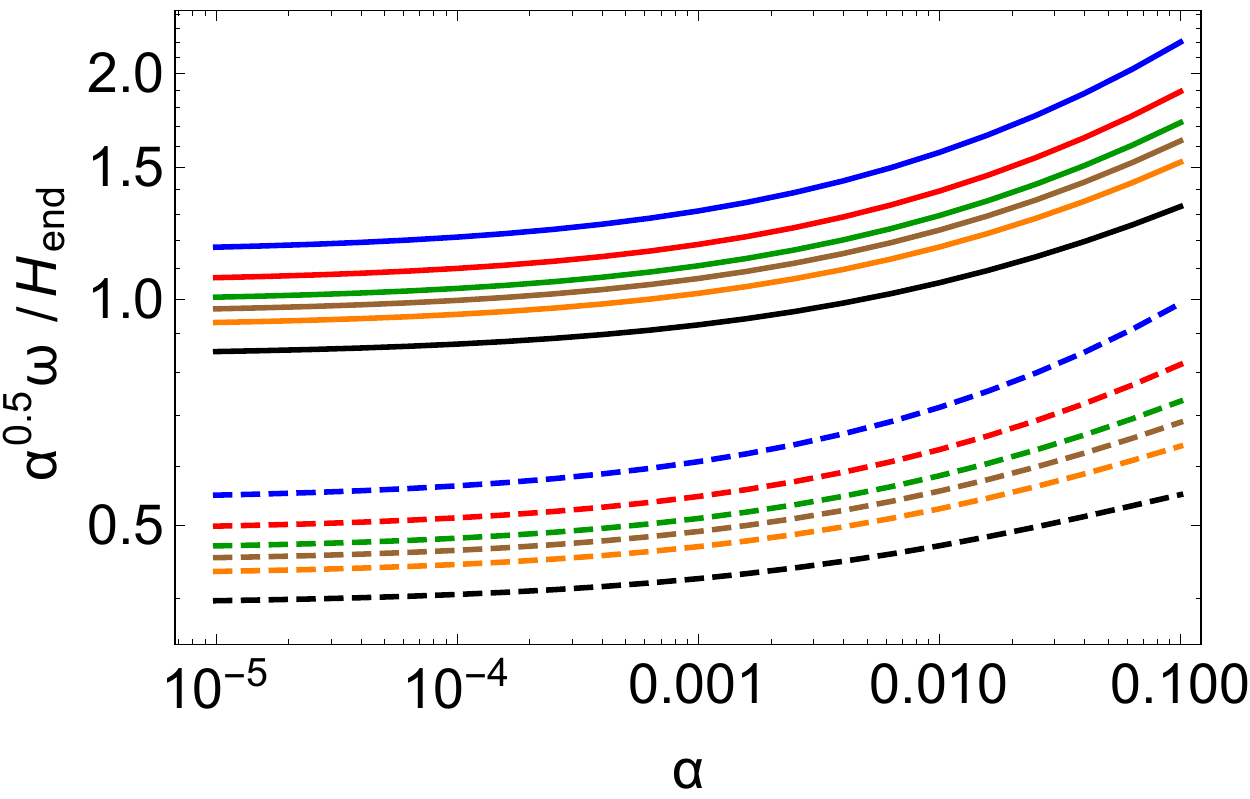}
\caption{
{\it Left:}
The period of background oscillations of the inflaton field  (in units of $\mu^{-1}$) for $H=0$ and $\phi_{\rm max}=\phi_{\rm end}$  as a function of $\alpha$ for $n=1, 1.5, 2, 2.5, 3, 5$ (blue, red, green, brown, orange and black respectively). The solid curves correspond to the period of the E-model, while the dotted ones correspond to the period of the  T-model  divided by $2$.
We see that the period of the T-model is twice that of the E-model to a high degree of accuracy. Both frequencies are largely insensitive to changes in $\alpha$.
{\it Right:}
The frequency of background oscillations $\omega =2\pi/T$ divided by the Hubble scale at the end of inflation, rescaled by $\sqrt{\alpha}$. The solid / dotted curves correspond to the E- and T-model respectively and the color-coding is the same.
It is evident that for small values of $\alpha$, the hierarchy between the background oscillation frequency and the Hubble scale grows as $1/\sqrt{\alpha}$.
}
 \label{fig:periodvsa}
\end{figure}

After inflation, the background field undergoes oscillations with a decaying amplitude, due to Hubble friction. In order to define a characteristic period of oscillations, we neglect Hubble friction and set the field to its value at the end of inflation, as given in Fig.~\ref{fig:phiendvsa}. The results are shown in Fig.~\ref{fig:periodvsa}, where both the period of oscillations $T$ as well as the scale hierarchy $\omega/H_{\rm end}$ is shown.
We see a strong hierarchy between the frequency of background oscillations and the Hubble scale, which gets stronger for smaller values of $\alpha$ (higher field-space curvature), scaling as $\omega  \propto  H_{\rm end}/\sqrt{\alpha}$. This means that for small $\alpha$ the Hubble scale can be neglected, to a good approximation, as it takes a large number of background oscillations for any considerable red-shifting to occur.
We also see that the hierarchy between the oscillation frequency and the Hubble scale is somewhat stronger for the T-model,
 hence we expect more damping of the background motion per oscillation for the E-model. In order to understand the relation between the period of the two models $T_T \simeq 2T_E$ that can be immediately extracted from Fig.~\eqref{fig:phiendvsa} we take a closer look at the single-field potential of the two models and compute one characteristic evolution for $\alpha=10^{-3}$ and $n=1$.

\begin{figure}
\centering
\includegraphics[width=0.45\textwidth]{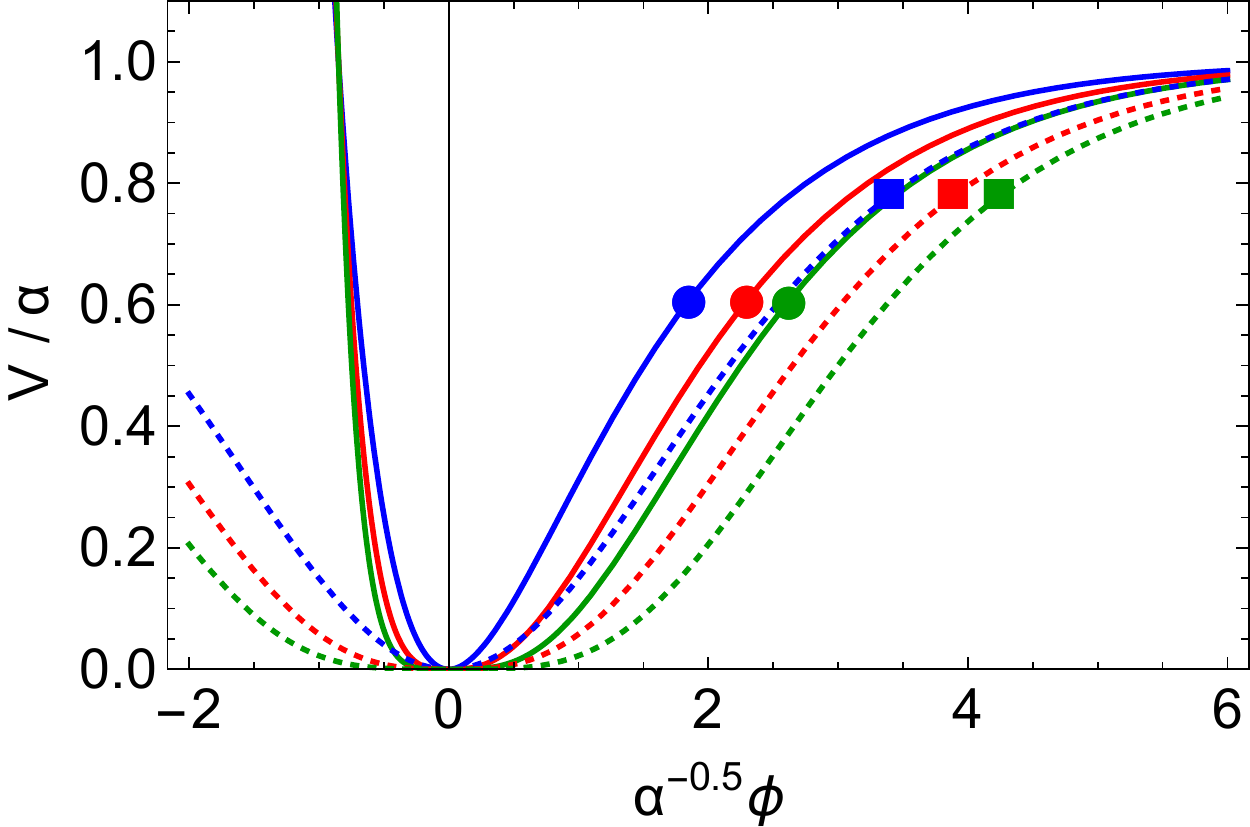} \quad
\includegraphics[width=0.45\textwidth]{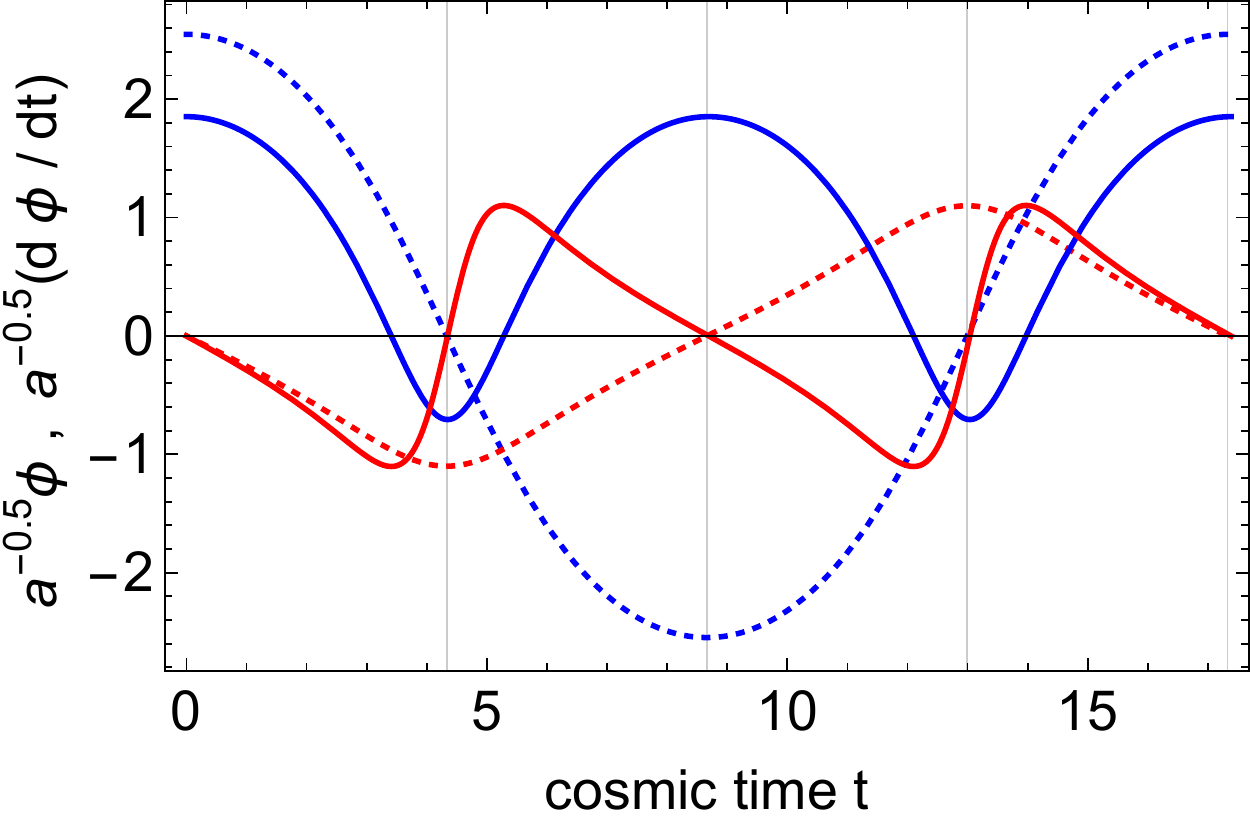}
\caption{
{\it Left:}
The single field potential rescaled by $\alpha$ for $n=1, 1.5, 2$ (blue, red and green respectively). The solid curves correspond to the E-model, while the dotted ones correspond to the  T-model. The dots / squares show $\phi_{\rm end}$ for the E- and T-model respectively for $\alpha=10^{-3}$.
{\it Right:}
The rescaled background motion ($\phi/\sqrt\alpha$ in blue and $\dot\phi/\sqrt\alpha$ in red) for $n=1$ and $\alpha=10^{-3}$ for the E- and T-models (solid / dotted), by neglecting the Hubble friction term.
}
 \label{fig:potential}
\end{figure}

The T-model potential is symmetric with respect to the origin, while the E-model potential is highly asymmetric, consisting of a flat plateau on one side (akin to the T-model) and a steep potential ``wall'' on the other side. One thus expects that the background motion will be equally asymmetric, spending much more time near the plateau ($\phi>0$) and far less time near the steep potential wall ($\phi<0$). This is indeed the case as shown in Fig.~\ref{fig:potential}. Given that the plateau behaviour is similar between the T- and E-models, one would expect that the T-model period would be larger, almost double that of the E-model. If one considers the difference in $\phi_{\rm end}$, the fact that the T-model starts ``higher up on the plateau'' at the end of inflation, the relation $T_T\simeq 2T_E$ ends up being an excellent description of the relation between the background motion of the two models.

\begin{figure}
\centering
\includegraphics[width=0.475\textwidth]{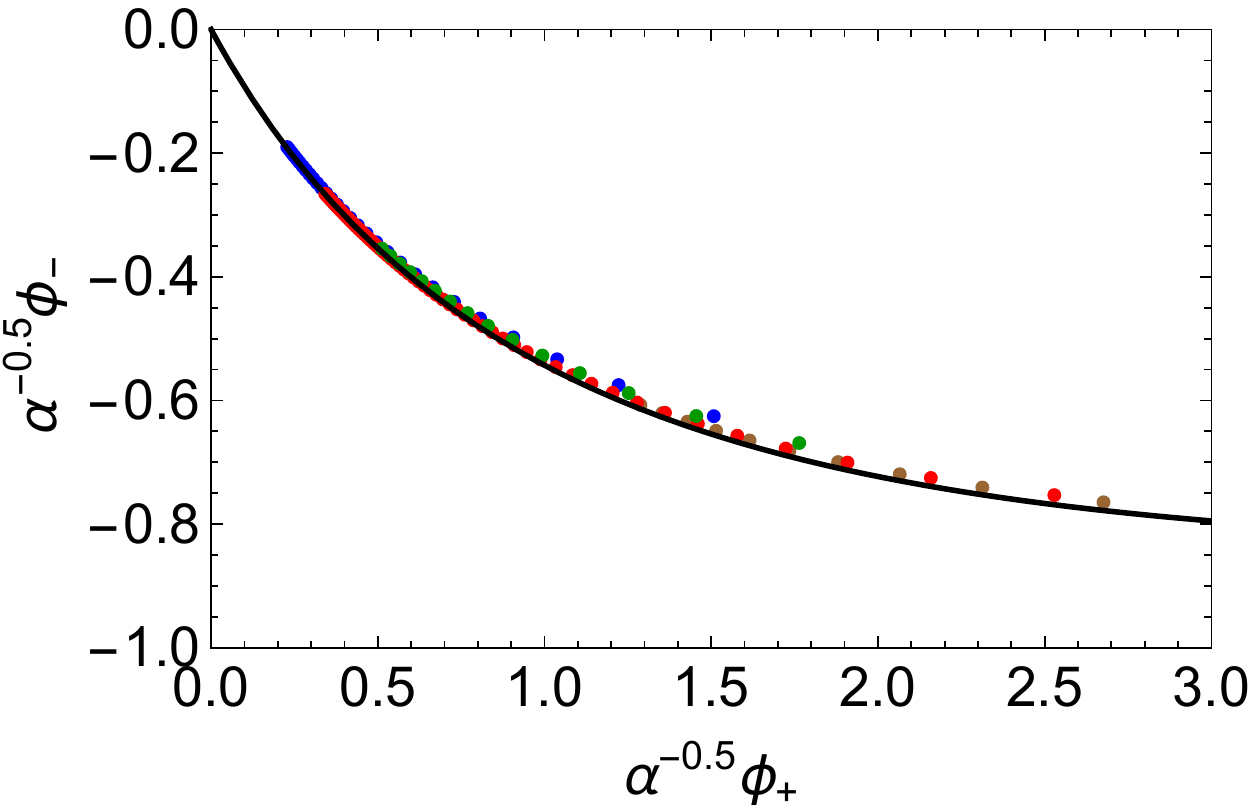} \quad
\includegraphics[width=0.45\textwidth]{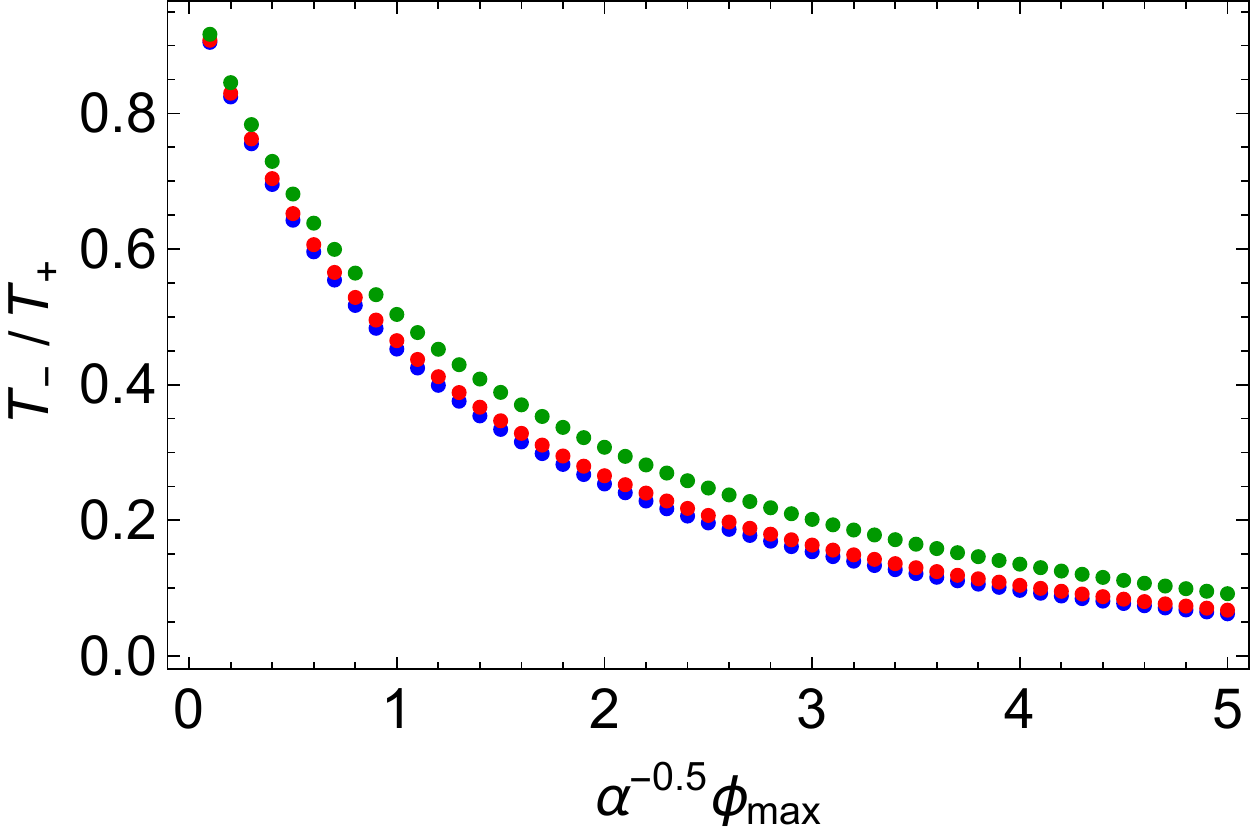}
\caption{
{\it Left:}
The relation of $\phi_+$ to $\phi_-$ for $\{ \alpha ,n \} =\{ 10^{-3} ,1 \} ,\{ 10^{-4} ,1 \} ,\{ 10^{-3} ,3/2 \} ,\{ 10^{-3} , 5\} $ (blue, red, green and black respectively). The black solid curve shows the analytic result of Eq.~\eqref{eq:phiplusphiminus}.
{\it Right:}
 The ratio of the negative to positive half-period as a function of the rescaled field amplitude for $n=1,3/2, 5$ (blue, red and green respectively).  }
 \label{fig:period}
\end{figure}

A simple measure of the asymmetry of the background motion of the E-model seen in Fig.~\ref{fig:potential} can be analytically captured, by computing the ratio of the period for the positive and negative half-cycle. By neglecting the effect of Hubble friction on the background motion, the relation of the two maximum field values $\phi_{\pm }$ ($\phi_+$ being the maximally positive value and $\phi_-$ the  maximally negative value) are given  by
\beq
1-e^{-\beta \phi_{+ }} = -1+e^{-\beta \phi_{- }} \, .
\label{eq:phiplusphiminus}
\eeq
This is independent of the parameter $n$ and is derived through simple conservation of energy for $\dot \phi_{\pm}=0$. We see that the effect of $\alpha$ is trivially given through the rescaling of $\phi_\pm $ by $ \sqrt{\alpha}$.
Fig.~\ref{fig:period} shows that Eq.~\eqref{eq:phiplusphiminus} accurately captures the behaviour of the system for a wide range of parameters. The half-period is then computed as
\beq
T_\pm = \pm \int_{0}^{\phi_{\pm}} {1\over \sqrt{2(V_{\rm max}-V)}}d\phi
\, .
\eeq
Fig.~\ref{fig:period} shows the ratio $T_-/T_+$ for different values of $\phi_+/\sqrt{\alpha}$. As expected, the ratio  approaches unity for small field values, since the field only probes the first (symmetric) term in a Taylor expansion $V \propto |\phi|^n$. For large field-values the asymmetry of the background motion can be very pronounced. Furthermore, the effect of $n$ on the period ratio is not important.

As an interesting remark, we must note that  the first oscillation is larger in amplitude than what one would naively compute by using $\phi_+=\phi_{\rm end}$. By including the kinetic energy at the end of inflation, Eq.~\eqref{eq:phiplusphiminus} becomes
\beq
{3\over 2}\left (1-e^{-\beta \phi_{\rm end }} \right)= -1+e^{-\beta \phi_{- }}
\label{eq:phiendphiminus}
\eeq
for the first half-oscillation. This is especially important for low values of $\alpha$, where the Hubble scale is much smaller than the frequency of oscillation, hence the Hubble damping per oscillation is negligible (at least initially).

\begin{figure}
\centering
\includegraphics[width=0.45\textwidth]{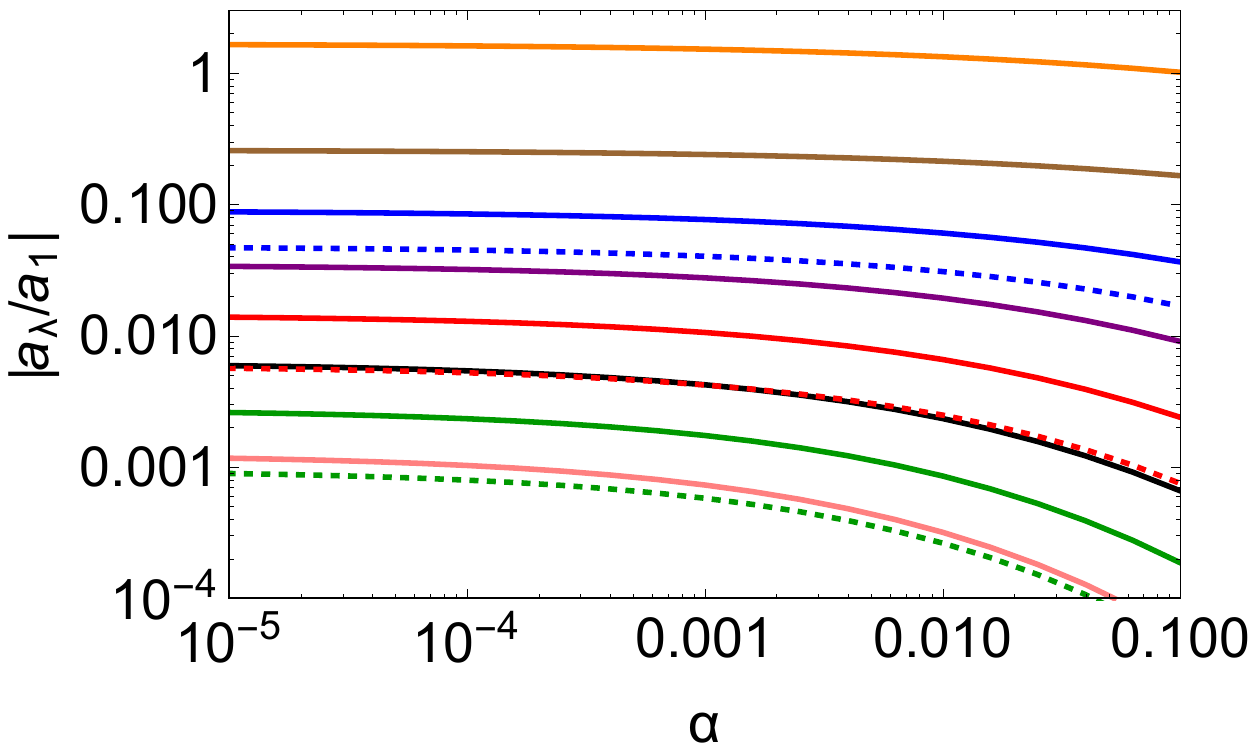} \quad
\includegraphics[width=0.45\textwidth]{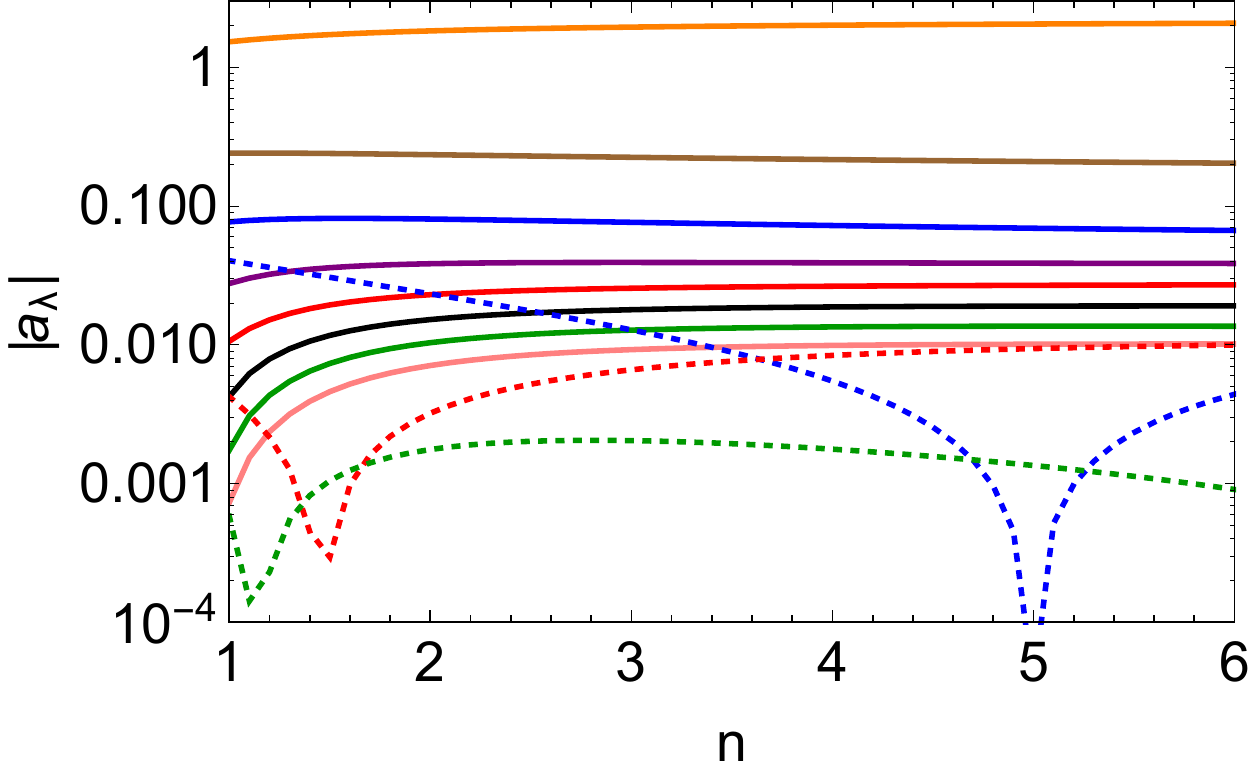}
\caption{
{\it Left:}
The magnitude of the normalized Fourier coefficients $\left | a_\lambda / a_1 \right |$ for $\lambda=0,2,3,4,5,6,7,8$ (orange, brown, blue, purple, red, black, green and pink respectively) for the E-model (solid) and the T-model (dotted) as a function of $\alpha$ with $n=1$.
{\it Right:}
 $\left | a_\lambda  \right |$ with the same color-coding as a function of $n$ with $\alpha=10^{-3}$.
Both panels show that the background motion of the E-model has a richer harmonic structure than that of the T-model. 
For the T-model $a_n\ne 0$ for $n=3,5,7$ as explained in the main text.
 }
 \label{fig:fourier}
\end{figure}

In light of the difference between the background trajectories of the E- and T-model and the highly asymmetric nature of the former, it is interesting to examine the spectral content of $\phi(t)$ in both cases as a function of the  parameters $n$ and $\alpha$. By neglecting the Hubble drag term, the background evolution of the inflaton field is a periodic function, and thus can be written as a Fourier series 
\beq
\phi(t) = a_0 + \sum_{\lambda=1} ^{\infty} a_\lambda  \cos\left ( { 2\pi \lambda\over T} t\right ) + \sum_{\lambda=1} ^{\infty} b_\lambda  \sin\left ( { 2\pi \lambda \over T} t\right )
\eeq
with the Fourier coefficients
\beq
a_0 = {2\over T} \int_0^T \phi(t) dt \, ,\quad a_\lambda = {2\over T} \int_0^T \phi(t) \cos\left ( { 2\pi \lambda\over T} t\right )  dt  \, ,\quad b_\lambda=0
\eeq
We compute the background motion in the static universe approximation ($H=0$)  by setting the initial conditions $\{\phi,\dot \phi\} =\{\phi_{\rm end},0\}$ at $t=0$, where $\phi_{\rm end}$ is the field value at the end of inflation, and numerically solving the Minkowski-space background equation of motion. In this context the coefficients of the sinusoidal terms $\{ b_\lambda \}$ vanish identically for both the E- and T-model, while the Fourier series for the T-model consists of only odd terms: $\{ \alpha_\lambda\}$ with $\bmod(\lambda,2)=1$. Fig.~\ref{fig:fourier} shows the richer spectral content of the E-model as opposed to the T-model.

\subsection{Multi-field effects during inflation}
\label{sec:multi-field-Inflation}

\begin{figure}[h]
\centering
\includegraphics[width=0.45\textwidth]{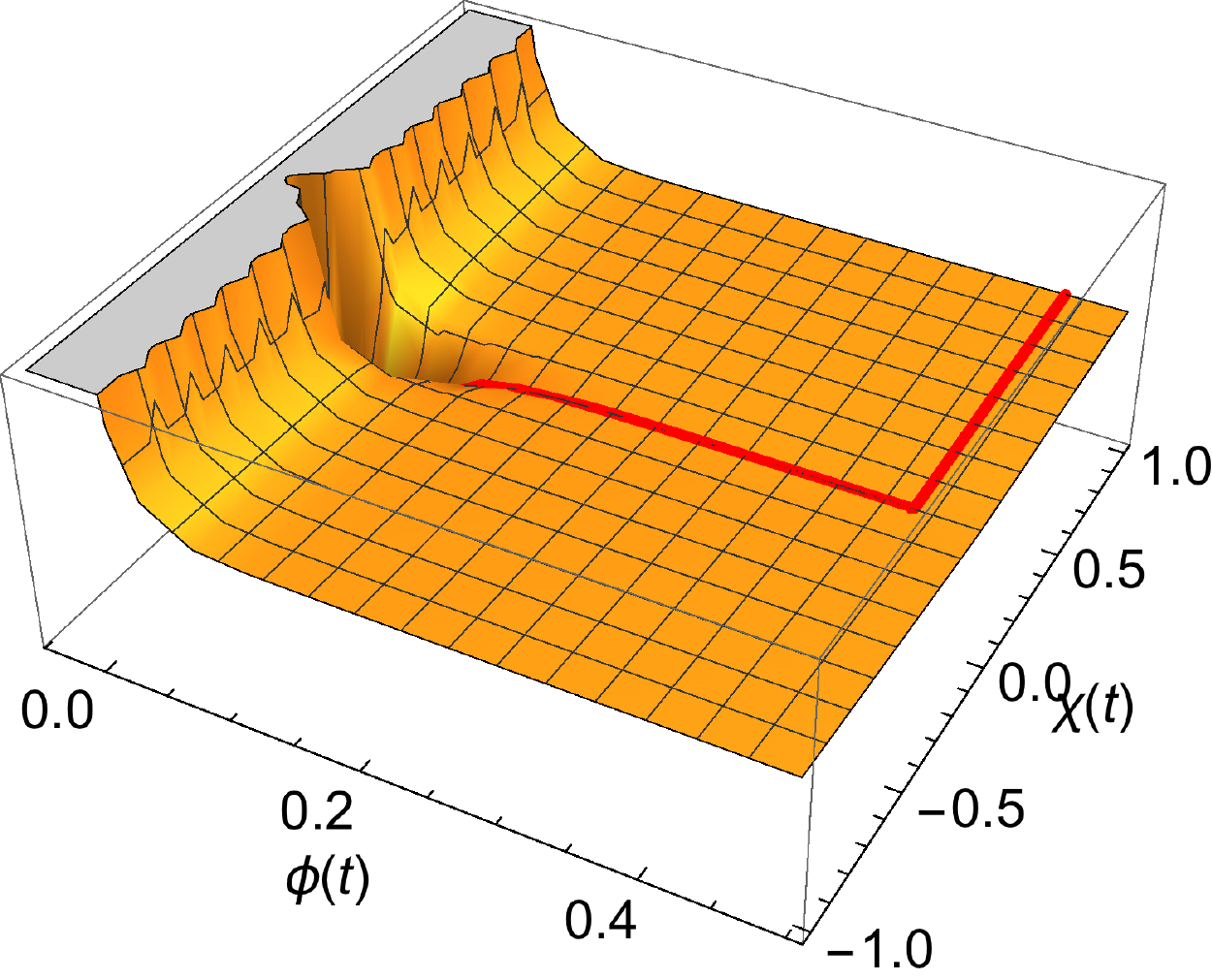} \quad
\includegraphics[width=0.45\textwidth]{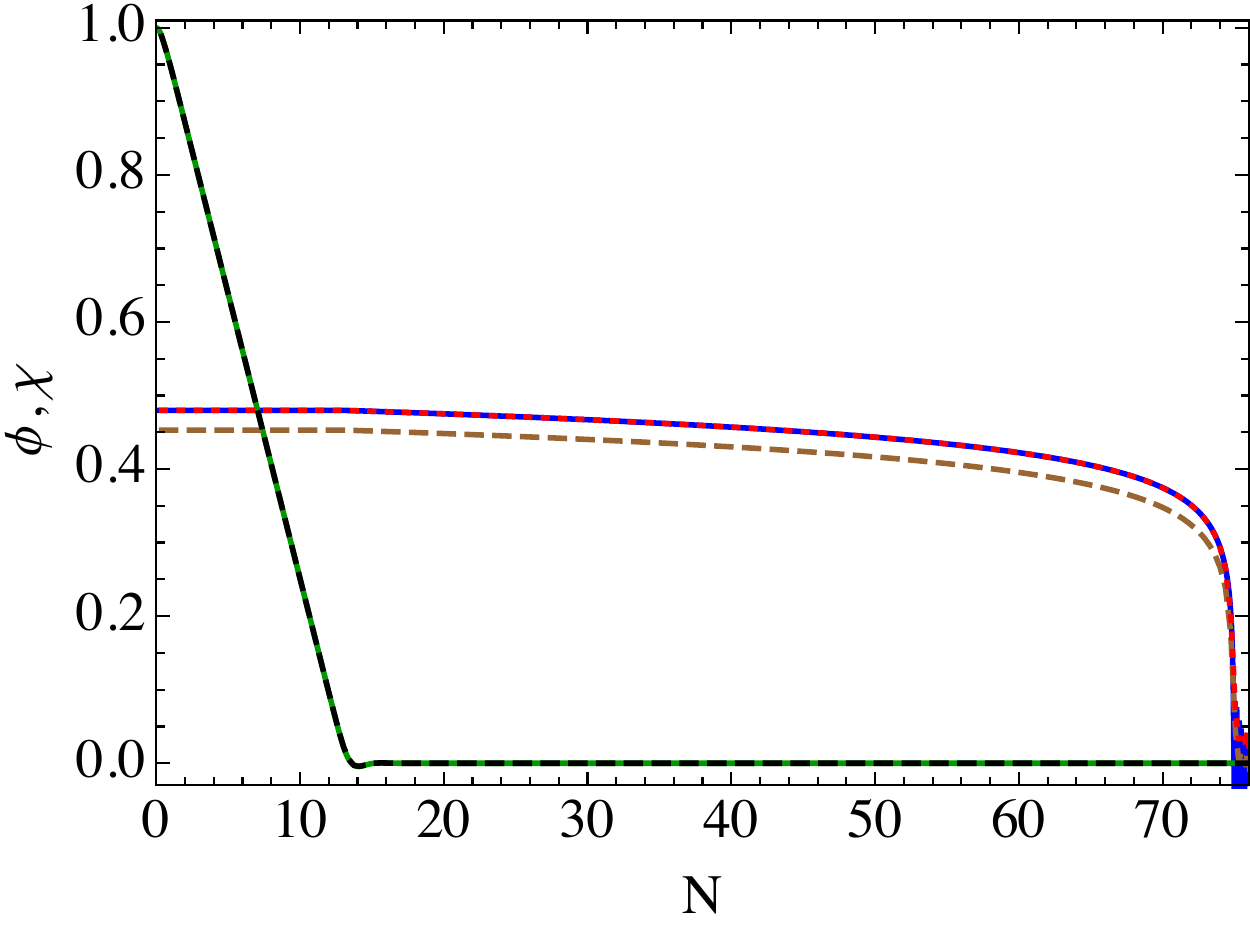}
\caption{
{\it Left:} The three-dimensional plot of the  E-model potential $V(\phi,\chi)$ 
given in Eq.~\eqref{eq:potential}
for $n = 3/2$ and $\alpha = 0.001$ along with a characteristic trajectory computed by choosing the initial conditions $\phi_0={1\over \beta}\log\left ({4 n N \over 3\alpha }\right )$ and $\chi_0=1$.
{\it Right:} The evolution of $\phi$ (blue), $\chi$ (green) for the same parameters. The brown-dashed and black-dashed curves correspond to the T model with the same parameters and the initial conditions chosen as  $\chi_0=1$ and $\phi_0={1\over \beta}\log\left ({8 n N \over 3\alpha }\right )$. The red-dashed curve is $\phi$ for the T-model shifted vertically by $\log(2)/\beta$, following Eq.~\eqref{eq:phiTvsE}.
All field values are measured in units of $M_{\rm Pl}$. 
It is worth noting that the blue-solid and red-dashed curves are indistinguishable, as are the green-solid and black-dashed ones. This demonstrates the identical multi-field behavior of the T- and E-models during inflation, that is derived in the main text.
}
 \label{fig:2fieldtrajectory}
\end{figure}

Similarly to the generalized T-model \cite{Iarygina:2018kee, Carrasco:2015rva, Krajewski:2018moi}, the generalized E-model  exhibits a single valley along $\chi=0$, as shown in Fig.~\ref{fig:2fieldtrajectory}. 
Analogously to Ref.~\cite{Iarygina:2018kee}, we find that by starting away from the $\phi$ axis, inflation will proceed along a single-field trajectory with $\phi$ being effectively constant until $\chi=0$. After that, inflation will proceed along the valley of the potential, as shown in Fig.~\ref{fig:2fieldtrajectory}.
By using the single-field slow-roll equations of motion, we can express the field $\phi$ as a function of the $e$-folding number $N_{{\rm sf},\phi}$ on a single-field trajectory along $\phi$
\beq
V(N_{{\rm sf},\phi},\chi) = \alpha\mu^2 \left [ 1-{2\over \cosh(\beta\chi) } {3\alpha \over 4 n N_{{\rm sf},\phi}} +\left ( {3\alpha \over 4 n N_{{\rm sf},\phi}} \right )^2 \right ]^nn
\cosh^{2/\beta^2} (\beta\chi)
\eeq
By dropping the field value in favor of the $e$-folding number, we gain a more intuitive understanding of the size of each term. Before proceeding, we must stress that $N_{{\rm sf},\phi}$ is the $e$-folding number of a single field trajectory with $\chi=0$, not the full multi-field trajectory, and it is only used as a substitute for the field $\phi$. 
As Fig.~\ref{fig:2fieldtrajectory} shows, the sharp turn in field-space means that the substitution of $\phi$ by $N_{{\rm sf},\phi}$ has physical relevance beyond its mathematical convenience. It corresponds to the duration of inflation that  will take place after the sharp turn at $\chi=0$.
By considering large values of $N_{{\rm sf},\phi}$, such that we get a large number of  $e$-folds ($55$ or more) along a single field trajectory along $\phi$, we can keep the lowest order term in $N_{{\rm sf},\phi}$, which leads to
\beq
V(N_{{\rm sf},\phi},\chi) = \alpha \mu^2 \cosh^{2/\beta^2} (\beta\chi)
\left [
1-{3\alpha\over 2 N_{{\rm sf},\phi}} \text{sech}(\beta\chi)
+{\cal O}\left ( { \alpha^2 \over N_{{\rm sf},\phi}^{2}}\right)
\right ] \, .
\label{eq:seriesexpansion}
\eeq
The next to leading order term ${\cal O}\left ( { \alpha^2 / N_{{\rm sf},\phi}^{2}}\right) $ can be dropped if
\beq
\beta \chi  < \log \left ({16 n N_{{\rm sf},\phi}\over 3\alpha}\right) \, .
\label{eq:seriesvalidity}
\eeq  
We must note again that Eq.~\eqref{eq:seriesexpansion} represents a series expansion in $1/N_{{\rm sf},\phi}$ and holds for every value of $\chi$, within the limits of Eq.~\eqref{eq:seriesvalidity}.
By applying the same procedure to the two-field generalized T-model that was studied in Ref.~\cite{Iarygina:2018kee}, we arrive at the exact same series expansion, up to and including the term that is ${\cal O}\left (N_{{\rm sf},\phi}^{-1} \right )$. The two potentials are different at the level of the  ${\cal O}\left (\alpha^2 / N_{{\rm sf},\phi}^{2} \right )$ term. This clearly shows that the two potentials are not only equivalent during inflation at the single-field level, leading to the same predictions for $n_s$ and $r$, but that their two-field behaviour is also identical, up to slow-roll corrections, since the potential along the $\chi$ direction is the same up to ${\cal O}(N^{-2}) = {\cal O}(\epsilon)$ terms. The equivalence becomes increasingly better for smaller values of $\alpha$.
 Hence the approach to the $\chi=0$ attractor, which was examined in Ref.~\cite{Iarygina:2018kee} for a wide range of initial conditions, as well as the behaviour along the attractor, will be practically indistinguishable between the two models. We must note that the above analysis does not provide any guarantee that this equivalence will persist during preheating, since it has been obtained by using the slow roll analysis during the early (CMB-relevant) stages of inflation.

Both the above analysis and the more extended multi-field analysis shown in Ref.~\cite{Iarygina:2018kee}, was performed in the $\{\phi,\chi\}$ basis. However, for curved field-space manifolds, the magnitude of a field value does not always correspond to the physically relevant parameter. For that, we can look for intuitive criteria to check the strength of the single-field attractor and the two-stage structure of the inflationary trajectory shown in Fig.~\ref{fig:2fieldtrajectory}. One such criterion for testing the strength of the late-time single field attractor arises as we
vary the field values on equi-potential surfaces $V(\phi,\chi)={\rm const.}$

\begin{figure}
\centering
\includegraphics[width=0.45\textwidth]{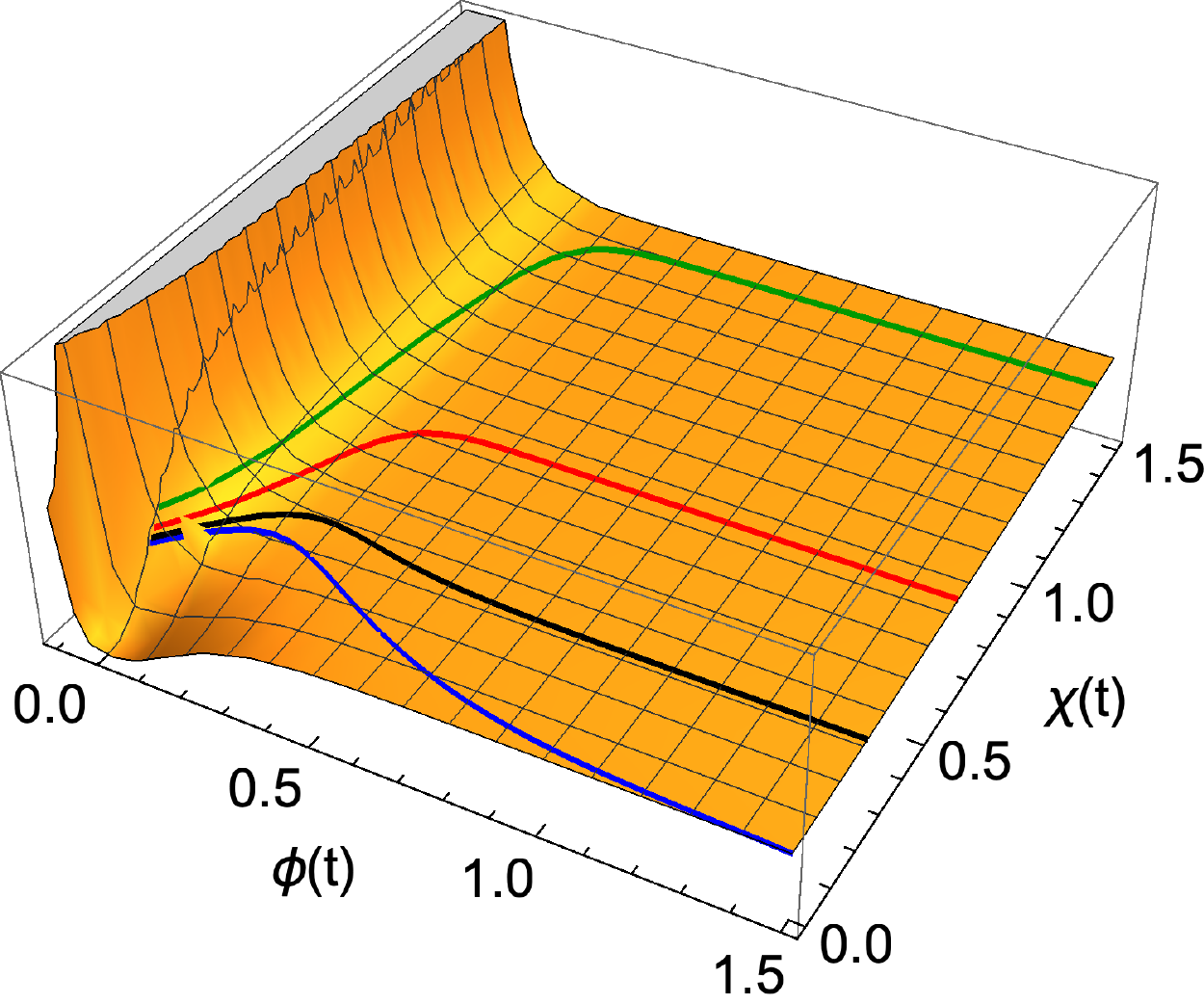} \quad
\includegraphics[width=0.45\textwidth]{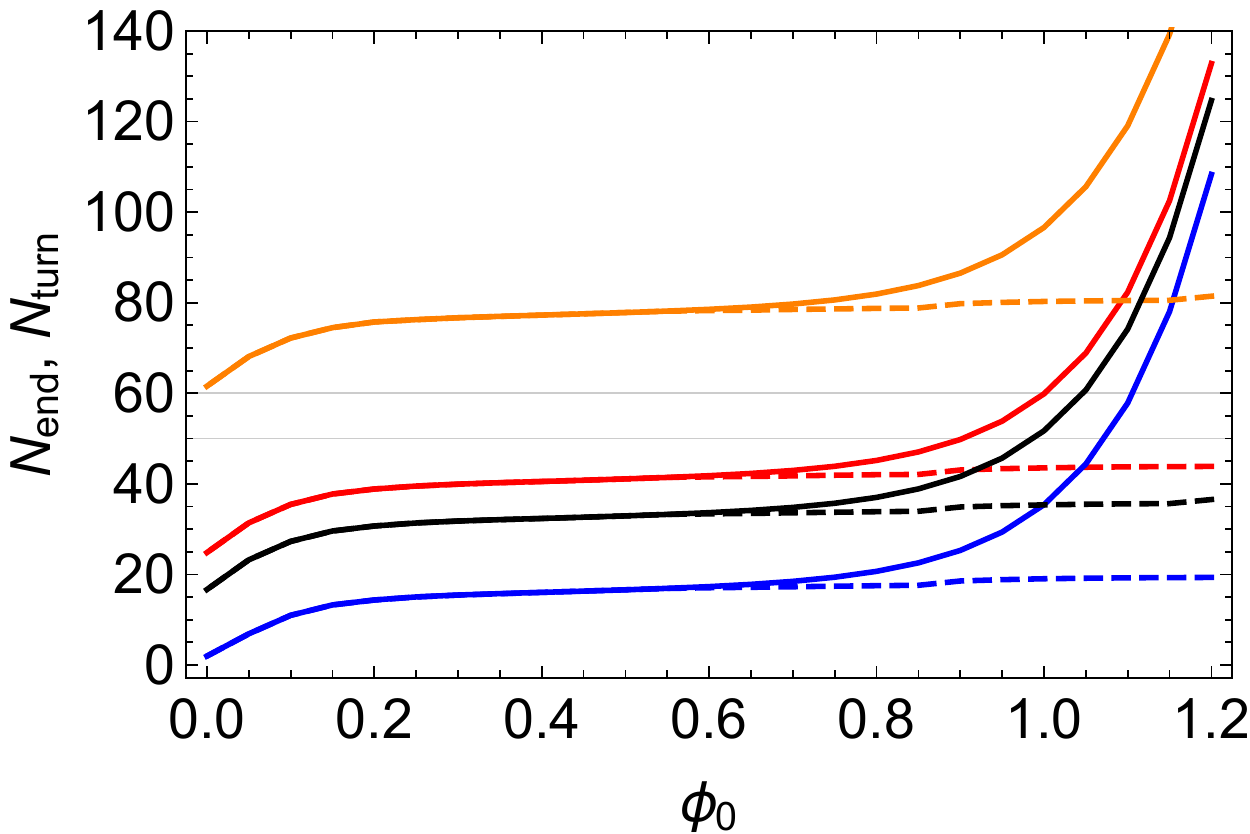}
\caption{
{\it Left:} The initial value lines for constant potential for $\phi_0$=1.5 and $\chi_0$=1.4 (green),  $\chi_0$=0.7 (red),  $\chi_0$=0.3 (black), $\chi_0$=0 (blue) (from top curves to bottom) for $n=3/2$ and $\alpha=0.01$.
\\
{\it Right:}
The total number of e-folds (solid lines) and the number of e-folds before the sharp turn (dashed lines) starting from the beginning of inflation for $\phi_0$=1.2 and  $\chi_0$=19 (orange), $\chi_0$=10 (red), $\chi_0$=8 (black), $\chi_0$=4 (blue),  (from top curves to bottom) for $n=3/2$ and $\alpha=0.01$. The two horizontal thin lines correspond to $50$ and $60$ $e$-folds, hence the range between them corresponds to the time, during which the CMB-relevant modes left the horizon.
All field values are measured in units of $M_{\rm Pl}$. 
}
 \label{fig:eqipotentialEfolds}
\end{figure}

 Given the fact that we know the behaviour and observables of the system, once it reaches the single field attractor at $\chi(t)\simeq 0$, we
examine the duration of inflation and the position of the sharp transition between the two single-field regimes as we fix the initial energy $V(\phi_0,\chi_0)$ for $\dot\phi_0=\dot\chi_0=0$. 
We start by fixing the potential energy of the initial conditions, which corresponds to the left panel of Fig.~\ref{fig:eqipotentialEfolds}. The initial potential energy is taken to be $V(\phi_0,\chi_0)$. We see that the lines are equidistant to each other before the turn of the trajectory happens. The total number of e-folds and the number of e-folds before the turn are sensitive to the change of initial values of $\phi_0$ and $\chi_0$. From the right panel of the Fig.~\ref{fig:eqipotentialEfolds} we see that to get 60 e-folds of inflation we must have $\phi_0\gtrsim 1.1$. With the increase of $\chi_0$ the number of e-folds before the turn increases as well. For the equi-potential choice of initial conditions the subtraction from the duration of inflation the position of the turn, i.e. $N_{\text{end}}-N_{\text{turn}}$, is the same for all parameters  $\phi_0,\chi_0$.

Intruiging phenomenology can arise if one puts the evolution of $\chi(t)$ into the observable range, i.e. let it evolve at least 30 e-folds before the turn. To make it happen for $\alpha=0.01$ we have to artificially tune $\chi_0$ to be $\chi_0\approx 10$, at the same time keeping $\phi_0 ={\cal O}(1)$, however for $\alpha=0.001$ both $\phi_0$ and $\chi_0$ can be of the same order ${\cal O}(1)$. Using the two-field potential of Eq.~\eqref{eq:potential}, we can compute the slow-roll quantities during the initial phase of inflation along $\phi\simeq {\rm const}$. We use the fact that the field trajectory proceeds with almost zero turn-rate, hence the projection vectors align with the coordinate system, $\hat \sigma^\phi\simeq 0$. This greatly simplifies the calculations (we use the notation of Ref.~\cite{KMS}), since the motion occurs along $\chi$, which is a canonically normalized field. The slow roll quantities along the adiabatic direction are
\beq
\epsilon = -{\dot H \over H^2} \simeq {M_{\rm Pl}^2\over 2 } \left( {V_\chi\over V}\right)^2
\, , \quad \eta_{\sigma\sigma} \simeq M_{\rm Pl}^2 {{\cal M}_{\sigma\sigma}\over V}
\eeq
where the adiabatic effective mass along the $\chi$ direction is
\beq
{\cal M}_{\sigma\sigma} \simeq {\cal G}^{\chi\chi} ({\cal D}_\chi{\cal D}_\chi V) = V_{\chi\chi}
\eeq
It is straighforward to compute the above quantities. Interestingly they both asymptote to a fixed value for $\{\phi,\chi\} \gtrsim {\cal O}(1)$, which reads
\beq
\epsilon \simeq 3\alpha\, , ~\eta_{\sigma\sigma} \simeq 2\epsilon \simeq 6\alpha
\eeq
This result is insensitive to the exact value of $\alpha$ and $n$ and it is identical for the E- and T-model.
The orthogonal direction, which in this case is the $\phi$ direction, controls the evolution of the isocurvature modes. It is straightforward to check that the isocurvature effective mass in this case is larger than the Hubble scale, hence the isocurvature modes decay. The curvature perturbation is thus controlled by the $\chi$ fluctations
that exit the horizon during this stage, which acquire a spectral tilt
\beq
n_s = 1-6\epsilon+2\eta_{\sigma\sigma} \simeq 1-6\alpha.
\eeq
This can be made compatible with the {\it Planck} data. However, the tensor to scalar ratio $r=16\epsilon\simeq 48\alpha$ is too large, $r>0.1$, for values of $\alpha$ that provide the correct scalar spectral index. These results use the asymptotic values of $\epsilon$ and $\eta_{\sigma\sigma}$ and a region of (almost) zero turn rate $|\omega| \ll H$.
The existence of a non-zero turn rate during this first stage of inflation can lower the tensor-to-scalar ratio (see e.g. Ref.~\cite{heavy2}). A full calculation of the power spectrum during the transition between the two (almost) single field trajectories requires a more thorough investigation, possibly focusing on a different parameter range than the ones associated with efficient preheating ($\alpha\ll 1$).

A full analysis of the initial condition dependence that defines the observables of the two-stage inflationary phase and the corresponding observational viability of two-stage $\alpha$-attractor inflation is
 beyond the scope of the present work. However Fig.~\ref{fig:eqipotentialEfolds} shows that if one wants to extract information about the probability distribution of the inflationary trajectory and the resulting spectral observables, one would need to choose a prior distribution for the initial values of the fields (and corresponding velocities). Our  intuitive choice for choosing initial conditions through iso-potential lines, shows that the choice of prior distribution is likely to affect the outcome (see e.g. Ref.~\cite{Christodoulidis:2019hhq}). Even though the single field attractor is strong enough to suppress multi-field signatures, the size of the part of parameter space that would showcase them is non-trivial to compute.

\bigskip

\section{Fluctuations}
\label{sec:Fluctuations}

In principle, the analysis of fluctuations in models of inflation that involve multiple fields on a curved manifold requires the use of a covariant formalism. This has been developed for preheating in Ref.~\cite{DeCross:2015uza} and extensively used in Refs.~\cite{DeCross:2016fdz, DeCross:2016cbs, Sfakianakis:2018lzf} for studying preheating in multi-field inflation with non-minimal couplings to gravity. Our current parametrization of the hyperbolic field-space manifold (see Appendix A) makes the equations for the gauge-invariant perturbations
\beq
Q^I \equiv \delta\phi^I + {\dot \phi^I\over H}  \psi
\eeq
particularly simple along the single-field attractor $\chi=0$. Their equations of motion are
 \beq
\ddot Q^I + 3 H \dot Q^I + \left[ \frac{ k^2}{a^2} \delta^I_{\> J} + {\cal M}^I_{\> J} \right] Q^J = 0 ,
\label{eq:eomQ}
\eeq
where the mass-squared matrix is defined as
\beq
{\cal M}^I_{\> J} \equiv {\cal G}^{IK} \left( {\cal D}_J {\cal D}_K V \right) - {\cal R}^I_{\> LMJ} \dot{\varphi}^L \dot{\varphi}^M - \frac{1}{M_{\rm pl}^2 a^3} {\cal D}_t \left( \frac{ a^3}{H} \dot{\varphi}^I \dot{\varphi}_J \right)
\label{MIJdef}
\eeq
and ${\cal R}^I_{\> LMJ}$ is the Riemann tensor constructed from ${\cal G}_{IJ} (\varphi^K)$.

By rescaling the perturbations as $Q^I (x^\mu) \rightarrow X^I (x^\mu) / a(t)$ and
 working in terms of conformal time, $d \eta = dt / a(t)$, we write the second order action in a form that resembles Minkowski space
\beq
S_2^{(X)} = \int d^3x d\eta \left [ -{1\over 2} \eta^{\mu\nu} \delta_{IJ} \partial_\mu X^I \partial_\nu X^J -{1\over 2}\mathbb{M}_{I J} X^I X^J \right ] \, ,
\eeq
where
\beq
\mathbb{M}_{I J} = a^2 \left ( {\cal M}_{IJ} - {1\over 6} \delta_{IJ} R\right)
\eeq
This makes  quantization  straightforward, by promoting the fields $X^I$ to operators $\hat X^I$
and expanding $\hat{X}^\phi$ and $\hat{X}^\chi$ in sets of creation and annihilation operators and associated mode functions
\beq
\hat X^I = \int {d^3k \over (2\pi)^{3/2}} \left [u^I(k,\eta)\hat a e^{i k\cdot x} +  u^{I*}(k,\eta)\hat a^\dagger e^{-i k\cdot x} \right ] \, .
\eeq
Since the modes decouple on a single-field background with vanishing turn-rate, the equations of motion are
\beqn
\begin{split}
\partial^2_\eta{v}_k & + \Omega_\phi^2(k,\eta) v_k \simeq 0 \, , \quad \Omega_\phi^2(k,\eta)=k^2 + a^2 m_{\rm eff, \phi}^2 \, ,
 \\
\partial^2_\eta{z}_k & + \Omega_\chi^2(k,\eta)z_k \simeq 0 \, , \quad \Omega_\chi^2(k,\eta)=k^2 + a^2 m_{\rm eff, \chi}^2 \, ,
\end{split}
\label{eq:vzeom}
\eeqn
where we defined $u^\phi \equiv v$ and $u^\chi \equiv z$.
The effective masses of the two types of fluctuations, along  the background motion and perpendicular to it, consist in principle of four distinct contributions \cite{DeCross:2015uza}:
\beqn
 m^2_{{\rm eff},\phi}\equiv m_{1,\phi}^2+m_{2,\phi}^2+m_{3,\phi}^2+m_{4,\phi}^2
 \\
  m^2_{{\rm eff},\chi}\equiv m_{1,\chi}^2+m_{2,\chi}^2+m_{3,\chi}^2+m_{4,\chi}^2
 \eeqn
each of them corresponding to a different source. Full expressions for arbitrary ${\cal G}_{IJ}$ can be found for example in Ref.~\cite{DeCross:2015uza}. However using the fact that $\chi=0$ and ${\cal G}_{IJ} = \mathbb{I}$ along the single field attractor at background level the effective mass components become simple:
\begin{itemize}
\item The components $m^2_{2,I}$ are written as
\beq
m_{1,\phi}^2 = V_{\phi\phi} \, , \quad m_{1,\chi}^2 = V_{\chi\chi}
\eeq
corresponding to the local curvature of the potential.

\item The component $m^2_{2,\phi}$ vanishes identically, while
\beq
m^2_{2,\chi}={1\over 2} R \dot\phi^2
\eeq
arises from the field-space curvature and has no analogue in flat field-space models.

\item The component $m^2_{3,\phi}$ encodes the effects of the coupled metric perturbations and is written as
\beq
m_{3,\phi}^2= - {1\over M_{\rm Pl}^2 a^3} {\cal D}_t \left ( {a^3\over H} \dot \phi^2 \right ) \, .
\eeq
Since the metric perturbations are only related to the adiabatic perturbations and cannot affect the isocurvature modes, the term $m^2_{3,\chi}$ vanishes identically\footnote{During inflation, the adiabatic modes are fluctuations along the background trajectory and the isocurvature modes are fluctuations perpendicular to it. Due to the existence of a single-field attractor $\chi=0$, the adiabatic and isocurvature modes can be simply matched to $\delta\phi$ and $\delta\chi$ respectively.}. Furthermore, this contribution is subdominant for these models and parameter range of interest, as discussed in Refs.~\cite{Krajewski:2018moi, Iarygina:2018kee}.
\item Finally the terms
\beq
m_{4,\phi}^2=m_{4,\chi}^2= -{1\over 6} R \, ,
\eeq
where $R=6(2-\epsilon)H^2$ is the space-time Ricci scalar, arise from our choice of mode-functions in a  curved space-time.
\end{itemize}
It is straightforward to check that the potential components of the effective masses scale as $m_{1,I}^2 \sim \mu^2$, as does the field-space curvature component $\mu_{2,\chi}^2$. The coupled metric perturbations component is subdominant for $\alpha \ll 1$, since $m_{3,\phi}^2 \sim \mu^2 \sqrt{\alpha}$ (see Ref.~\cite{Iarygina:2018kee}). Finally, the term that encodes the space-time curvature is even smaller, scaling as $m_{4,I}^2\sim \mu^2 \alpha$. This is reminiscent of another family of plateau models,  $\xi$-attractors, which produce similar CMB spectra to $\alpha$-attractors \cite{KS, DeCross:2015uza,  Galante:2014ifa}.

Before we proceed with preheating calculations, we must revisit the claims made in Sections.~\ref{sec:multi-field-Inflation} about the existence and stability of a single-field attractor along $\chi=0$. The analysis made so far relies on background quantities. However, it has been shown that negatively curved manifolds can lead to unstable fluctuations during inflation and a subsequent destabilization of the inflationary trajectory. 
After the system has settled into the attractor at $\chi=0$, the effective super-horizon mass of $\chi$ fluctuations is given by
\beqn
m_{\chi,{\rm eff}}^2 &=& V_{\chi\chi}(\chi=0) +{1\over 2} R \dot\phi^2
\nonumber
\\
&=&2 \alpha  e^{-2 \beta  \phi } \left(e^{-2 \beta  \phi } \left(e^{\beta  \phi }-1\right)^2\right)^{n-1} \left(e^{\beta  \phi } \left(e^{\beta  \phi }+\beta ^2 n-2\right)+1\right) - {2\over 3\alpha }\dot\phi^2
\label{eq:meffinfl}
\eeqn
By using the slow-roll equations of motion this becomes
\beq
m_{\chi,{\rm eff}}^2 \simeq \left ( 2+{1\over N} \right ) \alpha
\eeq
for small $\alpha$ and  large $N$. This means that until close to the end of inflation, where the slow-roll expressions break down, the $\chi$ fluctuations exhibit a  positive effective mass and are suppressed. The E-model is thus safe from ``geometric destabilization'' effects during the inflationary stage along $\chi=0$ \cite{Renaux-Petel:2015mga, Renaux-Petel:2017dia}, even for highly curved field-space manifolds. This arises because the potential also depends on the curvature parameter $\alpha$. We verified this claim by numerically evaluating Eq.~\eqref{eq:meffinfl} for various choices of $n$ and $\alpha$.

Eqs.~\eqref{eq:vzeom} allow for a simple connection to the Bunch-Davies vacuum during inflation
\beqn
v_k \sim {1\over \sqrt{2 \Omega_\phi(k,\eta)} }e^{-i \int \Omega_\phi(k,\eta) d\eta}
\, , \quad
z_k \sim {1\over \sqrt{2 \Omega_\chi(k,\eta)} }e^{-i \int \Omega_\chi(k,\eta) d\eta}
\eeqn
We will perform our numerical computations of the mode evolution during preheating 
in cosmic rather than conformal time.
The corresponding equations of motion are
\beqn
\begin{split}
\ddot {v}_k & +H \dot v_k+
 \omega_\phi^2(k,\eta) v_k \simeq 0 \, , \quad \omega_\phi^2(k,\eta)={k^2 \over a^2}+  m_{\rm eff, \phi}^2 \, ,
 \\
\ddot {z}_k & + H \dot z_k +
\omega_\chi^2(k,\eta)z_k \simeq 0 \, , \quad \omega_\chi^2(k,\eta)={k^2 \over a^2}+  m_{\rm eff, \chi}^2
\end{split}
\label{eq:vzeomt}
\eeqn
for the rescaled fluctuations and
\beqn
\begin{split}
\ddot Q_{\phi,k} & +3H \dot Q_{\phi,k}+ \left (
{k^2\over a^2}+V_{\phi\phi}
\right )Q_{\phi,k}=0
 \\
\ddot Q_{\chi,k} & +3H \dot Q_{\chi,k}+ \left (
{k^2\over a^2}+V_{\chi\chi} + {1\over 2}{\cal R} \dot \phi^2
\right )Q_{\chi,k}=0
\, .
\end{split}
\label{eq:vzeomt}
\eeqn
for the gauge invariant Mukhanov-Sasaki variables, where we neglected $m_{3,\phi}^2$ for simplicity. Also note that $m_{4,\{\phi,\chi\}}^2$ does not appear when the equations are written in terms of the original fluctuations or the Mukhanov-Sasaki variables instead of the re-scaled ones.

We define the energy density in each mode as
\beqn
\rho_{\delta\phi}(k,\eta) = {1\over 2a^4} \left (
|\partial_\eta v_k(\eta)|^2 + \Omega^2_\phi(k,\eta) |v_k(\eta)|^2
\right )
\\
\rho_{\delta\chi}(k,\eta) = {1\over 2a^4} \left (
|\partial_\eta z_k(\eta)|^2 + \Omega^2_\chi(k,\eta) |z_k(\eta)|^2
\right )
\eeqn
where we ignored interaction terms, since we are working in the linear approximation. The  expressions can be easily written in cosmic rather than conformal time.

 We focus primarily on the parametric excitation of $\delta\chi$ modes, since the analysis of single field parametric resonance can be found in the literature (see e.g. Refs.~\cite{Lozanov:2017hjm, Turzynski:2018zup}). For  field-space manifolds with $\alpha \gtrsim {\cal O}(10^{-3})$
  the corresponding instability factors are much smaller for $\delta\phi$ as compared to $\delta\chi$, with the exception of the E-model for $n=1$. Furthermore, the analysis of $\phi$ fluctuations is in principle identical, with the exception that the curvature term is missing from the effective mass.
  We  provide further results for the growth of $\phi$ and $\chi$ fluctuations in Section~\ref{sec:numerics}.

\subsection{Effective frequency}
\label{sec:frequency}

Before we proceed to construct the Floquet charts and numerically compute the evolution of $\chi$ fluctuations,
 we focus on the  effective mass $\omega^2_{\chi}$ and its dependence on the parameters $n$ and $\alpha$.
 This will guide our intuition about the system, so that we can recognize the interesting parameter regimes and important factors that will ultimately determine the preheating efficiency.

We start with the  Riemann component $m_{2,\chi}^2$, which  does not depend strongly on the potential and field-space parameters $n$ and $\alpha$
\beq
{1\over 2} {\cal R}\dot \phi^2 \sim -{\cal O}(1)
\label{eq:tachyonicmass}
\eeq
similarly to the behaviour found in the context of the T-model \cite{Iarygina:2018kee}. This scaling can be simply understood as follows: the field-space curvature is ${\cal R}=-{4\over3\alpha}$. At the same time, the time derivative $\dot\phi^2$ has its maximum value at the minimum of the potential $\phi=0$. Following the analysis of Section~\ref{sec:Model}, we see that the field amplitude scales as $\phi\sim \phi_{\rm end} \sim \sqrt{\alpha}$, while the relevant oscillation time scale $T$ is essentially independent of $\alpha$ for sufficiently small $\alpha$. These scalings have been numerically verified in Figs.~\ref{fig:phiendvsa} and \ref{fig:periodvsa} respectively and lead to $\dot\phi^2 \sim \alpha \sim {\cal R}^{-1}$. The ${\cal O}(1)$ amplitude of the Riemann term in Eq.~\eqref{eq:tachyonicmass} has been numerically evaluated and is maximized close to unity for several parameter choices, especially for $\alpha \ll 1$.

\begin{figure}
\centering
\includegraphics[width=0.45\textwidth]{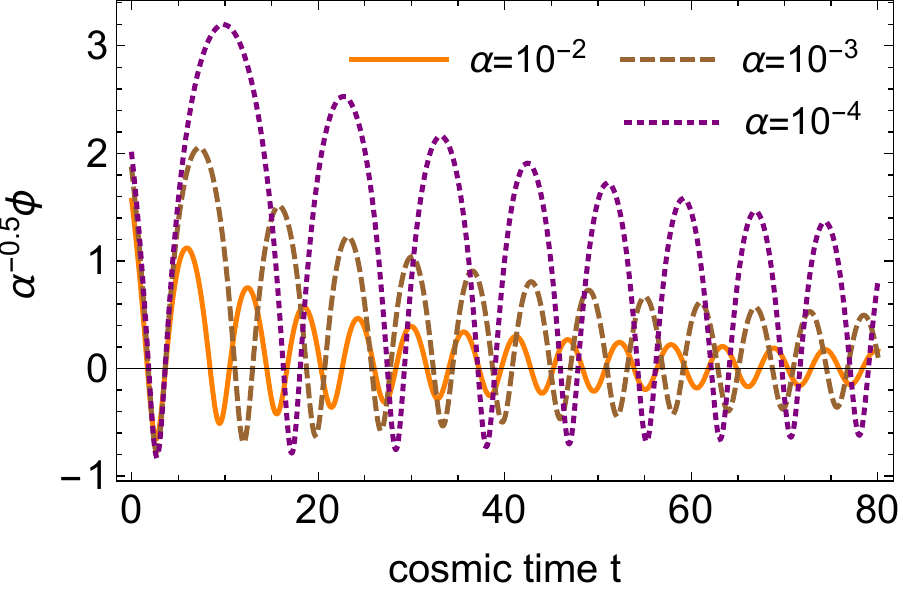}
\includegraphics[width=0.45\textwidth]{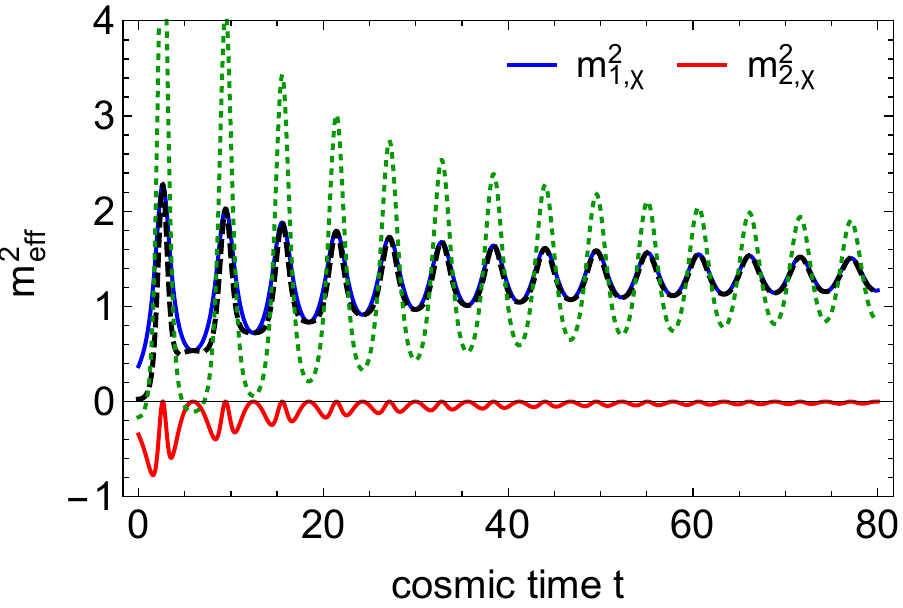}
\includegraphics[width=0.45\textwidth]{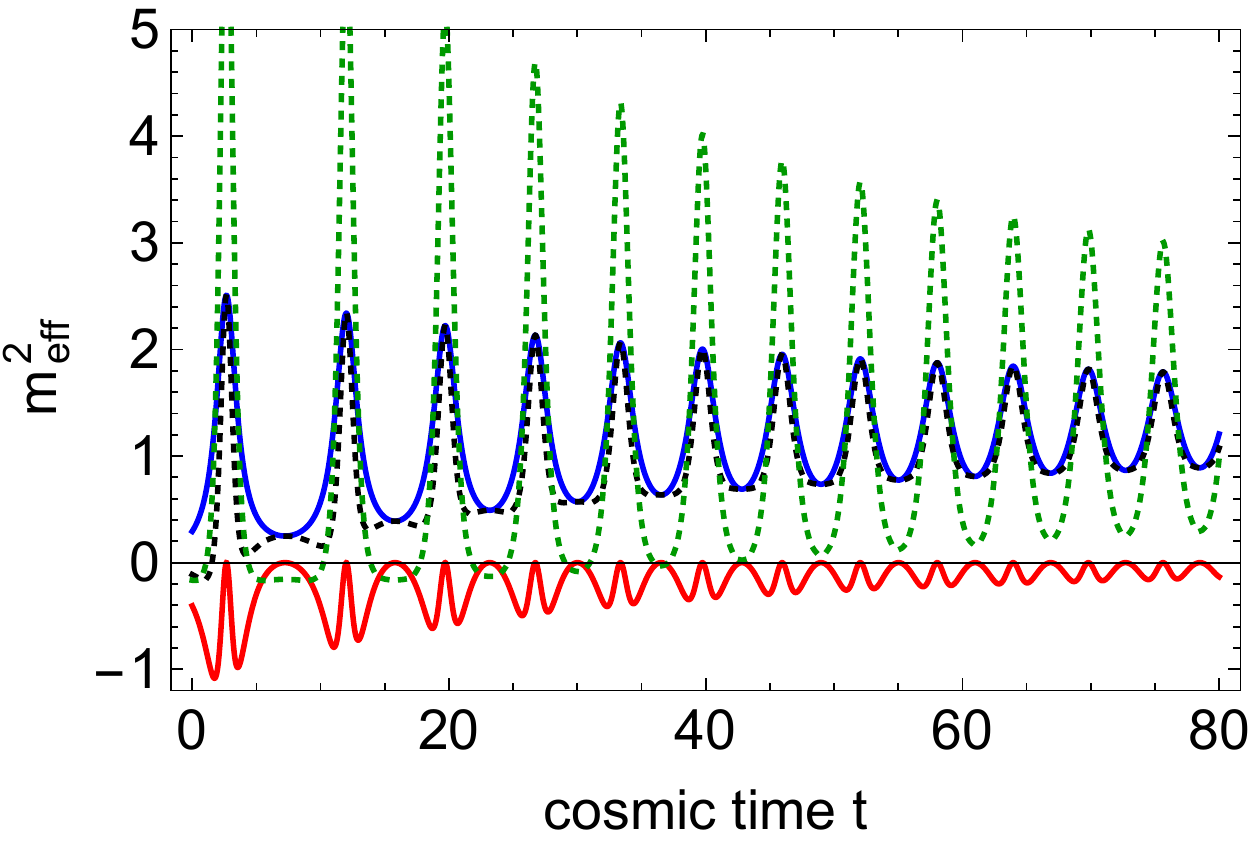}
\includegraphics[width=0.45\textwidth]{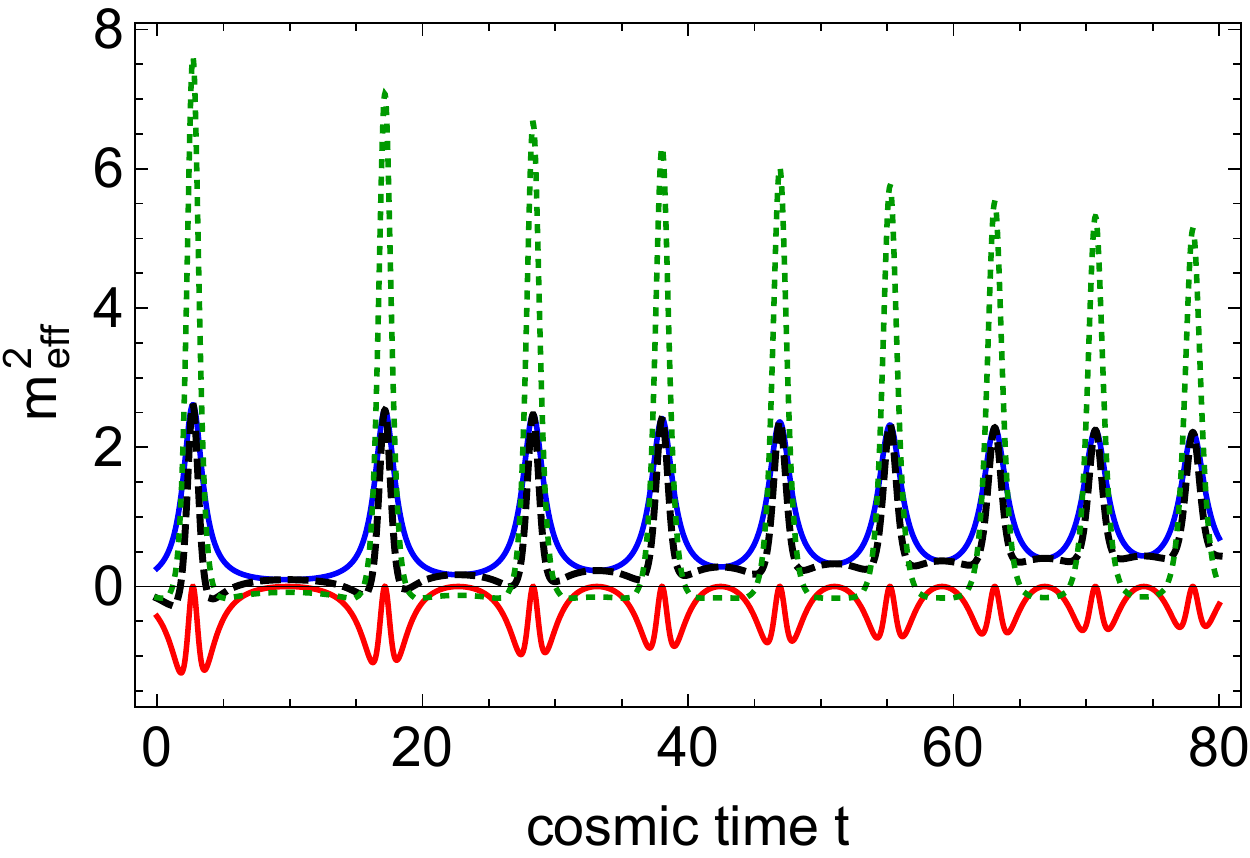}
\caption{{\it Top left:} The rescaled background amplitude of $\phi$ (in units of $M_{\rm Pl}$) for $n=1$ and $\alpha=10^{-2},10^{-3} , 10^{-4}$ (orange, brown-dashed and purple-dotted respectively).
{\it Top right} \& {\it bottom panels:} The effective mass (in units of $\mu^2$) of the $\phi$ and $\chi$ fluctuations (green-dotted and black-dashed) along with the components
  of  $m_\chi^2$ for the same parameters.
 The effective mass components $m^2_{3,\phi}$ and $m^2_{4,\{\phi,\chi\}}$ are not shown, because they are subdominant for $\alpha\ll1$. 
  }
 \label{fig:n1meff}
\end{figure}

We next move to the component $m_{1,\chi}^2$ of the effective $\chi$ mass, which is due to the potential. This can be written for small $\alpha$ as
\beq
V_{\chi\chi}(\chi=0) \simeq \frac{4}{3} n e^{-\beta  \phi } \left(\left(1-e^{-\beta  \phi }\right)^2\right)^{n-1} \, .
\eeq
For $n=1$, this term is simplified as $V^{n=1}_{\chi\chi}(\chi=0) \simeq \frac{4}{3}  e^{-\beta  \phi } $ and oscillates between two extremum values at $\phi=\phi_{\pm}$ (shown in Fig.~\ref{fig:period}),
while having a constant, time-independent value of $4/3$ when the  inflaton field crosses the origin $\phi=0$. The  maximum of $V_{\chi\chi}^{n=1}$ can be easily computed using Eq.~\eqref{eq:phiendphiminus}
\beqn
\left . V^{n=1}_{\chi\chi}\right |_{{\rm max},(1)} \simeq {2\over 3} \left (
5-3 e^{-\beta \phi_{\rm end}} \right )  \, ,
\label{eq:Vchichimaxn1}
\eeqn
where we neglected the effect of the Hubble drag. This is an increasingly good approximation for small values of $\alpha$.
The behaviour of the $\delta \chi$ effective mass is shown in Fig.~\ref{fig:n1meff} for $n=1$ and several values of $\alpha \ll 1$. It is simple to see that the Riemann term red-shifts as $a^{-2}$
, while the potential derivative oscillation amplitude red-shifts as $a^{-1}$. This follows trivially from $\Delta V_{\chi\chi} \sim (1-e^{-\beta\phi}) \sim \sqrt{V}$, 
where $\Delta V_{\chi\chi}$ is the amplitude of the oscillation of $V_{\chi\chi}$ shown in Fig.~\ref{fig:n1meff}, while from the equipartition theorem $\dot\phi^2\sim V$. Hence, both the wave-number contribution $k^2/a^2$, as well as the Riemann contribution become subdominant after the first $e$-fold, which lasts for more oscillations for smaller values of $\alpha$. Using Eq.~\eqref{eq:Vchichimaxn1} and the results $\phi_{\rm end} \lesssim 2 \sqrt{\alpha}$, shown in Fig.~\ref{fig:phiendvsa}, we arrive at $\left . V^{n=1}_{\chi\chi}\right |_{{\rm max},(1)} \lesssim 2.9$. This simple approximation is able to capture the exact (numerical) result shown in Fig.~\ref{fig:n1meff}.

E-model potentials with larger values of the potential parameter $n$ can be analyzed in a similar way. By Taylor-expanding the potential around its global minimum at $\phi=\chi=0$, it is straightforward to see that $V_{\chi\chi}\propto \phi^2 \chi^{2(n-1)}/\alpha^{n-1}$ for $n>1$. Thus all $\chi$ derivatives of the potential vanish for $\phi=0$, contrary to the case of $n=1$. Simply put, potentials with $n>1$ describe massless fields in the small-amplitude regime.

\begin{figure}
\centering
\includegraphics[width=0.45\textwidth]{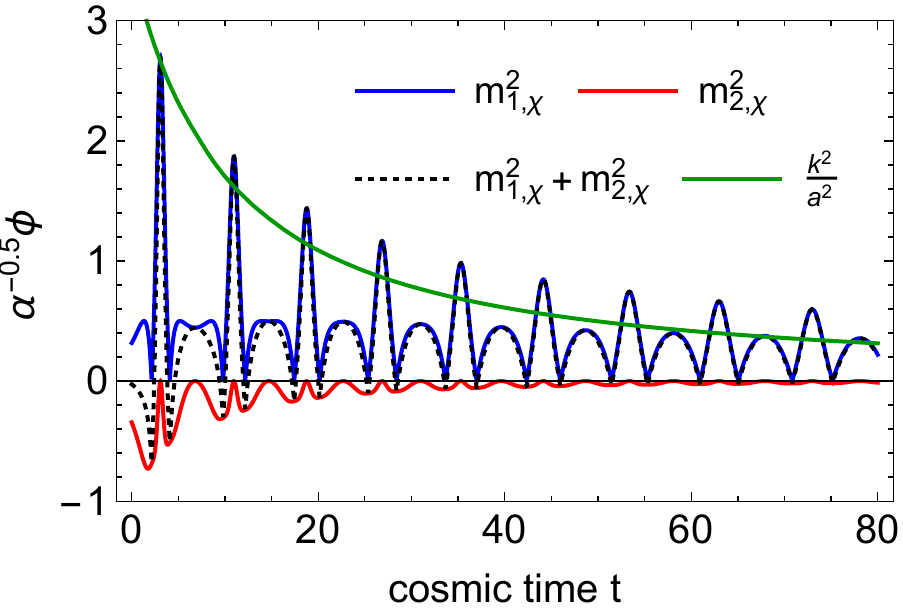}
\includegraphics[width=0.45\textwidth]{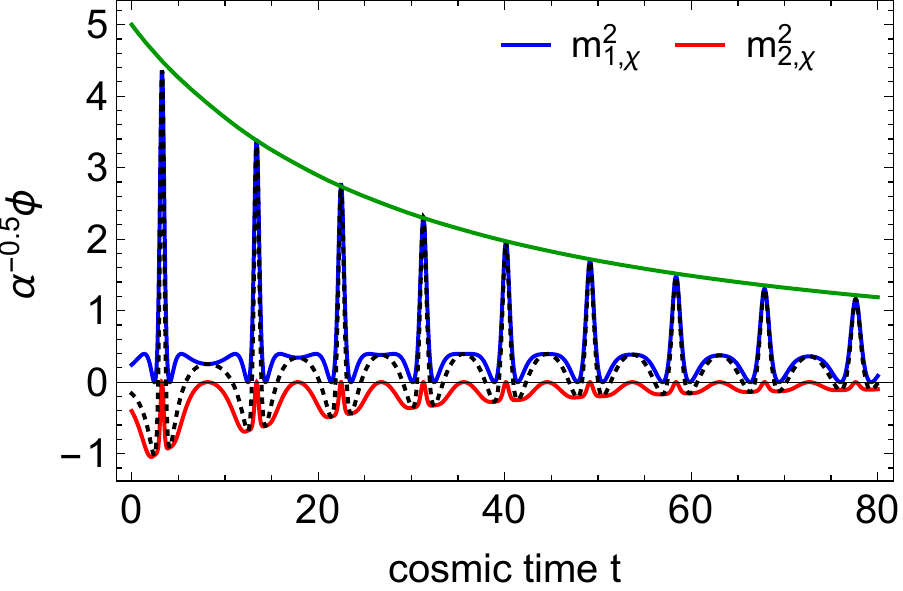}
\\
\includegraphics[width=0.45\textwidth]{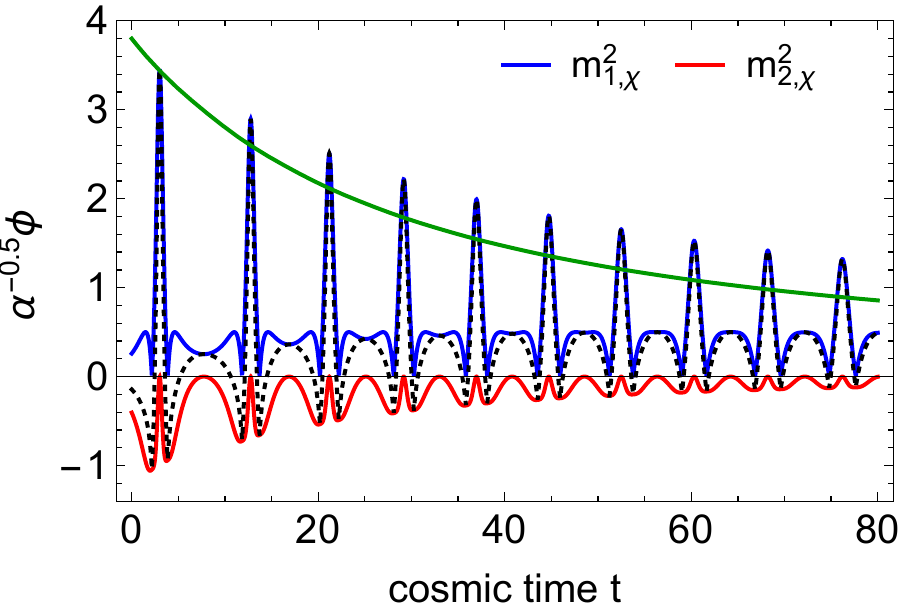}
\includegraphics[width=0.45\textwidth]{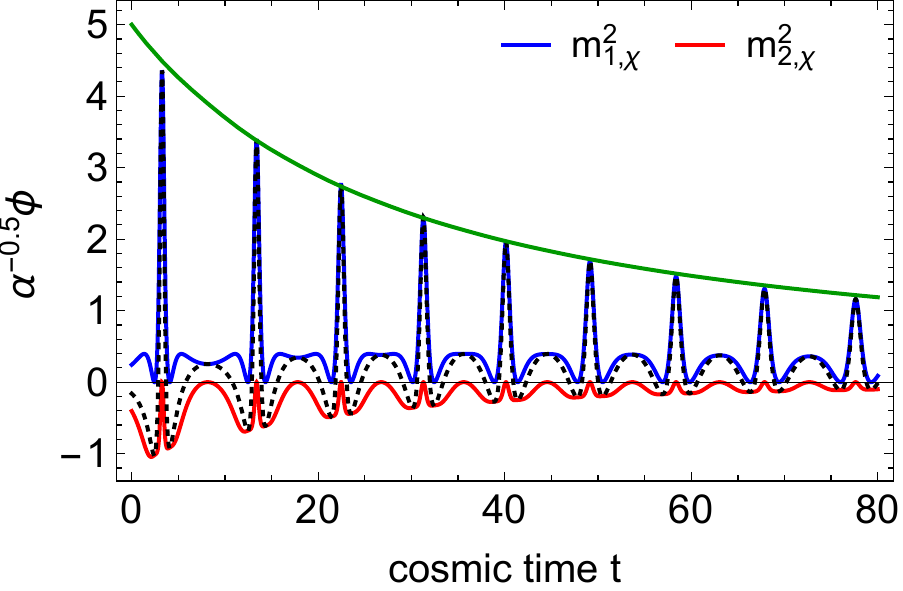}
\\
\includegraphics[width=0.45\textwidth]{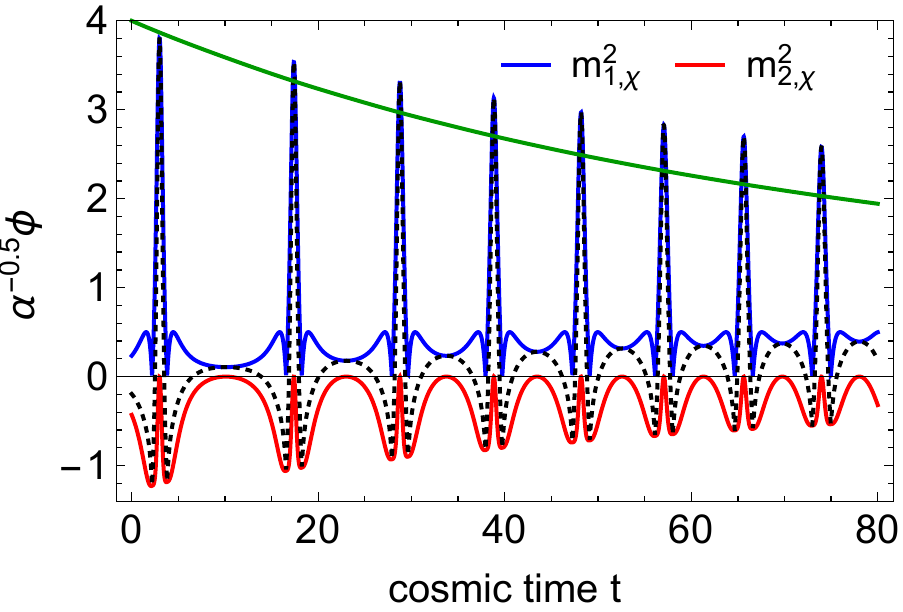}
\includegraphics[width=0.45\textwidth]{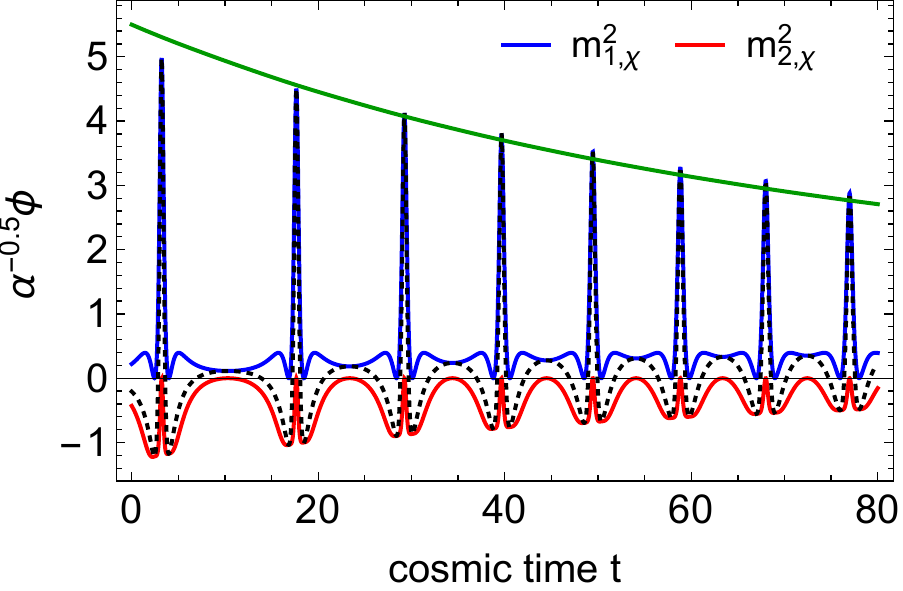}
\caption{Effective frequency $\omega_\chi^2(k,t)$ 
for $n=3/2, 2$ (left and right respectively)  and $\alpha=10^{-2},10^{-3},10^{-4}$ (top to bottom).
The effective frequency is measured in units of $\mu^2 \simeq 3.6 \cdot 10^{-11}M_{\rm Pl}^2$.
We see that for $n=3/2$ the effect of the wavenumber term $k^2/a^2$ becomes progressively less important compared to the potential spike, which is not the case for $n=2$, where the two terms red-shift in tandem.
 }
 \label{fig:meffn3ov2andn2}
\end{figure}

The component of the effective $\chi$ mass that is due to the potential can be written for small $\alpha$, similarly to the $n=1$ case, as
\beq
V_{\chi\chi}(\chi=0) \simeq \frac{4}{3} n e^{-\beta  \phi } \left(\left(1-e^{-\beta  \phi }\right)^2\right)^{n-1}
\eeq
The height of the first ``spike'' can be computed using Eq.~\eqref{eq:phiendphiminus}
\beq
\left . V_{\chi\chi}^{n>1}\right |_{{\rm max},(1)} = \left ({3\over 2} \right)^{2n-3} \left (
5-3 e^{-\beta \phi_{\rm end}} \right )
\left (
1- e^{-\beta \phi_{\rm end}} \right )^{2(n-1)} \, .
\eeq
Fig.~\ref{fig:meffn3ov2andn2} shows the evolution of the effective frequency and its two main components, the potential and Riemann terms, for $n=3/2$ and $n=2$. An interesting feature of this model is the evolution of the height of the first spike, which scales approximately as
\beq
V_{\chi\chi}^{\rm max}(\chi=0) \sim \left ({1\over a}\right )^{\min(n,4)} \, .
\eeq
Simply put, for $n<2$ the wavenumber contribution to the effective frequency $k^2/a^2$ becomes less important after the first few oscillations, while for $n>2$ it comes to dominate over the potential at late times, for sufficiently large wave-numbers. For the marginal case of $n=2$ the relative size of the wave-number and potential terms remains roughly constant.

By examining  the general form of $m_{{\rm eff},\chi}^2$ for $n=1$, shown in Fig.~\ref{fig:n1meff}, we see that  the negative  part of the effective mass $m_{2,\chi}^2$ is largely cancelled by the positive contribution of $m_{1,\chi}^2$ 
(not affected by neglecting the subdominant term $m_{4,\chi}^2$).
It means that the tachyonic resonance in the E-model is completely damped for $n=1$  and preheating can only proceed by  parametric resonance alone. Parametric resonance in the simple case of the Mathieu equation $\ddot u(t) + [A+2q \cos(2 t)]u(t)$ is largely controlled by the relative size of $A$ and $q$ and is suppressed for $A\gg q$. Fig.~\ref{fig:n1meff} shows that while the offset $A$ remains constant, the oscillation amplitude $q$ is damped, hence we expect parametric resonance to quickly shut off, at least for $\alpha = {\cal O}(0.01)$. For smaller values of $\alpha$ the effective mass exhibits a highly oscillatory behaviour, where the amplitude of the oscillation is almost equal to the constant offset of $m_{{\rm eff},\chi}^2$. Furthermore, the anharmonic behaviour of the background (see Fig.~\ref{fig:fourier}) is mirrored in the anharmonic effective mass, particularly in the dominant component $V_{\chi\chi}$, where we see a ``spike'' appearing at the points where $\phi$ is maximally negative. This can lead to a violation of the adiabaticity condition $\left | \dot \omega / \omega^2 \right | \ll 1$.

The phenomenon of a spike in the effective frequency of  fluctuations driving the adiabaticity violation was observed in preheating of multi-field models  with non-minimal coupling to gravity \cite{DeCross:2015uza, DeCross:2016fdz, DeCross:2016cbs, Ema:2016dny, Nguyen:2019kbm}, including but not limited to Higgs inflation \cite{Sfakianakis:2018lzf, Ema:2016dny}. In that context, the field-space curvature is non-uniform: the manifold is asymptotically flat at large field values and the Ricci scalar exhibits a large positive spike at the origin\footnote{This description corresponds to the analysis performed in the Einstein frame as in Refs.~\cite{DeCross:2015uza, DeCross:2016fdz, DeCross:2016cbs}. The  analysis of these models in the Jordan frame was performed in Ref.~\cite{Ema:2016dny}, where the adiabaticity violation was a result of a spike in the background field velocity as it crossed the origin.}. In order to properly define an adiabaticity parameter and use a WKB-type analysis, the frequency of the fluctuations $\delta\chi$ must be (much) greater than the frequency of the background oscillations; simply put, $\delta\chi$ must oscillate multiple times between the ``spikes" shown in Figs.~\ref{fig:n1meff} and \ref{fig:meffn3ov2andn2}. We have shown using both analytical and numerical arguments (see Fig.~\ref{fig:periodvsa}) that the background frequency is $\omega_{\rm bg} \equiv 2\pi/T \sim 0.5$, with a mild dependence on the parameters $\alpha$ and $n$. While the maximum value of $m_{{\rm eff},\chi}$ is larger than $\omega_{\rm bg}$, the averaged value over one period is not, in fact $\langle m_{{\rm eff},\chi}\rangle _T \sim\omega_{\rm bg} $. Thus, in order to properly use the adiabaticity condition as a criterion for preheating, we should  restrict ourselves to cases where the wave-number contribution $k^2/a^2$ is non-negligible. For now, let us consider cases where $k\gtrsim \mu$ (we choose to measure $k$ in units of $\mu$, as in Ref.~\cite{Iarygina:2018kee}).

\begin{figure}
\centering
\includegraphics[width=0.45\textwidth]{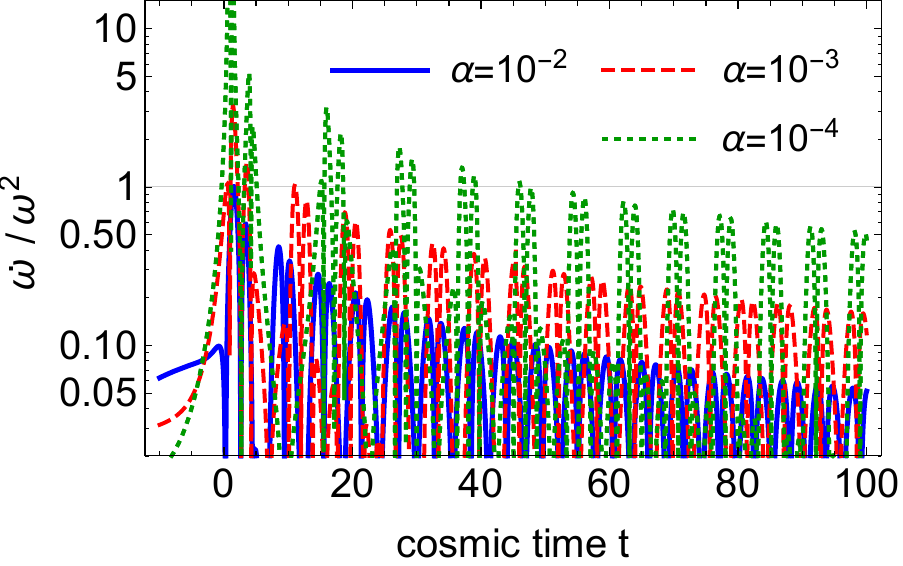}
\includegraphics[width=0.45\textwidth]{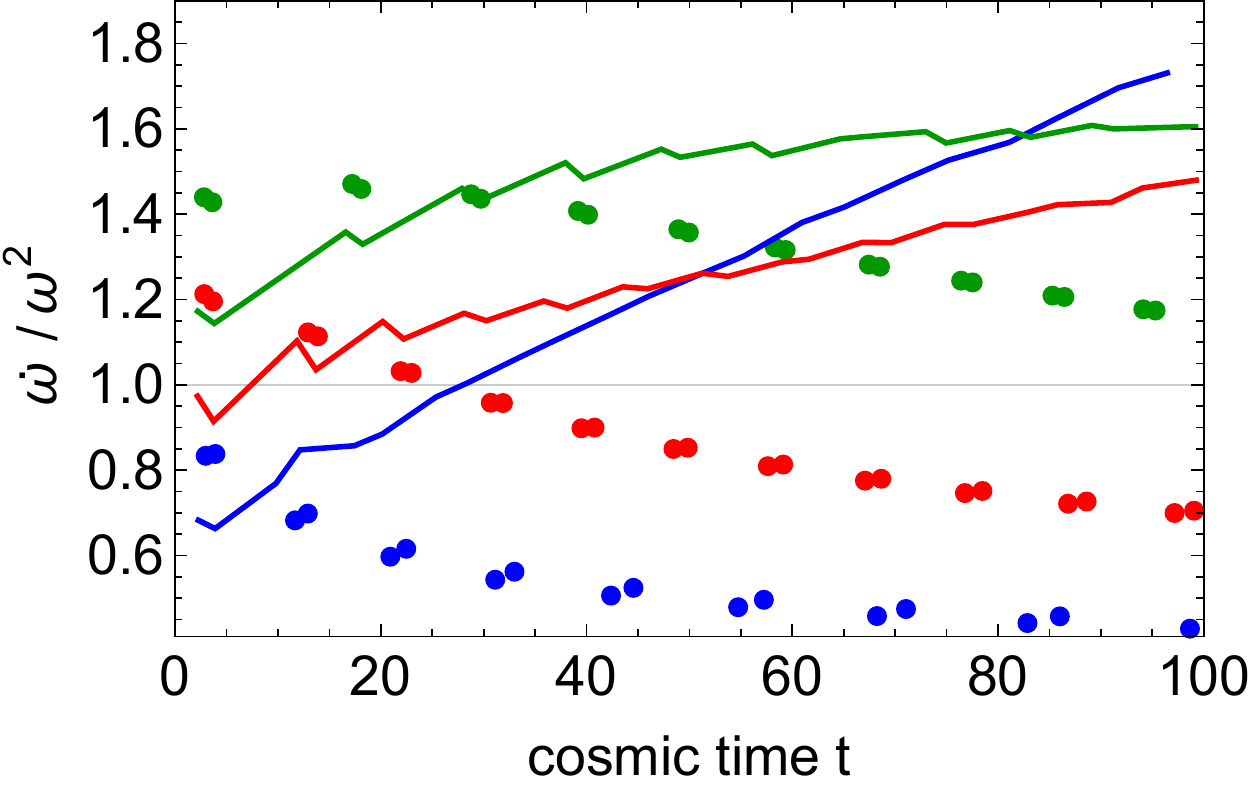}
\caption{{\it Left:} The adiabaticity condition $\left| {\dot \omega/ \omega^2}  \right|$ for $n=1$ and $\alpha=10^{-2},10^{-3},10^{-4}$ (blue solid, red dashed and green dotted respectively) and $k=0.5\mu$. {\it Right:} The peaks of the adiabaticity condition $\left| {\dot \omega / \omega^2}  \right|$ for $n=3/2$ (solid curves) and $n=2$ (dots) for $\alpha=10^{-2},10^{-3},10^{-4}$ (blue, red and green respectively)  and $k=1.5\mu$.
 }
 \label{fig:nad}
\end{figure}

 Fig.~\ref{fig:nad} shows the evolution of the adiabaticity condition for $n=1$ and $k=0.5\mu$, where we see adiabaticity violation for only a few oscillations at $\alpha\le 10^{-3}$. If we consider larger wave-numbers, $k\simeq \mu$, we find $\left | \dot\omega/\omega^2 \right |<1$ for $ n=1$.
 The situation is however different for $n\ge 3/2$, where we find instances of $\dot \omega/\omega^2>1$ for $k\ge \mu$. Fig.~\ref{fig:nad} shows the evolution of the peaks in the adiabaticity parameter, occurring around the maximally negative value of $\phi(t)$. For $n=2$ we see that the adiabaticity parameter is violated (for $\alpha \le 10^{-2}$) initially, but $\left |\dot\omega/\omega^2\right |$ decreases with time. The situation is reversed for $n=3/2$, where we see that the adiabaticity parameter grows with time.  Finally, the value of $\left | \dot \omega / \omega^2 \right |$ grows with decreasing $\alpha$ for all values of $n$ that we examined, signifying a common trend.

Before we conclude this section, it is important to distinguish two different types of sharp features in the effective frequency of fluctuations. The field-space induced spikes that were found in non-minimally coupled models \cite{DeCross:2015uza, DeCross:2016fdz, DeCross:2016cbs, Ema:2016dny, Nguyen:2019kbm, Sfakianakis:2018lzf} arise when the fields pass through the origin and have their maximal velocity. They can lead to significant adiabaticity violation over a large range of wave-numbers and thus can drive very efficient particle production. Contrary to this, Ref.~\cite{He:2018mgb} found sharp features in a model of unitarized Higgs inflation (mixed Higgs-Starobinsky inflation). This feature however arises from a sharp potential barrier and is thus similar to the feature found in the E-model and completely different than the sharp feature found in ``regular'' Higgs inflation preheating~\cite{Sfakianakis:2018lzf}. The potential-driven spike in the effective frequency leads to typically weaker preheating than the field-space-driven one, at least for the models mentioned here. It would be interesting to perform an EFT-type analysis for preheating models with sharp features, but this goes beyond the scope of our present analysis and is left for future work.

\section{Mass-scales and Preheating}
\label{sec:massscales}

Due to the construction of the E-model, which arises by defining the K\"ahler potential and superpotential for  a complex field $Z$,  the $\phi$ and $\chi$ dependence of the potential  are related to each other. The second derivative of the potential with respect to $\chi$, which is one of the two main components in the effective mass-squared $m^2_{{\rm eff},\chi}$, can be related to the potential value itself as
\beq
\left . V_{\chi\chi}\over V\right |_{\chi=0} = 2+ \frac{n}{3 \alpha  \sinh ^2\left(
{\beta\phi\over 2}
\right)}
\, .
\eeq
This diverges at $\phi=0$ for all values of $\alpha$ and $n$, which is easy to understand by Taylor expanding the two terms for $\chi=0$ as $V(\chi=0)\simeq {2^n\over 3^n a^{n-1}}\phi^{2n} +{\cal O}(\phi^{2n+1})$ and
$V_{\chi\chi}(\chi=0)\simeq {2^{n+1} n\over 3^n a^{n-1}}  \phi^{2n-2}+{\cal O}(\phi^{2n-1})$. We see that for all values of $n$ the potential $V$ vanishes faster than the derivative $V_{\chi\chi}$ for $\chi=0$ and $\phi\to 0$. For asymptotically large values of $\phi$ the ratio becomes constant and equal to $2$. However for $\phi = {\cal O}(1)\sqrt{\alpha}$, which is the relevant parameter range for preheating, the ratio is $V_{\chi\chi} /V = {\cal O}(1) \times \alpha^{-1}$, where the proportionality factor depends on $n$ and $\phi$.

\begin{figure}
\centering
\includegraphics[width=0.45\textwidth]{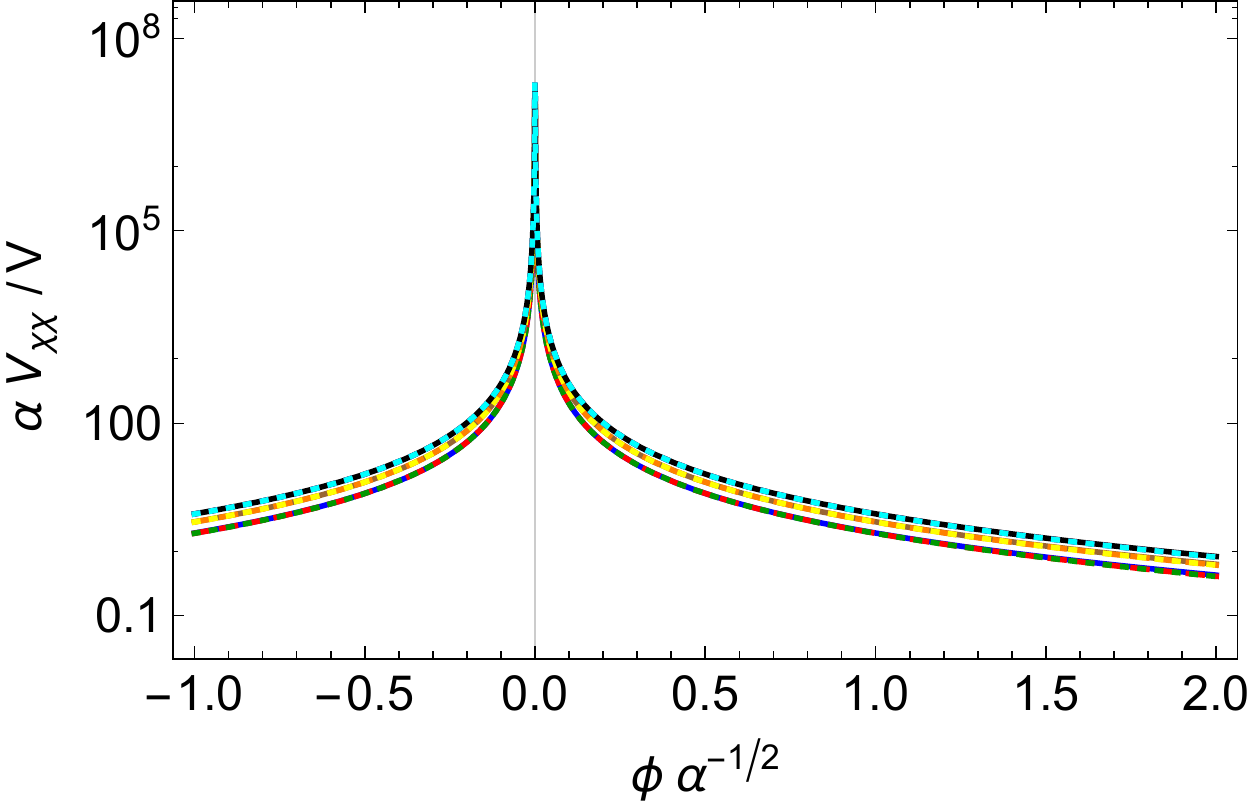}
\includegraphics[width=0.45\textwidth]{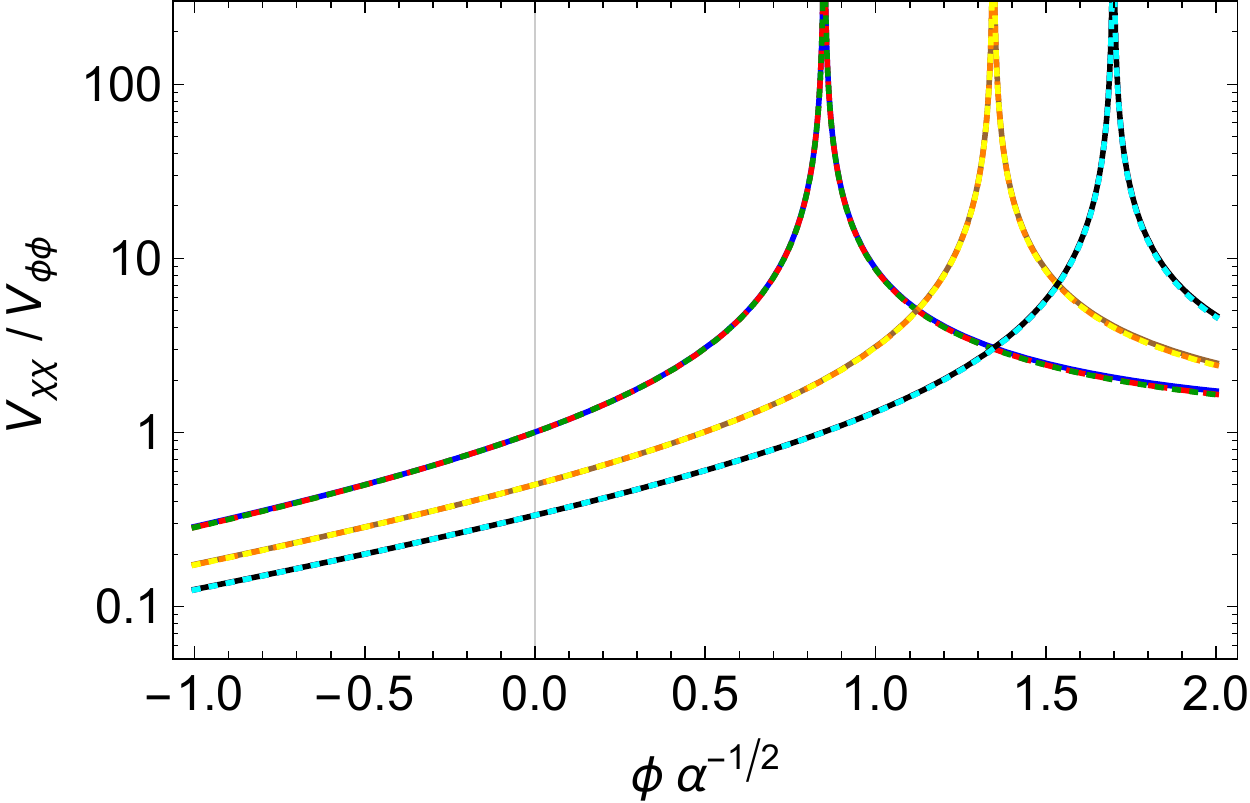}
\caption{{\it Left:} The ratio $V_{\chi\chi}/V$ multiplied by the factor $\alpha$ as a function of the rescaled inflaton field for $n=1,3/2,2$ and $\alpha=10^{-2},10^{-3},10^{-4}$.
The color-coding is as follows:
$\{n,\alpha \} =\{1,10^{-2} \},\{1,10^{-3} \},\{1,10^{-4} \}$: blue, red-dashed, green-dotted,
$\{n,\alpha \} =\{3/2,10^{-2} \},\{3/2,10^{-3} \},\{3/2,10^{-4} \}$: brown, orange-dashed, yellow-dotted,
$\{n,\alpha \} =\{2,10^{-2} \},\{2,10^{-3} \},\{2,10^{-4} \}$: black, purple-dashed, cyan-dotted.
 We see that for field values relevant for preheating the $\alpha$ dependence is canceled out when multiplying by $\alpha$ and the $n$ dependence is weak.
{\it Right:} The ratio $V_{\chi\chi}/V_{\phi\phi}$  for the same parameters and color-coding. 
In both panels, the three curves corresponding to the same value of $n$ and different values of $\alpha$ are  indistinguishable.
 }
 \label{fig:Vratio}
\end{figure}

Furthermore, the $\phi$ and $\chi$ mass-scales are also related to each other as
\beq
\left . V_{\chi\chi}\over V_{\phi\phi} \right |_{\chi=0} = -\frac{\left(e^{\beta\phi }-1\right)^2 \left(6 \alpha +n \text{csch}^2\left({\beta\phi\over 2}\right)\right)}{4 n \left(e^{\beta \phi }-2 n\right)}
\simeq -\frac{\left(e^{\beta\phi }-1\right)^2 \text{csch}^2\left(
{\beta\phi\over 2}
\right)}{4 \left(e^{\beta\phi }-2 n\right)}
\, ,
\eeq
where the last equation holds for $\alpha\ll1$. We see that $V_{\phi\phi}$ changes sign, since the potential is concave during inflation. For (large) negative values of $\phi$ the scaling of the ratio $V_{\chi\chi}/V_{\phi\phi}$ simplifies as $V_{\chi\chi}/V_{\phi\phi} \sim e^{\sqrt{2}\phi/\sqrt{3\alpha}}/(2n)$. The full behavior is shown in Fig.~\ref{fig:Vratio}.

\begin{figure}
\centering
\includegraphics[width=0.45\textwidth]{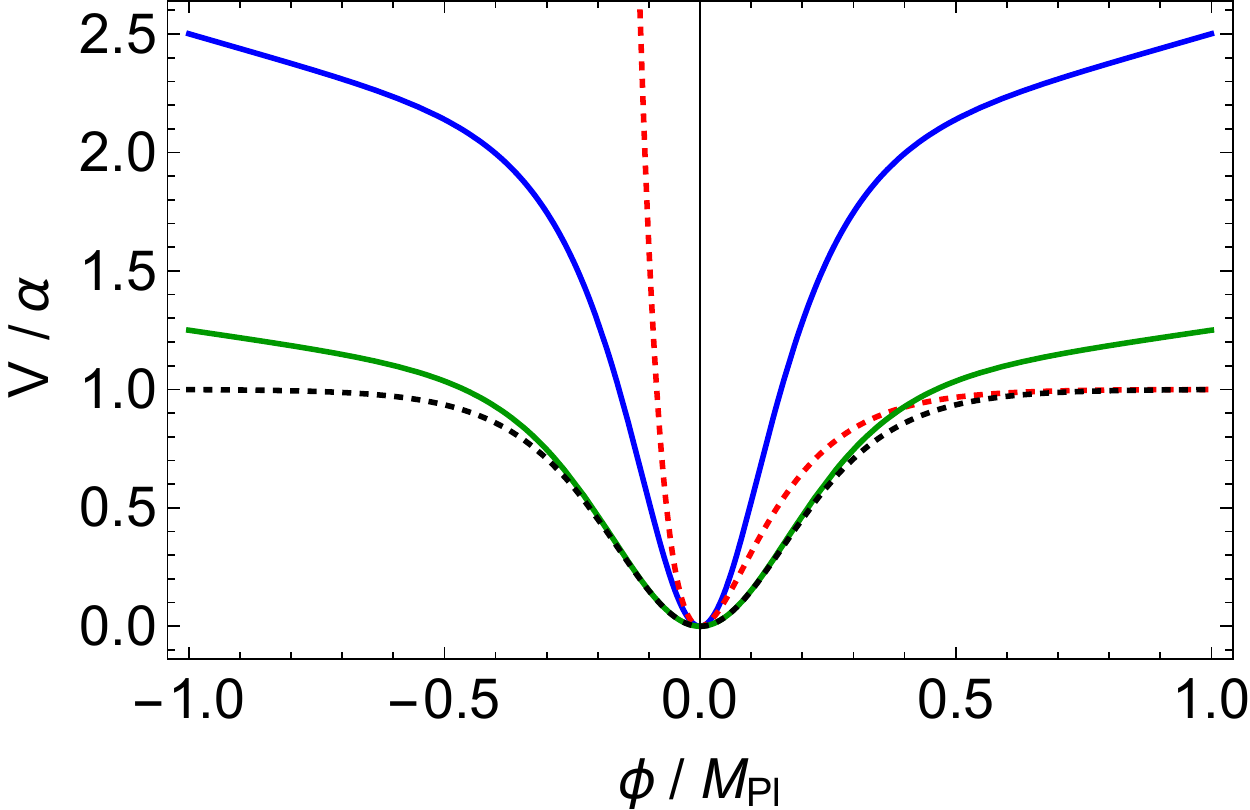}
\includegraphics[width=0.45\textwidth]{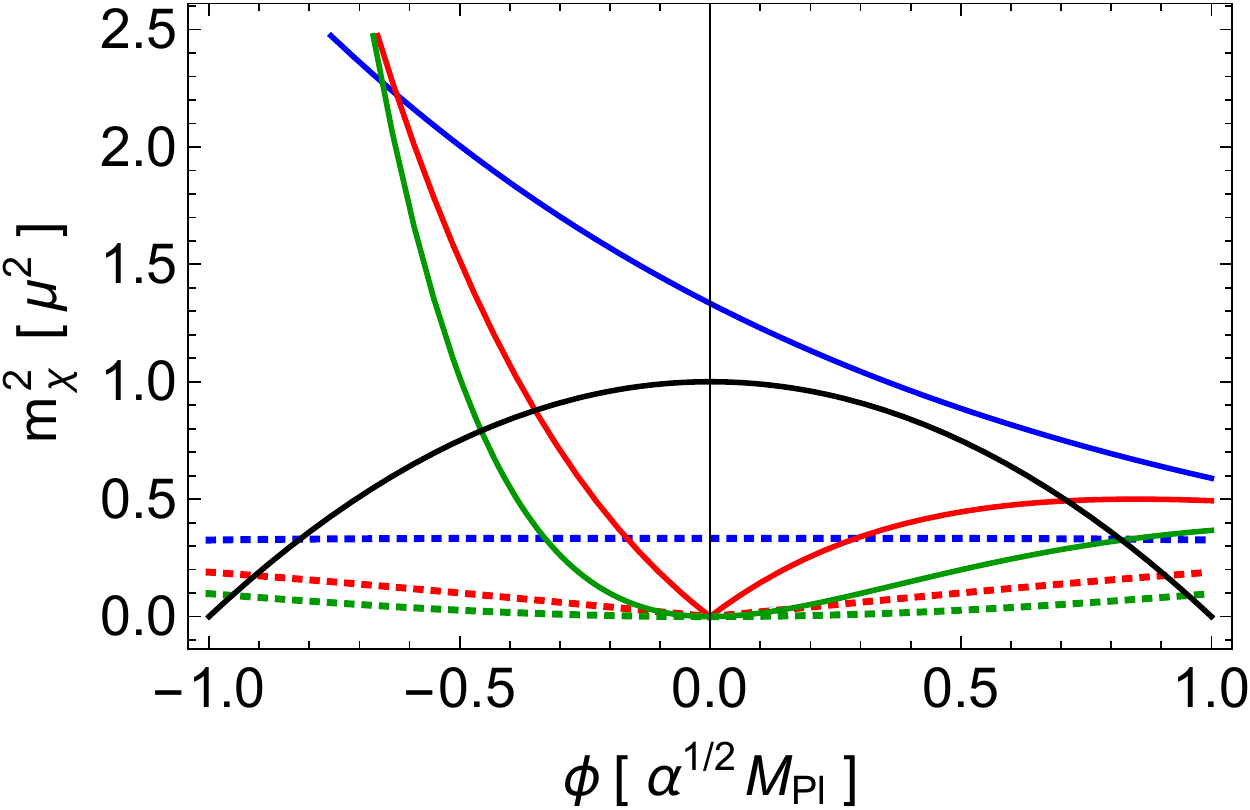}
\caption{
{\it Left:} The field $V(\phi, \chi=0)$ (dotted) and $V(\phi=0,\chi)$ (solid) for the massive case $n=1$ of the E- and T-models (red/blue and black/green respectively).
We see that in each model the $\phi$ and $\chi$ masses are equal to each other. However the masses of the fields in the T-model are larger than the ones in the E-model. 
{\it Right:}  The potential contribution to the $\delta\chi$ effective mass for the E- and T- model (solid and dotted curves respectively) for $n=1,3/2,2$ (blue,red green) and $\alpha \ll 1$. The black line shows an estimate of the tachyonic field-space contribution.
We see that in the E-model for $n=1$, the potential term can dominate over the tachyonic field-space curvature term, consistent with the behavior shown in Fig.~\ref{fig:n1meff}. 
}
 \label{fig:Vzero}
\end{figure}

Finally, Fig.~\ref{fig:Vzero} shows  the potential close to the origin for each of the fields $V(\phi, \chi=0)$ and $V(\phi=0,\chi)$ for $n=1$. We see that the mass of the $\phi$ and $\chi$ particles is equal. This can have important phenomenological consequences, since the inability of the particles to decay into each other opens the way for the emergence of composite oscillons, comprised of both fields \cite{Sfakianakis:2012bq}. 
Oscillons appear when the potential of a scalar field is shallower than quadratic away from the origin. Intuitively, this makes the frequency of large oscillations smaller than the mass of the particle, creating a potential barrier that keeps the particles bound inside the oscillon. Two-field oscillons are more complicated and only a few examples have been found in the literature (see e.g. Refs.~\cite{Sfakianakis:2012bq, Farhi:2005rz, Graham:2007ds, Graham:2006vy}. A feature of two-field systems exhibiting oscillons must be the inability of the scalar field comprising the oscillon to decay into lighter fields. In the $\alpha$-attractor case it is reasonable to expect that, since the two fields have the same mass, decays and scatterings will be kinematically suppressed, possibly leading to long-lived oscillons.
The study of oscillons in $\alpha$-attractors is beyond the scope of the present work. One further interesting observation can be made when one compares the mass of particles in the E- and T-model. In the latter case, the small field excitations of the $n=1$ potential have a mass of $\mu/\sqrt{3}$, half of the E-model case. This leads to a simple criterion for tachyonic resonance in $\alpha$-attractors. The maximally negative contribution to the effective mass of the $\chi$ fluctuations is  related to the Hubble scale at the end of inflation through energy conservation (neglecting Hubble drag after inflation)
\beq
{1\over 2} {\cal R} \dot\phi^2 = 3 {\cal R} M_{\rm Pl}^2 H^2 \simeq -\mu^2
\eeq
Fig.~\ref{fig:phiendvsa} shows that the Hubble scale at the end of inflation differs by about $10\%$ for the E- and T-models for small values of $\alpha$, regardless of the potential steepness $n$.  In the EFT language, $\alpha$-attractors in the small $\alpha$ regime show a strong hierarchy of scales, where the Hubble scale is almost constant and much smaller than the background oscillation frequency \cite{Giblin:2017qjp}. Fig.~\ref{fig:Vzero} shows the potential contribution to the effective mass for the E- and T-model. We see that while the tachyonic contribution is similar in the two models,  the potential contribution is larger for the E-model. Thus, a quick calculation of the energy density at the end of inflation and the mass of the spectator field in any $\alpha$-attractor model can provide a strong indication for the efficiency of tachyonic preheating. The case of asymmetric $\alpha$-attractors is slightly more involved, because of the introduction of one further mass-scale, in which case one must check the possibility of non-adiabatic behavior due to it, as discussed in Section~\ref{sec:frequency}.

\subsection{ Floquet charts}
\label{sec:Floquet charts}

In order to compare the efficiency of  particle production (mode amplification) during preheating, we will use
Floquet theory, by working in the static universe approximation, where the inflaton field oscillates periodically without Hubble friction. We use the algorithm  described in Ref.~\cite{Amin:2014eta}.  The equation of motion for the $\chi_k$ modes (similarly for the $\phi_k$ ones) for $H=0$ and $a(t)=1$ is written as
\beq
{d \over dt} \begin{pmatrix}
\chi_k \\
\dot{\chi}_k
\end{pmatrix} =  \begin{pmatrix}
0 & 1 \\
-(k^2 + m_{{\rm eff},\chi}^2) & 0
\end{pmatrix} \begin{pmatrix}
\chi_k \\
\dot{\chi}_k
\end{pmatrix} ,
\label{eqn:floquetmatrix}
\eeq
where $m_{\rm eff,\chi}^2 = m_{1,\chi}^2 + m_{2,\chi}^2$. This equation is of the form
\beq
\dot{x}(t) = {\cal P}(t) \> x(t) \, ,
\label{eq:1stordereq}
\eeq
where $ {\cal P}(t)$ is a periodic matrix.
The solutions are of the form
\beq
\chi_k(t) = e^{\mu_k t}g_1(t) + e^{-\mu_k t}g_2(t)
\eeq
where $g_1, g_2$ are periodic functions and $\mu_k$ is the Floquet exponent. If $\mu_k$ has a non-zero real component, one of the two solutions will be exponentially growing, signaling an instability and efficient amplification for this specific wavenumber.

\begin{figure}
\centering
\includegraphics[width=0.45\textwidth]{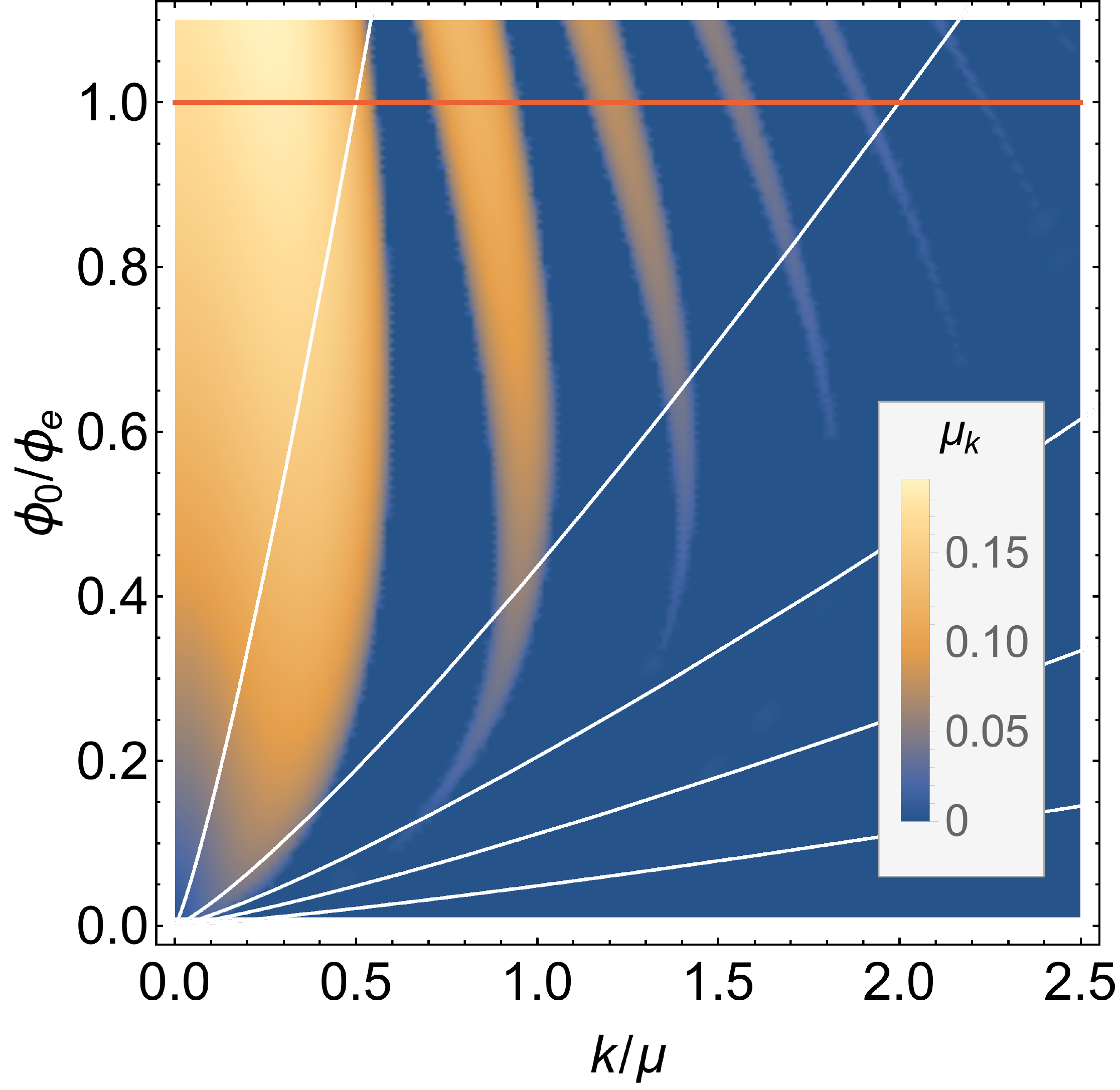}
\includegraphics[width=0.45\textwidth]{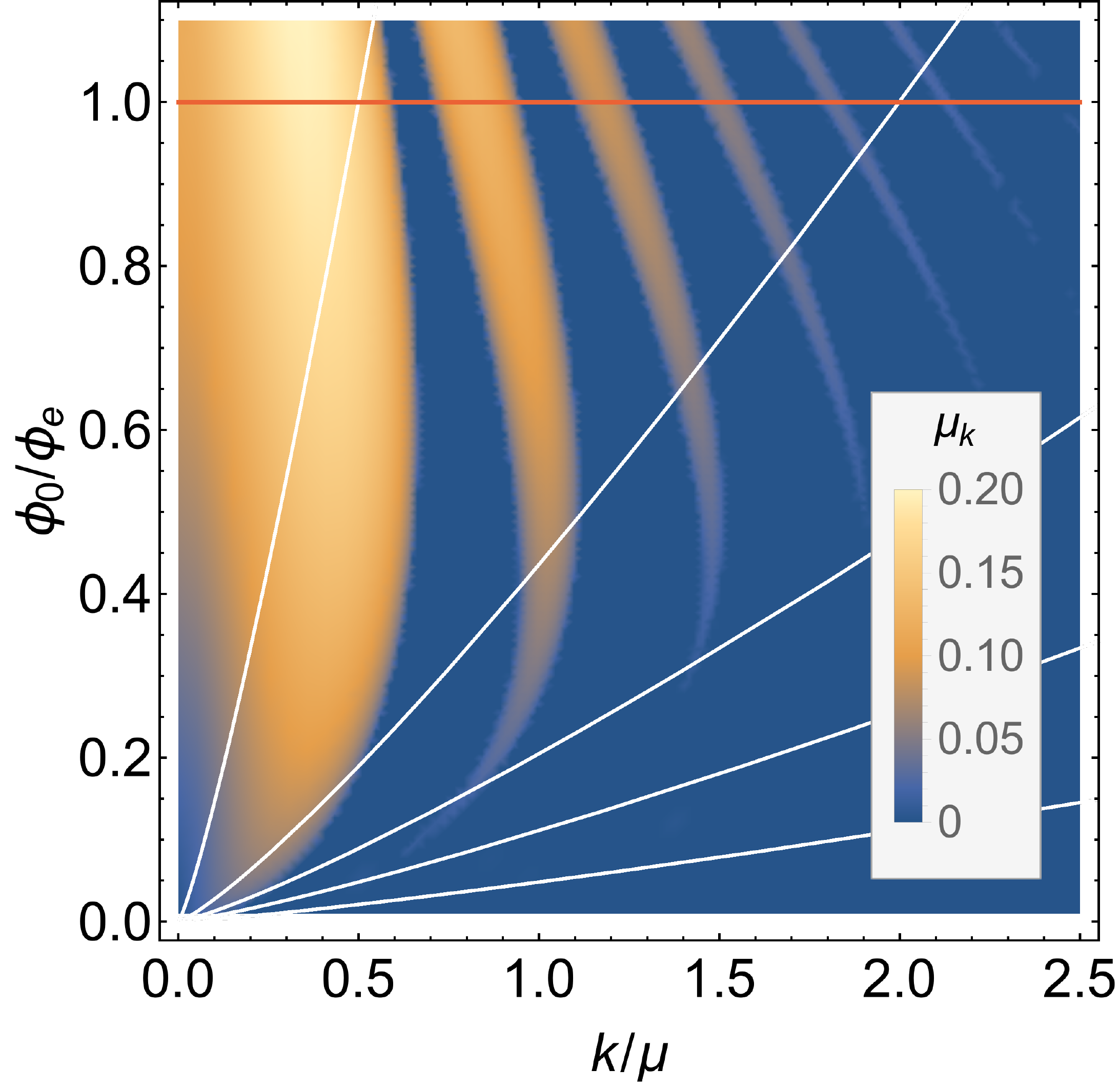}
\\
\includegraphics[width=0.45\textwidth]{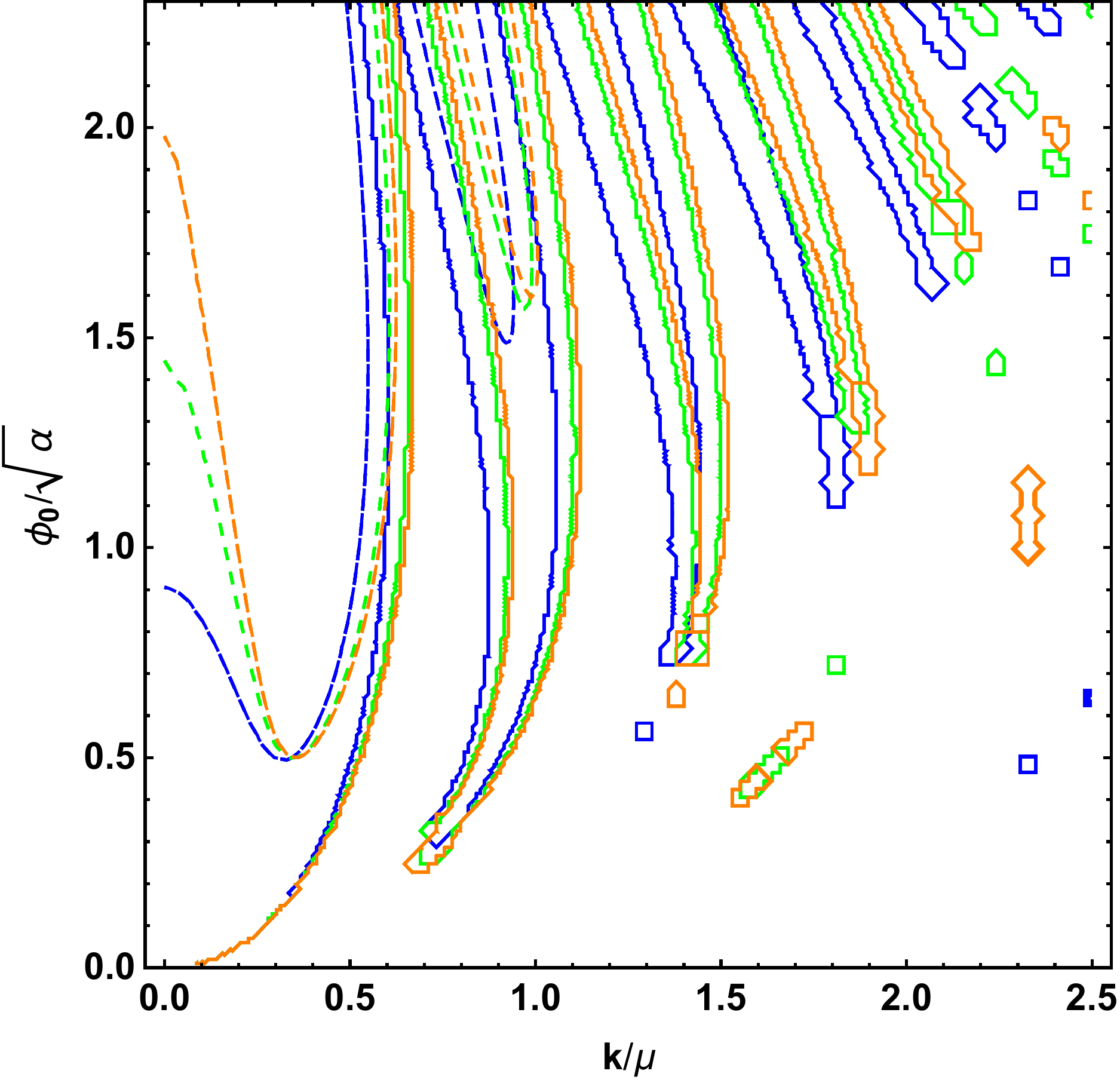}
\includegraphics[width=0.45\textwidth]{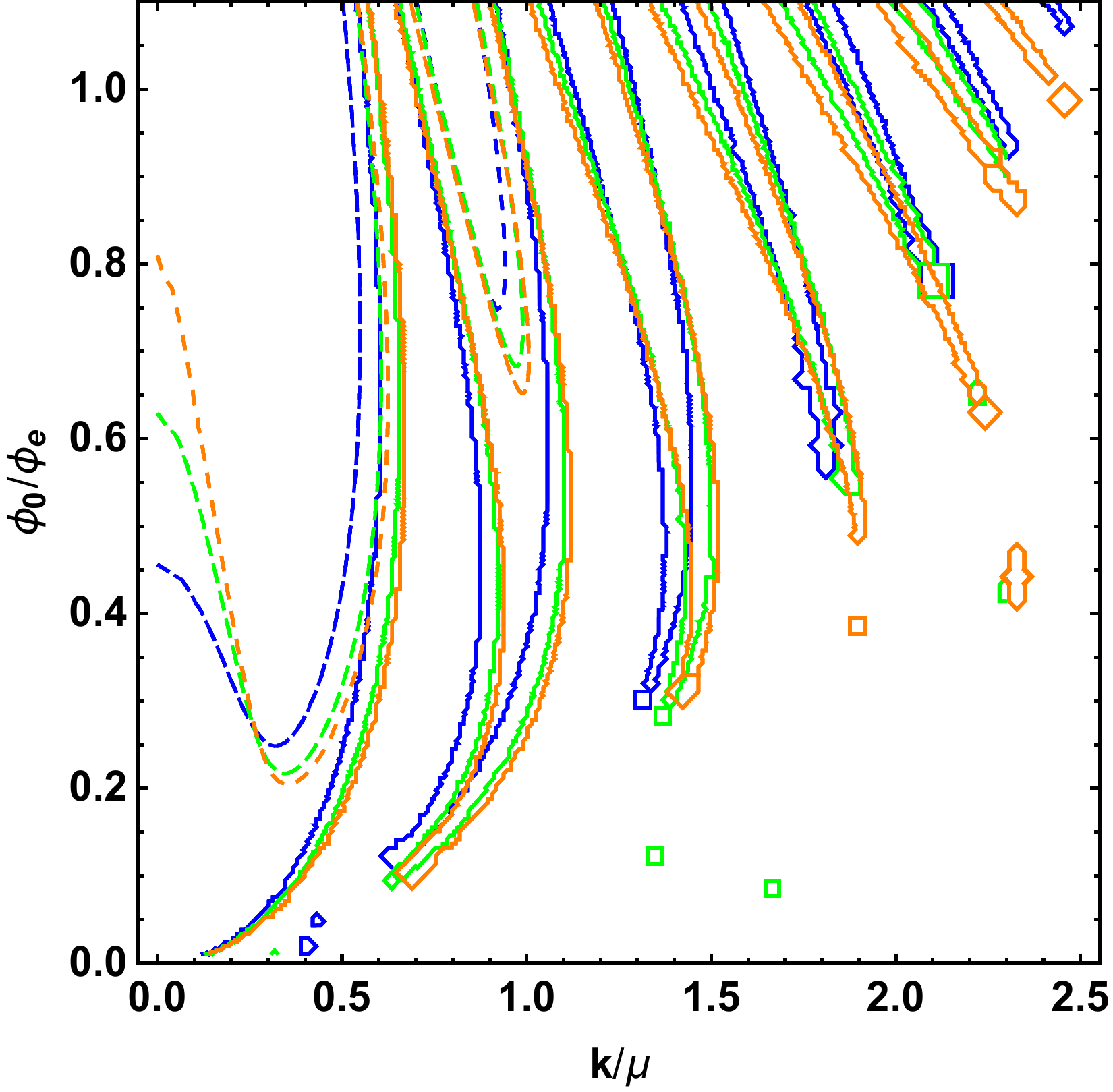}
\caption{
{\it Upper row:} The 3-D Floquet charts for $n = 3/2$ and
$\alpha=10^{-2},10^{-4}$  (left and right panels respectively).
{\it Bottom row:} The contour plots for ${\mu_k = 0} $ (solid lines) and
$ {\mu_k = 0.1}$  (dashed lines)  in units of $\mu$. The background field oscillation amplitude $\phi_0$ is rescaled either by $\sqrt{\alpha}$ (left) or by the field value at the end of inflation $\phi_{\rm end}$ [which is denoted here as $\phi_e$] (right). The blue, green and orange curves are for $\alpha=10^{-2},10^{-3},10^{-4}$
respectively. We see that, when properly rescaled, the Floquet charts asymptote to a ``master diagram'' for $\alpha \ll 1$.}
 \label{fig:Floquetn3ov2}
\end{figure}

Figure~\ref{fig:Floquetn3ov2} shows the Floquet charts for the generalized E-model for the case of $n=3/2$. We see that, when properly rescaled, the Floquet charts for different values of $\alpha\ll1$ are similar to each other.
However, unlike the case of the T-model \cite{Iarygina:2018kee}, the Floquet charts do not exactly reach a ``master diagram'' for $10^{-2}\le \alpha\le 10^{-4}$. This can be traced back to the existence of two mass-scales: the field-space curvature and the steep potential at maximum negative  $\phi$. While the former does not scale with $\alpha$ for $\alpha\ll 1$, the latter does in a non-trivial way, albeit weakly, as shown in Fig.~\ref{fig:n1meff}. Furthermore, the existence of multiple instability bands, unlike in the T-model case, can be an indication of the richer spectral content of the background field.

By constructing the Floquet diagram using the field amplitude $\phi_0$ rescaled by the field value at the end of inflation $\phi_{\rm end}$ rather than $\sqrt{\alpha}$, the approach to a master diagram becomes better, especially for the higher $k$ instability bands.
 This is due to the high sensitivity of $V_{\chi\chi}^{\rm max}$ on $\phi_0$, as shown in Fig.~\ref{fig:meffn3ov2andn2}.
 Furthermore, the value of $V_{\chi\chi}^{\rm max}$  mostly affects the higher instability bands, as we will discuss in Section~\ref{sec:massscales2}.

\begin{figure}
\centering
\includegraphics[width=0.45\textwidth]{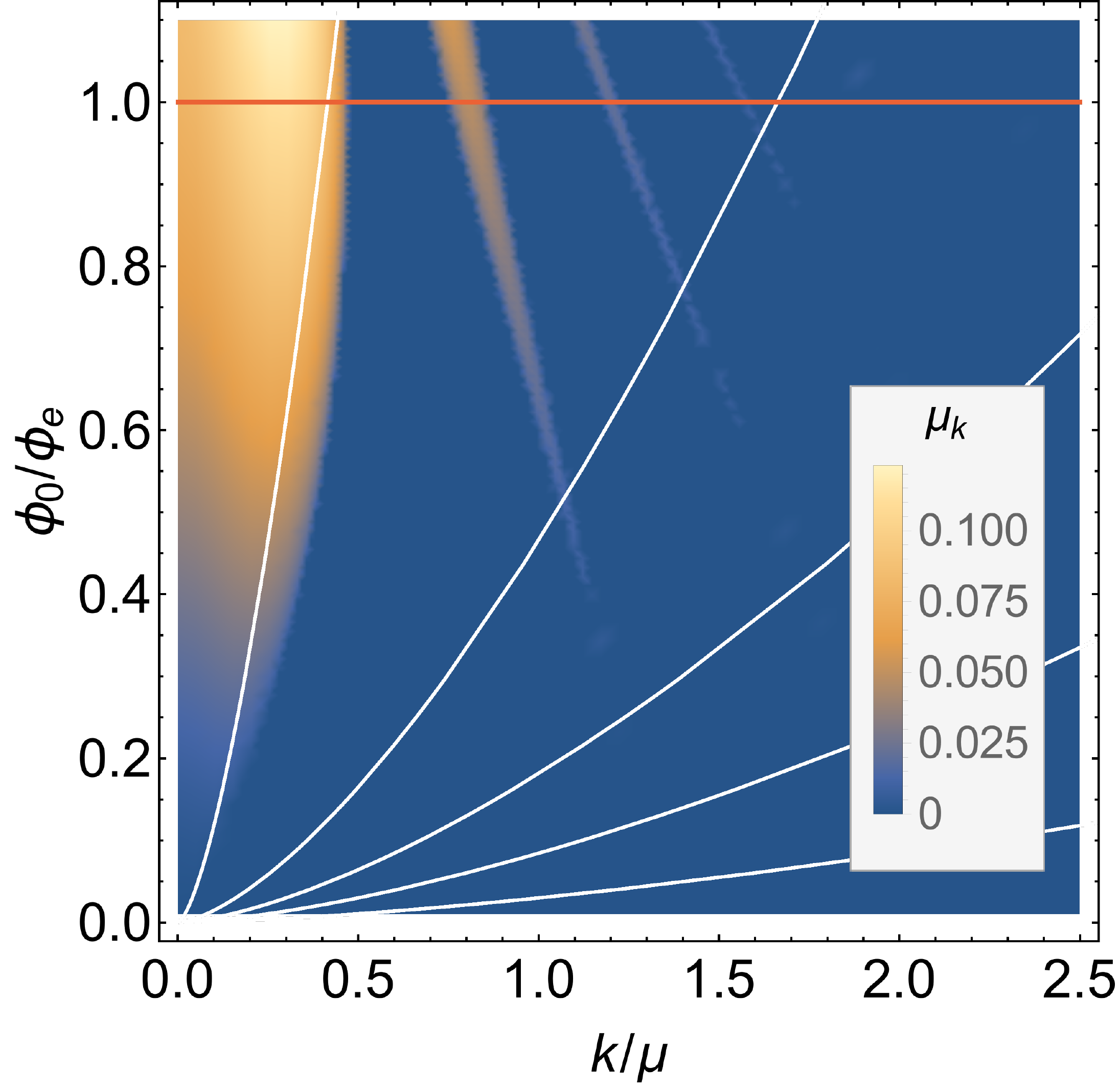}
\includegraphics[width=0.45\textwidth]{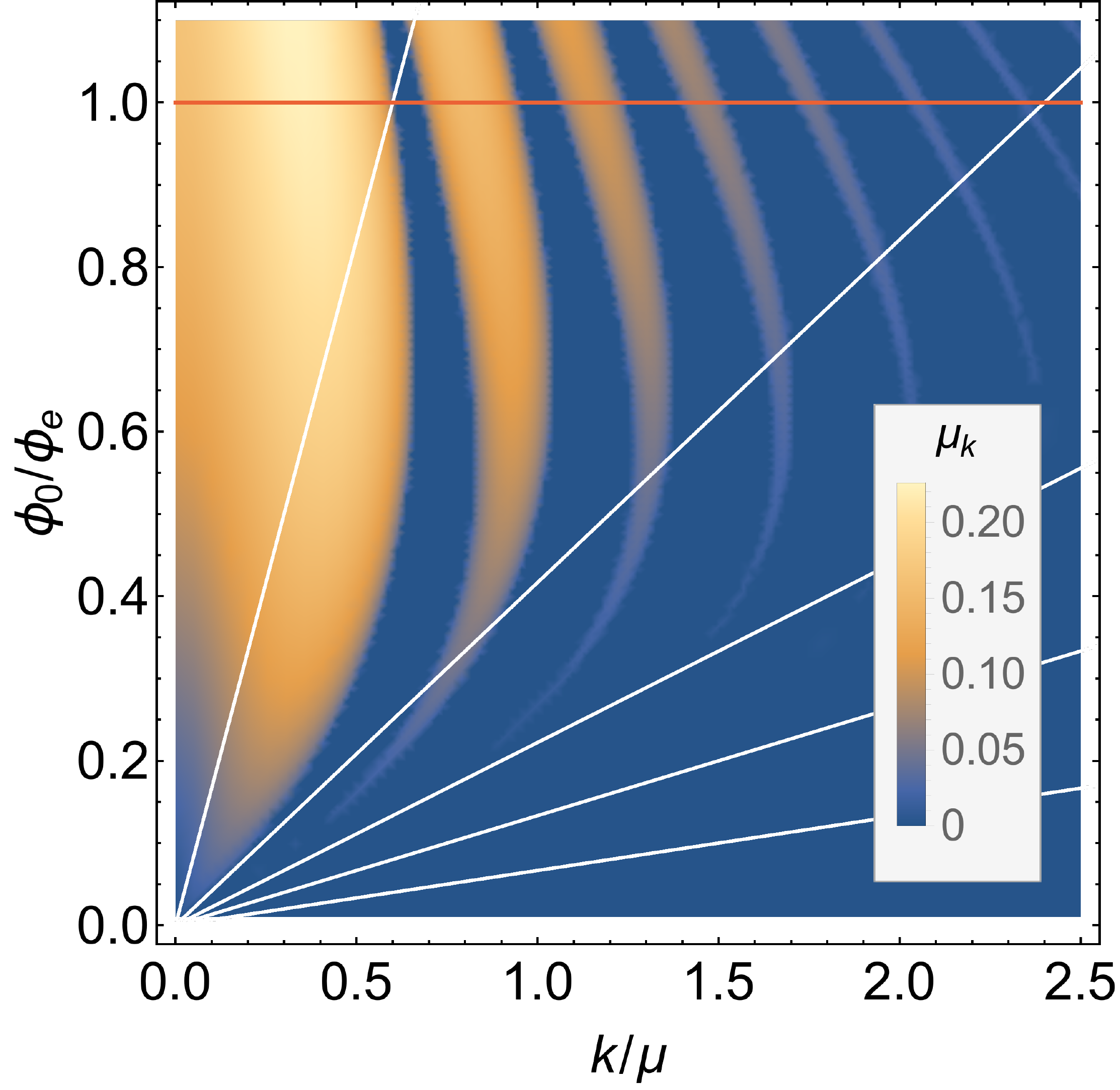}
\\
\includegraphics[width=0.45\textwidth]{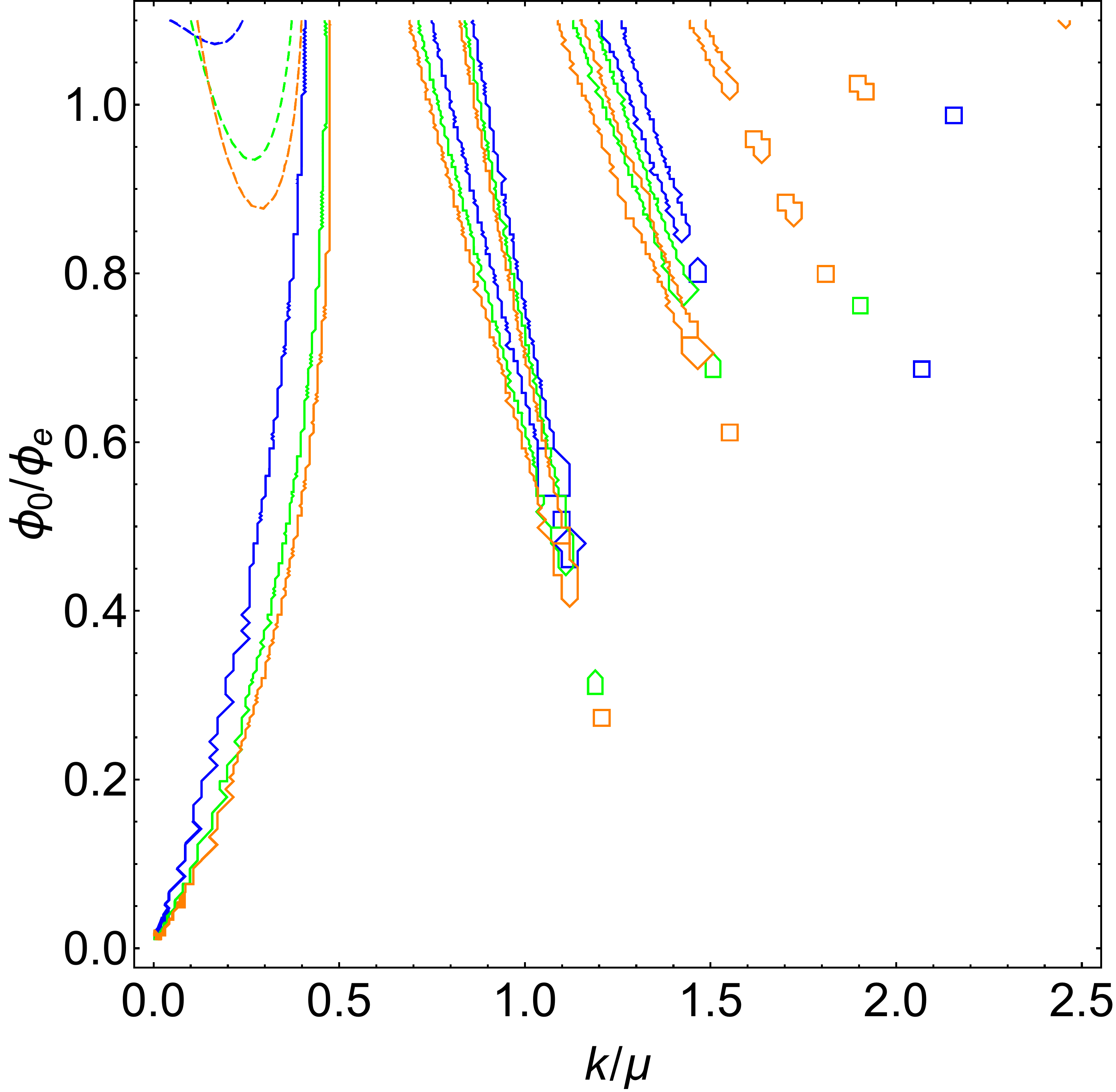}
\includegraphics[width=0.45\textwidth]{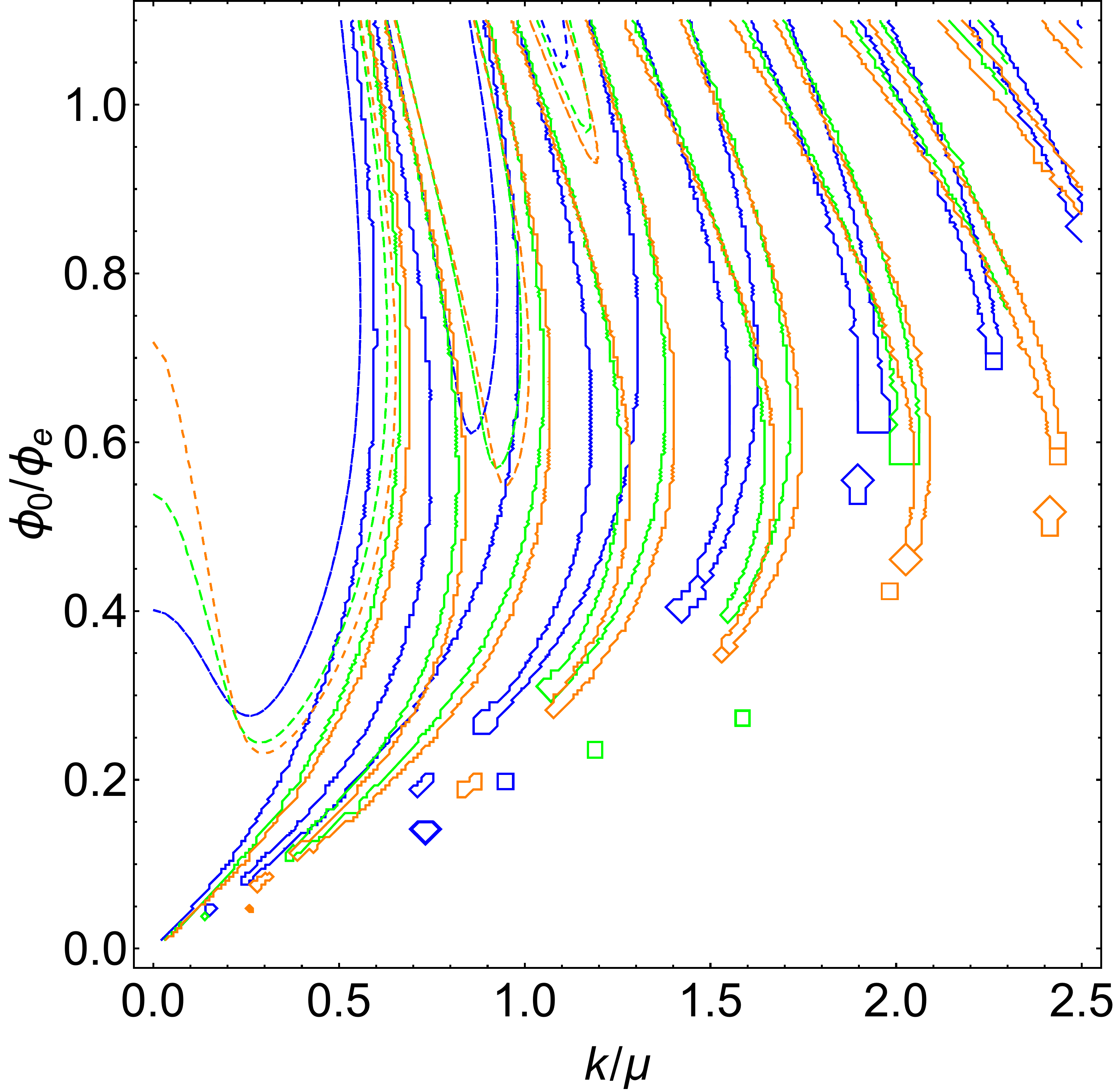}
\caption{
{\it Left column:}  The 3-D Floquet charts for $n = 1$ (upper panel) and
$\alpha=10^{-4}$. 
The contour plots for ${\mu_k = 0} $(solid lines) and
$ {\mu_k = 0.1}$   (dashed lines)  for $n = 1$ with the field amplitude rescaled by the field value at the end of inflation $\phi_e$ (bottom panel). The blue, green and orange curves are for $\alpha=10^{-2},10^{-3},10^{-4}$
respectively.
{\it Right column:} The same quantities for $n=2$.
We see the significantly suppressed parametric resonance for $n=1$, both in the number of instability bands, as well as in the width and magnitude of the main instability band.
}
 \label{fig:Floquetn1n2}
\end{figure}

Fig.~\ref{fig:Floquetn1n2} shows the Floquet charts for the cases of $n=1$ and $n=2$, which
correspond to a locally quadratic and quartic potential near the origin. The Floquet charts for $n=3/2$ and $n=2$ are visually similar exhibiting multiple, non-trivial, instability bands in the range $k\lesssim 2.5\mu$. However, the Floquet chart for $n=1$ has a completely distinct structure. The reason behind this discrepancy is that, as shown in Section~\ref{sec:frequency},  the structure of the effective mass of $\chi$ fluctuations is different for $n=1$ as compared to $n\ge 3/2$. For $n=1$ the tachyonic contribution of the field-space is entirely negated by the potential contribution. For $n\ge 3/2$ both the negative field space contribution and the positive potential term are visible. As we will show, the field-space effects are present for $n\ge 3/2$, especially for $k\lesssim\mu$, hence the dominant instability bands are similar amongst those models. This is different from the generalized T-model case \cite{Iarygina:2018kee}, where the Floquet charts for all values of $n$ show instability bands of similar shape and position, albeit not identical ones, exhibiting smaller Floquet exponent $\mu_k$ for $n=1$.

\subsection{Parametric resonance  and competing mass-scales}
\label{sec:massscales2}

As a way to encode the structure of preheating in the generalized E-model and make our results easily transferrable to other models, we examine the different mass-scales (and corresponding time-scales) that arise for the background motion and $\chi$ fluctuations.

The Hubble scale at the end of inflation $H_{\rm end}$ is proportional to the mass-scale $\mu$ and is defined by the requirement that the density fluctuations encoded in the CMB have the proper amplitude. It enters the calculation, by normalizing the amplitude of the Bunch-Davies vacuum, compared to the background energy density, hence it shows how much fluctuations must grow to dominate over the background energy density and lead to complete preheating. Furthermore, the Hubble scale controls the red-shifting of the mode wavenumbers and the amplitude of the inflaton condensate.

 The background frequency $\omega_{\rm bg}$ controls the period of background oscillations. This is related to the
  local curvature of the potential near the origin $V_{\phi\phi}(\phi=0,\chi=0)$. In simple polynomial models of inflation, for example quadratic inflation, these two time-scales, the Hubble scale and the background frequency, are connected. However in plateau models, like $\alpha$-attractors, a large hierarchy can exist between them (Fig.~\ref{fig:periodvsa}).
  The potential exhibits more mass-scales, including the local curvature of the spectator field potential $V_{\chi\chi}(\phi=0,\chi=0)$ and the average frequency of the $\chi$ fluctuations $\langle \omega_\chi \rangle$ over one background period. The former defines the mass of $\chi$ particles, while the latter is related to the existence of broad or narrow resonance.

 The field-space curvature ${\cal R} \propto \alpha^{-1}$ enters the effective frequency of $\chi$ fluctuations through the combination ${1\over 2} {\cal R} \dot\phi^2$. This drives the efficient tachyonic resonance. For both the T- and E-model, this combination peaks close to $-1$, when the background field crosses the origin.

 Finally, the maximum value of the potential curvature   $V_{\phi\phi}^{\rm (max)}$, as well as the width of the ``spike'' measured as $\Delta\phi$ or $\Delta t$, control the higher harmonic content of the background motion. Due to the structure of the E-model potential, $V_{\phi\phi}^{\rm (max)}$ is also related to the spike in the effective frequency of the $\chi$ fluctuations, $V_{\chi\chi}^{\rm (max)}$.

\begin{figure}
\centering
\includegraphics[width=0.45\textwidth]{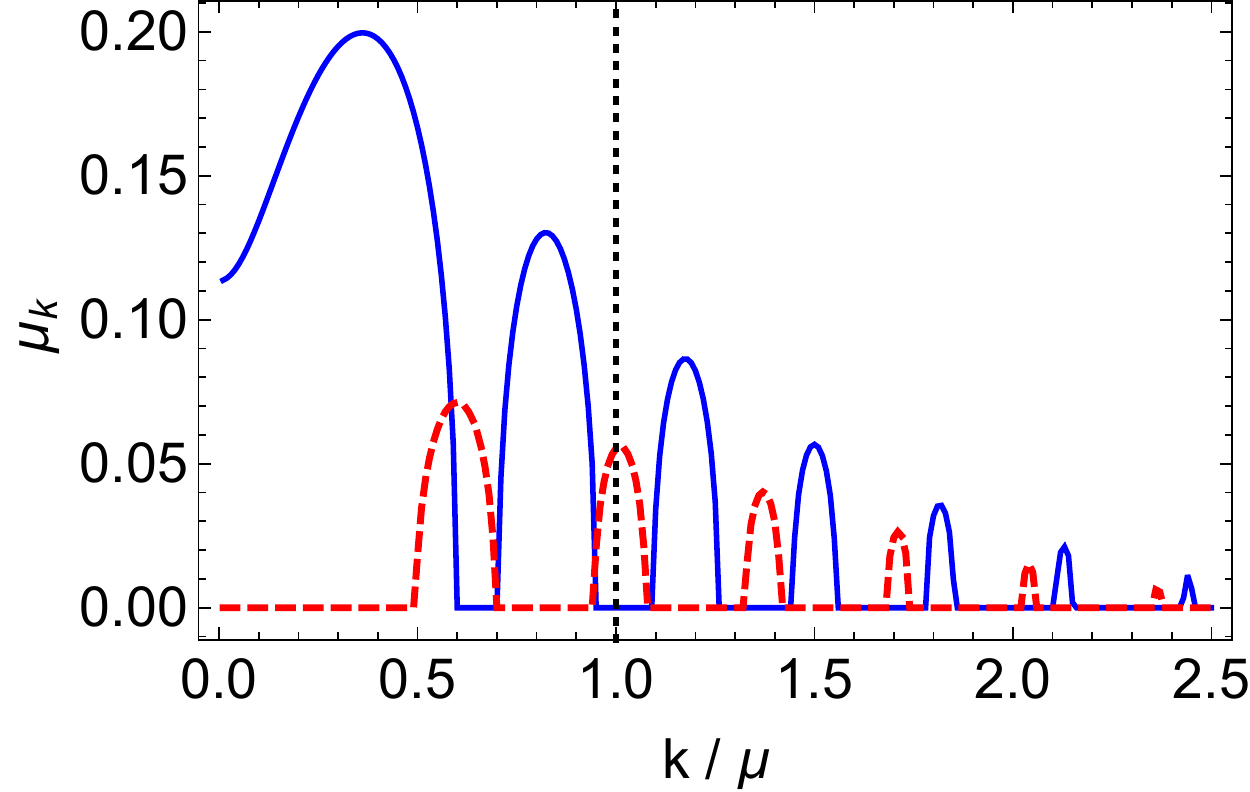}
\includegraphics[width=0.45\textwidth]{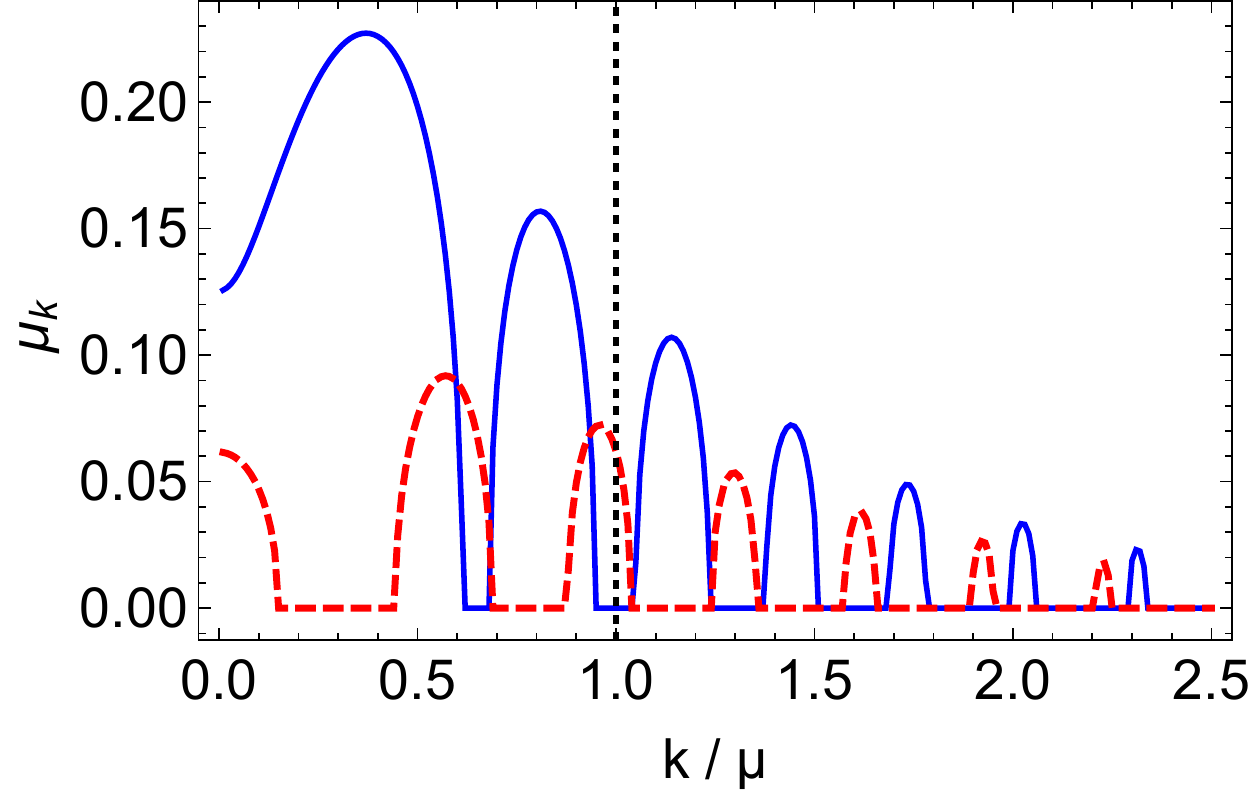}
\caption{The Floquet exponent $\mu_k$ for $\alpha=10^{-4}$, $\phi=\phi_{\rm end}$ and $n=3/2,2$ (left and right respectively). The blue curves correspond to the full Floquet exponent, while the red-dashed one correspond to the Floquet exponent computed by neglecting the field-space contribution. 
The Floquet exponents are measured in units of $\mu^{-1}$. The vertical dotted lines distinguish the regimes $k<\mu$ and $k>\mu$. The regime $k<\mu$  is controlled by the Ricci term in the effective mass, since  the large field-space induced instability is absent in the case of the potential-only calculation.
The regime $k>\mu$ is populated by multiple instability bands in both cases, with minor differences in position and height. We can thus deduce that 
parametric resonance in this regime is dominated by the effects of the potential term.
 }
 \label{fig:floquetVR}
\end{figure}

Having seen that the field-space contribution is similar for the E- and T-models and also similar among different parameter choices $n$ and $\alpha\ll1$, we turn our attention to disentangling the potential and background contributions to the parametric resonance.
For that we  construct the Floquet diagrams for $\delta\chi$ by neglecting the field-space contribution. Fig.~\ref{fig:floquetVR} shows the Floquet exponents for $n=3/2$ and $n=2$. We see that the exponents arising from the full $\delta \chi$ effective mass and those that are computed by considering only the potential and wavenumber contributions are very similar for $k> \mu$ and differ greatly for $k\lesssim \mu$, where the full system shows much more efficient particle production than the potential-only contribution.

We can thus conclude that the high-$k$ resonance bands are mostly controlled by the potential. By contrast, the resonance structure differs greatly for $k\lesssim \mu$. This is due to the fact that the tachyonic part strongly enhances modes with $k\lesssim \mu$, as shown extensively for the T-model in Ref.~\cite{ Iarygina:2018kee}, while it plays a subdominant role for large wavenumbers.

The existence for the multiple resonance bands for the E-model and not the T-model ~\cite{Iarygina:2018kee} is rooted in the existence of another mass-scale in the problem $V_{\phi\phi}^{\rm max}\propto V_{\chi\chi}^{\rm max} $, which leads the inflaton field and the effective mass of the $\chi$ fluctuations to acquire a large number of higher harmonics.

Before proceeding to compute particle production in an expanding universe, we wish to make a general comment in order to clear a common misconception in the literature. Frequent use of the term ``$\alpha$-attractors'' is made to describe single-field systems with flat potentials of the form $V=V_0 \left | 1-e^{-\phi/\Lambda}\right |^{2n}$ or $V=V_0 \left | \tanh(\phi/\Lambda) \right |^{2n}$. However the flattening of the potential is merely a by-product of a more general feature of $\alpha$-attractors: the existence of a hyperbolic field-space manifolds. As we have demonstrated in the present work and in Ref.~\cite{Iarygina:2018kee}, along with similar work by other authors,  the presence of a second field is crucial for the full dynamics of $\alpha$-attractors during preheating. The full two-field dynamics must be considered in order to properly extract the predictions of these models.

\subsection{Expanding Universe }
\label{sec:numerics}

Having extensively analyzed the parametric resonance structure of the generalized two-field E-model for any value of the potential steepness parameter $n$ and the field space curvature parameter $\alpha$, we now incorporate the effects of the non-zero expansion rate of the universe during preheating. While there are semi-analytic methods to incorporate the effects of the expansion in parametric resonance studies, using either Floquet theory or the WKB approximation (see e.g. Refs.~\cite{Krajewski:2018moi, Adshead:2015pva, WKBtachy}), we will not rely on them, since they do not provide anything further in this case, in terms of intuitive understanding, to the static universe analysis. We will instead numerically compute the evolution of fluctuations, taking into account the expansion of the universe and the red-shifting of the amplitude of the background inflaton oscillations. We will however neglect the back-reaction of the fluctuations onto the inflaton condensate and the non-linear mode-mode coupling of the fluctuations.

Our present study can be used as a strong indication for the parameter values that can lead to complete preheating, as well as elucidating the differences between the T- and E-models. Ultimately, the question of complete preheating and subsequent thermalization will have to be decided using lattice simulations, such as the ones presented in Ref.~\cite{Krajewski:2018moi} for the T-model and in Ref.~\cite{Nguyen:2019kbm}  for the related family of $\xi$-attractors. In the case of the generalized two-field T-model, our semi-analytical results were shown to agree with the full lattice computation for a broad range of parameters, while at the same time elucidating the underlying physics and demonstrating the scaling properties of the Floquet charts \cite{Krajewski:2018moi, Iarygina:2018kee}. In the present work, we show that single-field simulations are unable to capture the most important time-scales, which are controlled by the tachyonic growth of the spectator field in  both the E- and T-models of $\alpha$-attractors. Section~\ref{sec:massscales} suggests that this effect will carry over to other models with negatively curved field-space manifolds

\begin{figure}
\centering
\includegraphics[width=.325\textwidth]{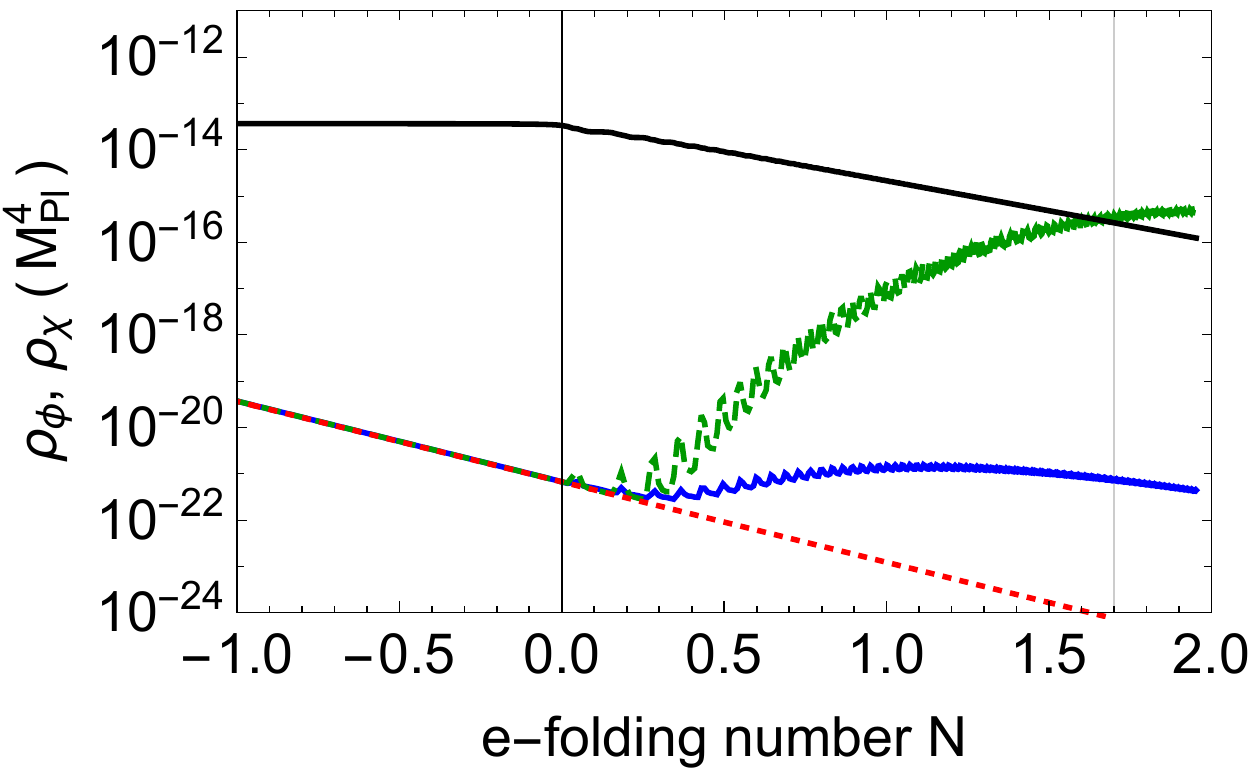}
 \includegraphics[width=.325\textwidth]{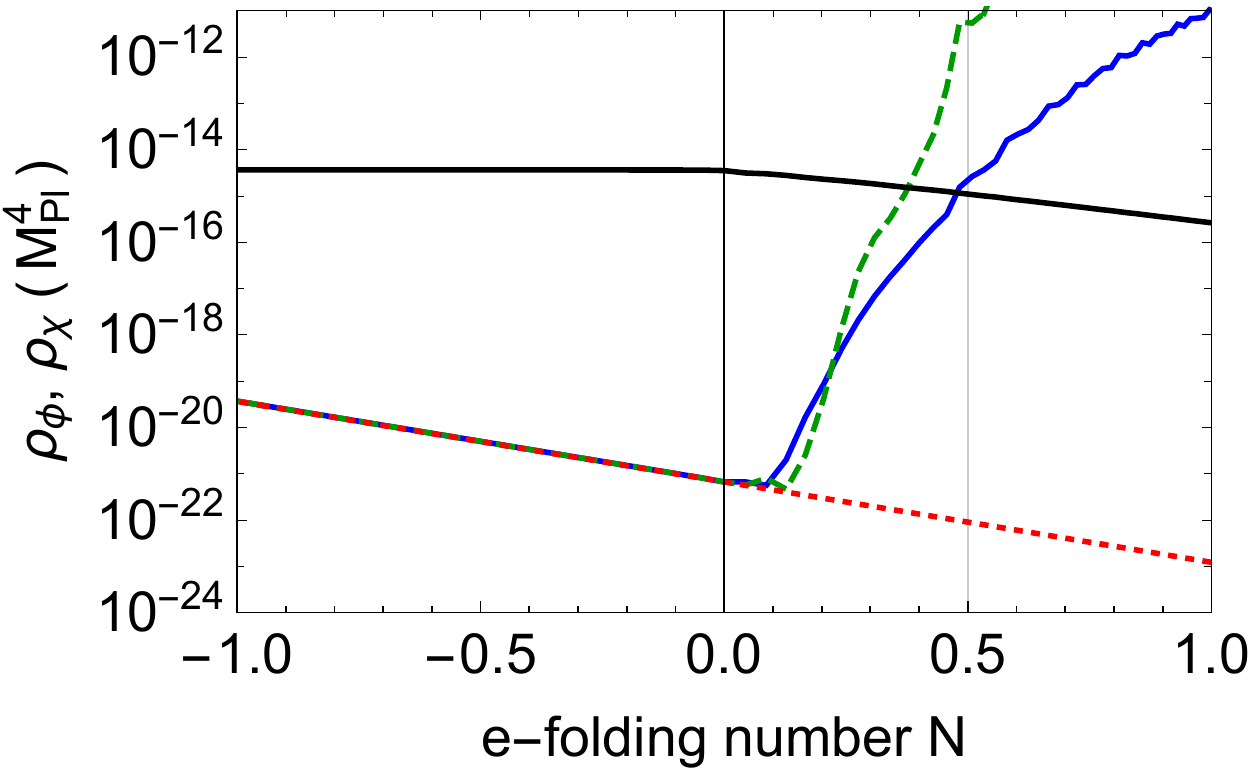}
\includegraphics[width=.325\textwidth]{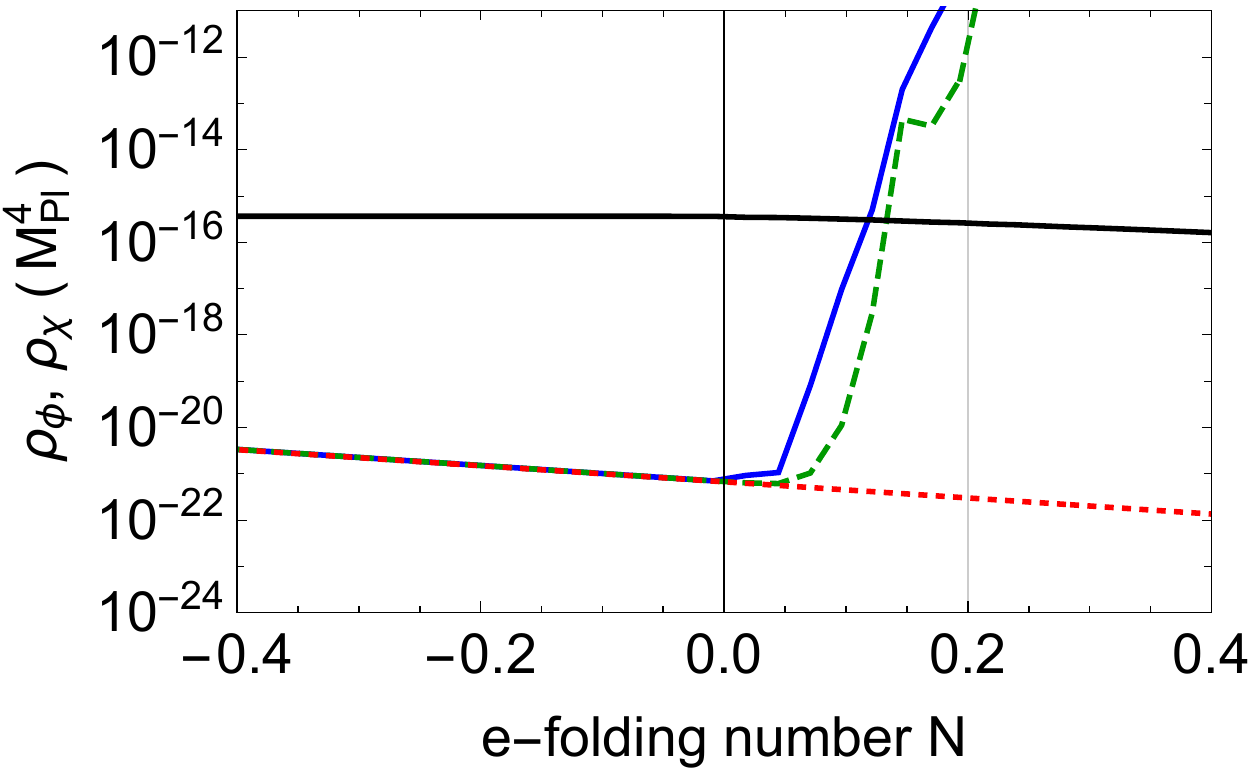}
\\
\includegraphics[width=.325\textwidth]{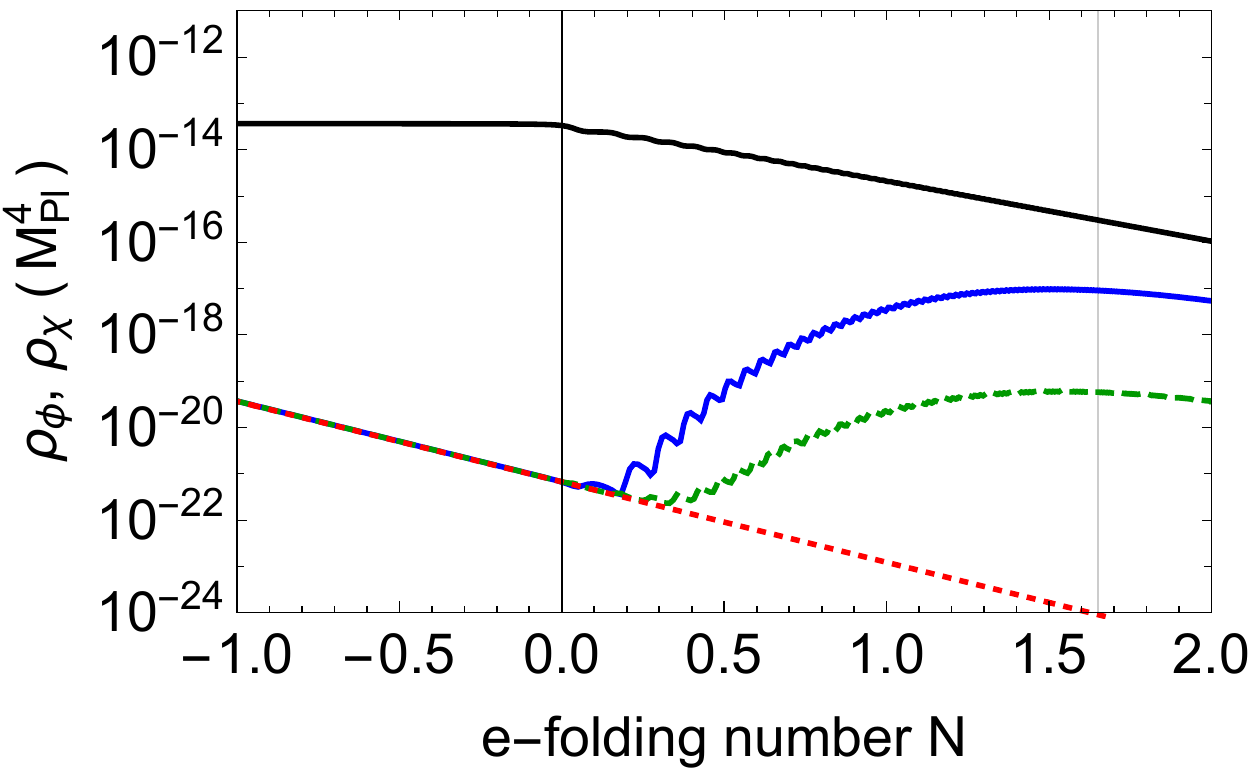}
\includegraphics[width=.325\textwidth]{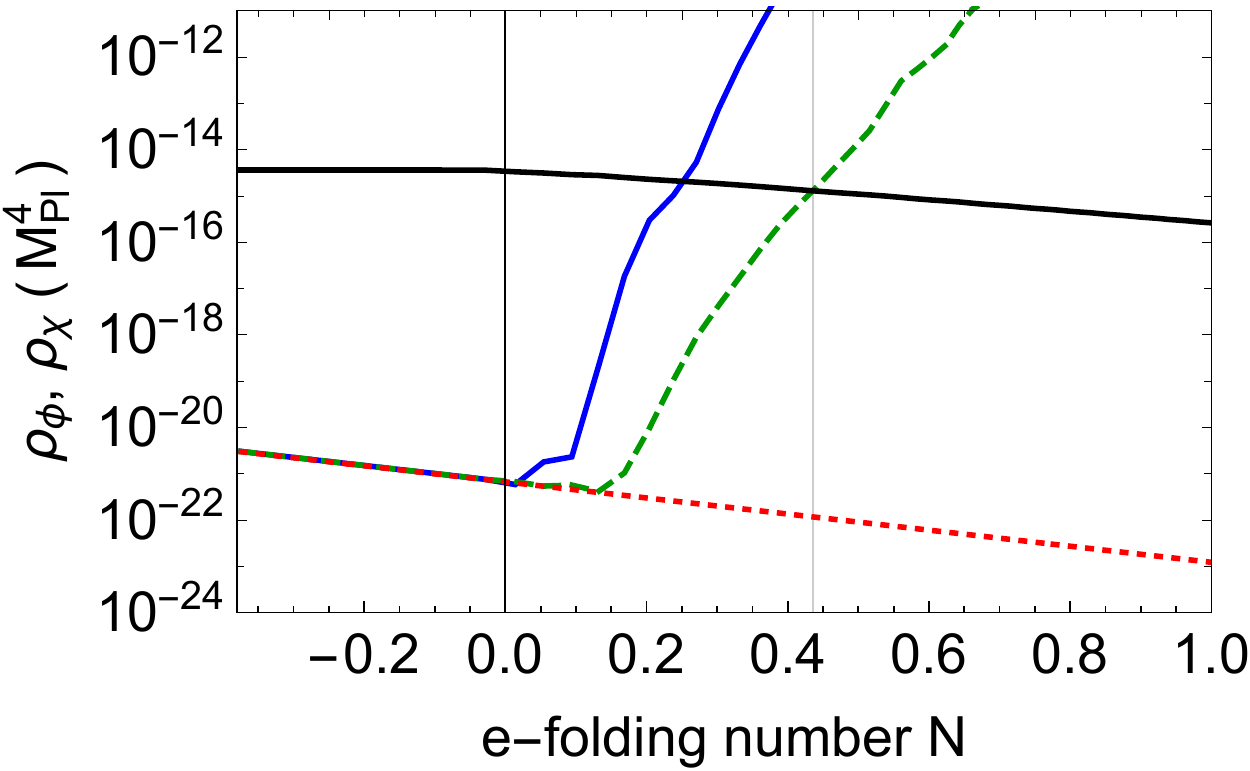}
 \includegraphics[width=.325\textwidth]{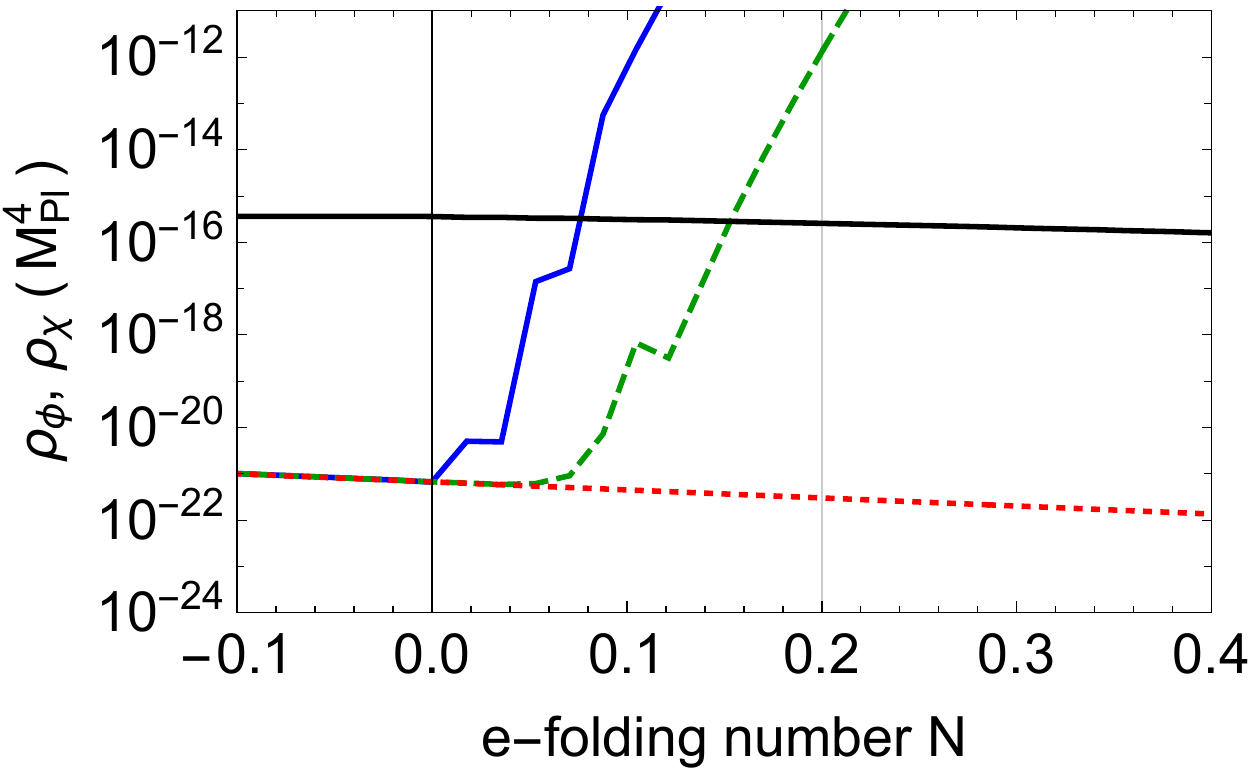}
\caption{Energy density in $\phi$ and $\chi$ fluctuations (green-dashed and blue) and the background energy density of the inflaton (black) as a function of $e$-folds for the E-model (upper panels) and the T-model (lower panels) with $n=1$ and $\alpha=10^{-3},10^{-4},10^{-5}$ (left to right).
We see efficient preheating for the E-model for $\alpha=10^{-3}$, which is absent for the T-model. Furthermore, the E-model for $n=1$ and $\alpha >10^{-5}$ preheats predominately through inflaton self-resonance, while in the case of the T-model tachyonic amplification of the spectator field is always stronger than inflaton self-resonance.
 }
 \label{fig:ETgrowth}
\end{figure}

Fig.~\ref{fig:ETgrowth} shows the growth of $\phi$ and $\chi$ fluctuations for the T- and E-models with n=1. This can be thought of as the physically ``generic'' case, since it describes massive particles  in the small field limit. We see that the behavior of the two models is qualitatively different. In the case of the E-model the $\phi$ resonance is stronger, leading to possibly complete preheating already at $\alpha=10^{-3}$, where the $\chi$ resonance is vastly subdominant. The two become comparable at  $\alpha\lesssim 10^{-3}$, where preheating can complete within less than an $e$-fold.

 In the case of the T-model, the $\chi$ resonance is always stronger than the $\phi$ resonance for $n=1$. We see that the T-model does not completely preheat for $\alpha=10^{-3}$. In the case of efficient parametric resonance in the $\phi$ field (for $\alpha\lesssim 10^{-4}$), lattice simulations have shown the fragmentation of the inflaton condensate and the subsequent formation of localized structures (oscillons) \cite{Lozanov:2017hjm}. It is interesting to consider whether
tachyonic resonance into the $\chi$ field can deplete the inflaton condensate before it has time to fragment. Even in the case of a fragmented inflaton, one must consider the two possibilities: either
  resonance of the $\phi$ field to $\chi$ modes can proceed within the oscillons leading to the decay of the localized structures
 or  composite oscillons consisting of both fields can form (see e.g. \cite{Sfakianakis:2012bq, Gleiser:2011xj}).
Parametric resonance of scalar fields in localized structures, such as oscillons \cite{Hertzberg:2010yz}, Q-balls \cite{Kawasaki:2013awa} or axion clumps \cite{Hertzberg:2018zte}, is similar to the homogenous field case with one important qualitative difference. If the Floquet exponent (computed by neglecting the spatial structure of the clump) is smaller than the time-scale on which the produced particles escape the clump, Bose enhancement is destroyed and the parametric resonance effectively shuts off  \cite{Hertzberg:2018zte}. In our case the maximum Floquet exponent is $\mu_k\sim \mu$, where $\mu= {\cal O}(10^{-6})M_{\rm Pl}$. The size of the oscillons formed in  single-field models with $\alpha$-attractor-like potentials is
  $L = {\cal O}(\mu^{-1})$. The comparison between the homogeneous field Floquet exponent and the escape time $\mu_{\rm esc} \approx1/(2L)$ shows that it is indeed possible for efficient production of $\chi$ particles to proceed within the oscillon, but a detailed calculation is needed to reach a definite conclusion, since non-trivial ${\cal O}(1)$ factors are involved in the calculation.

\begin{figure}
\centering
 \includegraphics[width=.45\textwidth]{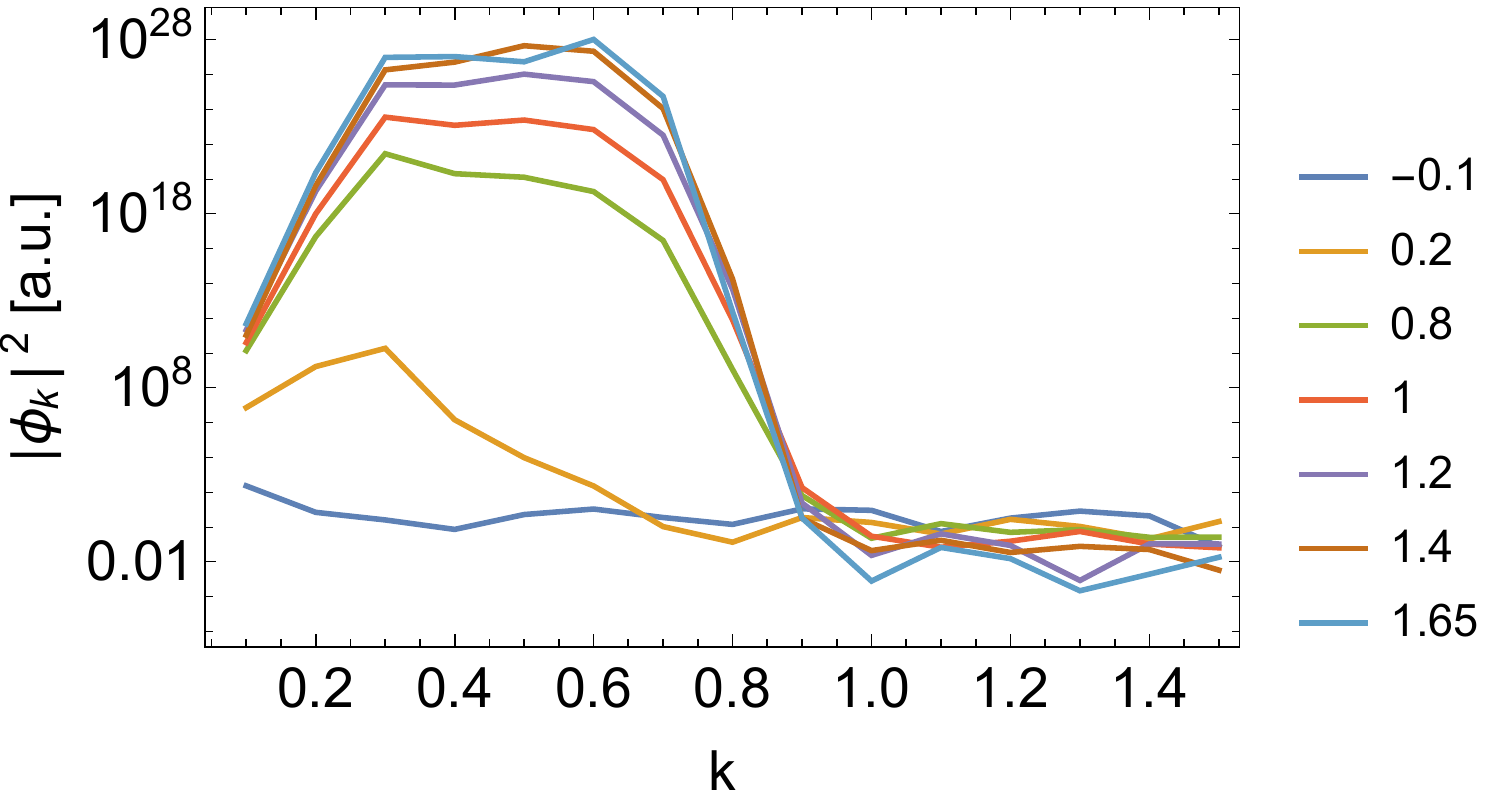}
 \includegraphics[width=.45\textwidth]{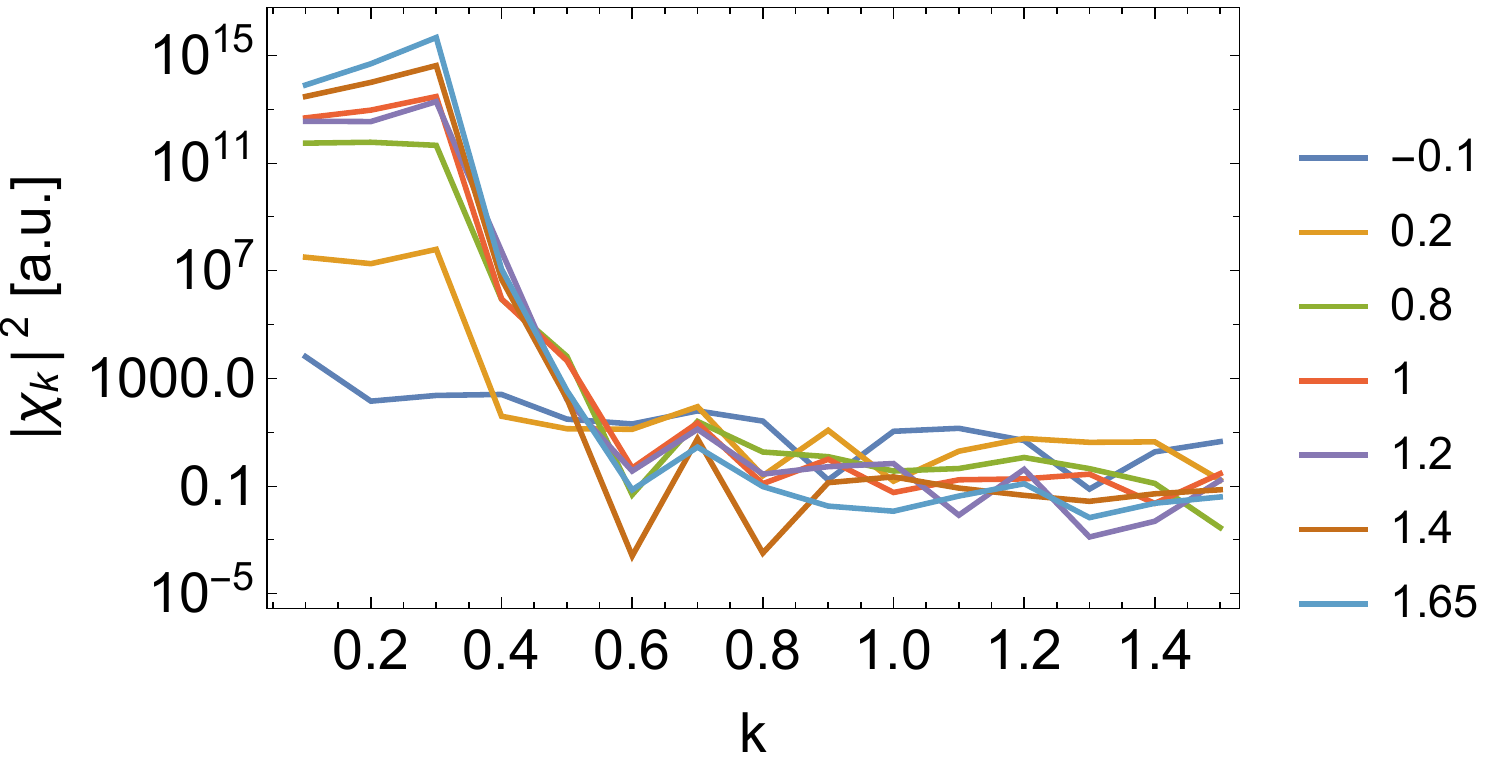}
 \\
 \includegraphics[width=.45\textwidth]{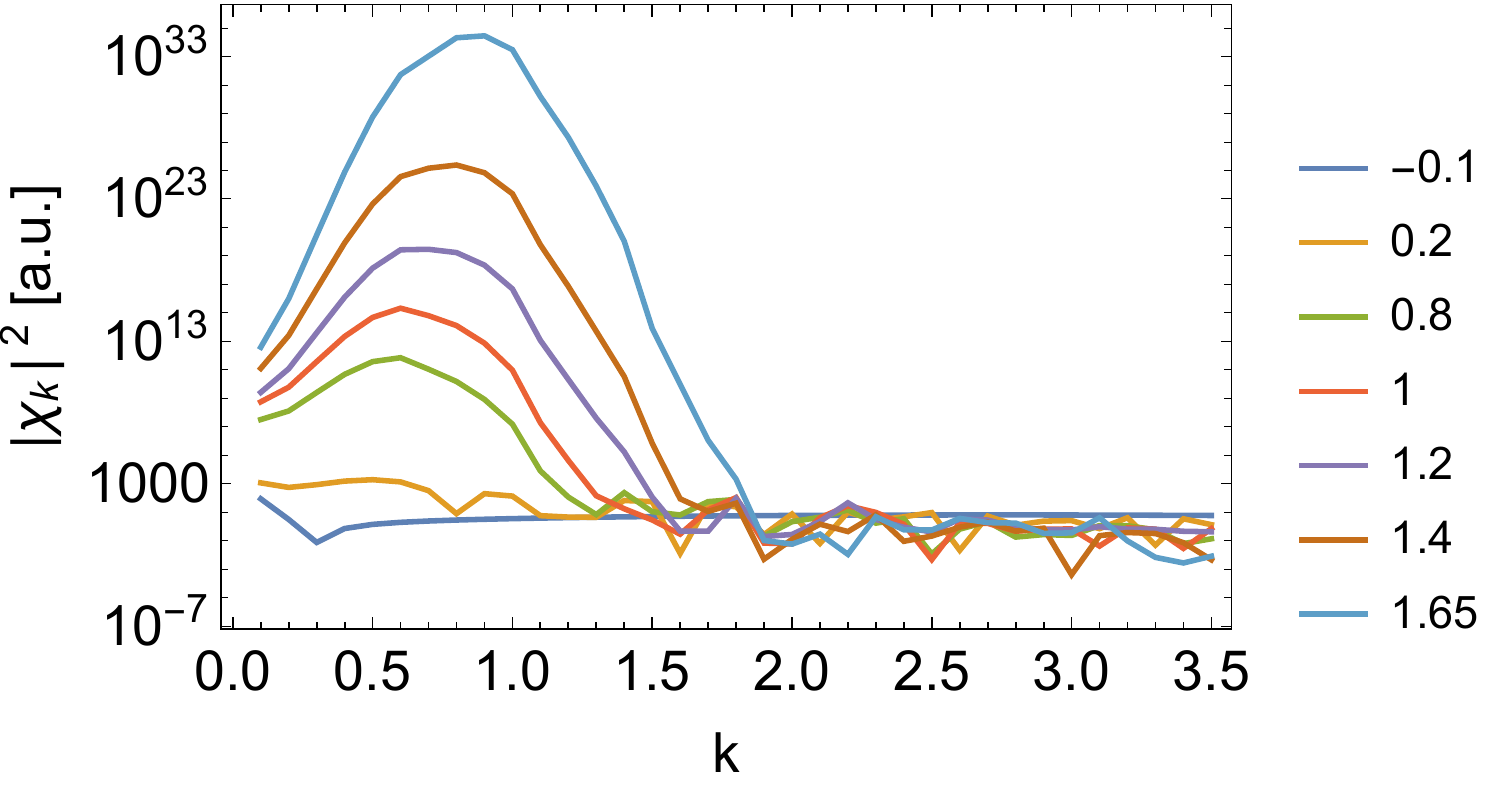}
  \includegraphics[width=.45\textwidth]{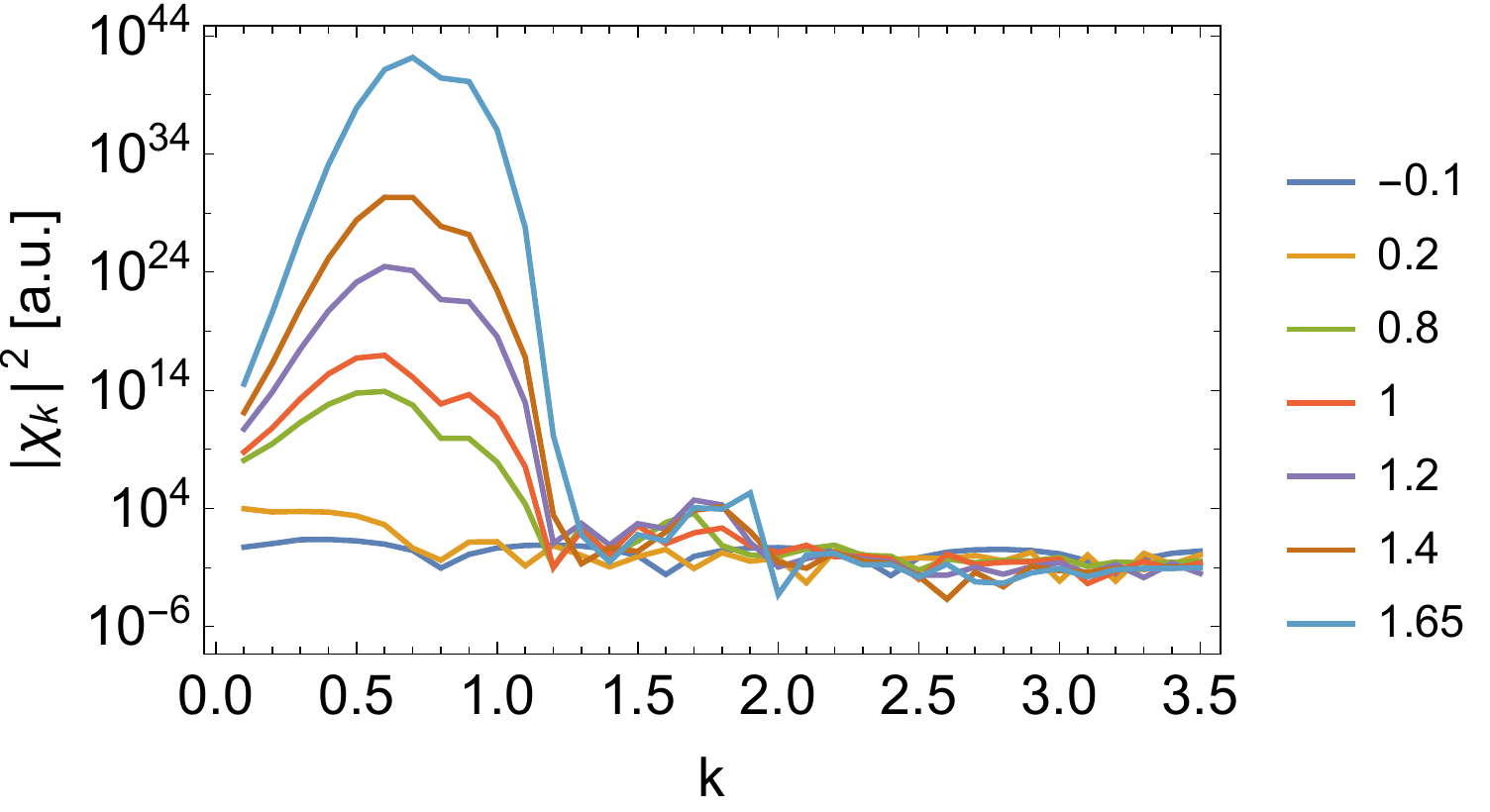}
\caption{The spectra of the $\phi$ fluctuations $\left | \phi_k \right |^2$ and $\chi$ fluctuations $\left | \chi_k \right |^2$
(in arbitrary units)  in the E-model as a function of
the wavenumber $k$ (in units of $\mu$)  at different times for $\{n,\alpha\} = \{1,10^{-4}\}$ (upper panels) and
 $\{n,\alpha\} = \{3/2,10^{-3}\}$, $\{n,\alpha\} = \{2,10^{-3}\}$ (lover panels, left and right respectively). The times corresponding to the various
curves are shown in the legend of each panel, measured in $e$-folds after the end of inflation
(negative values correspond to spectra during the last stages of inflation).
We see that for $n=1$ the amplification of the $\phi$ (inflaton) modes is much stronger than that of the $\chi$ (spectator) modes. 
For $n=3/2$ we see that at later times, the range of excited $\chi$ wavenumbers grows, while for $n=2$ it remains constant at $k_{\rm max}\simeq \mu$. This is in agreement with the behavior of the effective mass shown in Fig.~\ref{fig:meffn3ov2andn2}.
}
 \label{fig:Pk}
\end{figure}

Fig.~\ref{fig:Pk} shows the spectrum of produced $\phi$ and $\chi$ modes during the initial stages of preheating, before backreaction effects become important.
We see that for $n=1$ and $\alpha=10^{-4}$ the parametric resonance of the $\phi$ modes is stronger than that of the $\chi$ ones. This can be expected based on the results of Fig.~\ref{fig:n1meff}, where we see that the two effective masses oscillate around $m^2=4\mu^2/3$, while the oscillation amplitude for the $\phi$ effective mass is larger, leading to a stronger resonance (see e.g. Ref.~\cite{Amin:2014eta}).
The similarity of the two effective masses for $\phi$ and $\chi$ fluctuations is a direct consequence of the E-model potential, which arises from a supergravity construction, where one specifies the potential of a complex scalar field, whose components are related to $\phi$ and $\chi$, as shown in Appendix A.
 An interesting feature arises when we compare the $\chi$ spectrum for $n=3/2$ and $n=2$. For $n=2$ the maximum excited wavenumber is set by the initial amplification and is found to be $k_{\rm max}\simeq 1.2\mu$. For $n=3/2$  the value of $k_{\rm max}$ grows with time. This can be traced back to the behavior we saw in Fig.~\ref{fig:nad}, where the adiabaticity violation for $n=3/2$ was shown to grow with time, contrary to $n=2$. This behavior is explained by using the results of Fig.~\ref{fig:meffn3ov2andn2}, where it was demonstrated that the height of the effective mass spike --which controls the large $k$ resonance-- red-shifts slower than $a^{-2}$ for $n=3/2$, hence it becomes progressively more important compared to the wavenumber term $k^2/a^2$.

\begin{figure}
\centering
 \includegraphics[width=.45\textwidth]{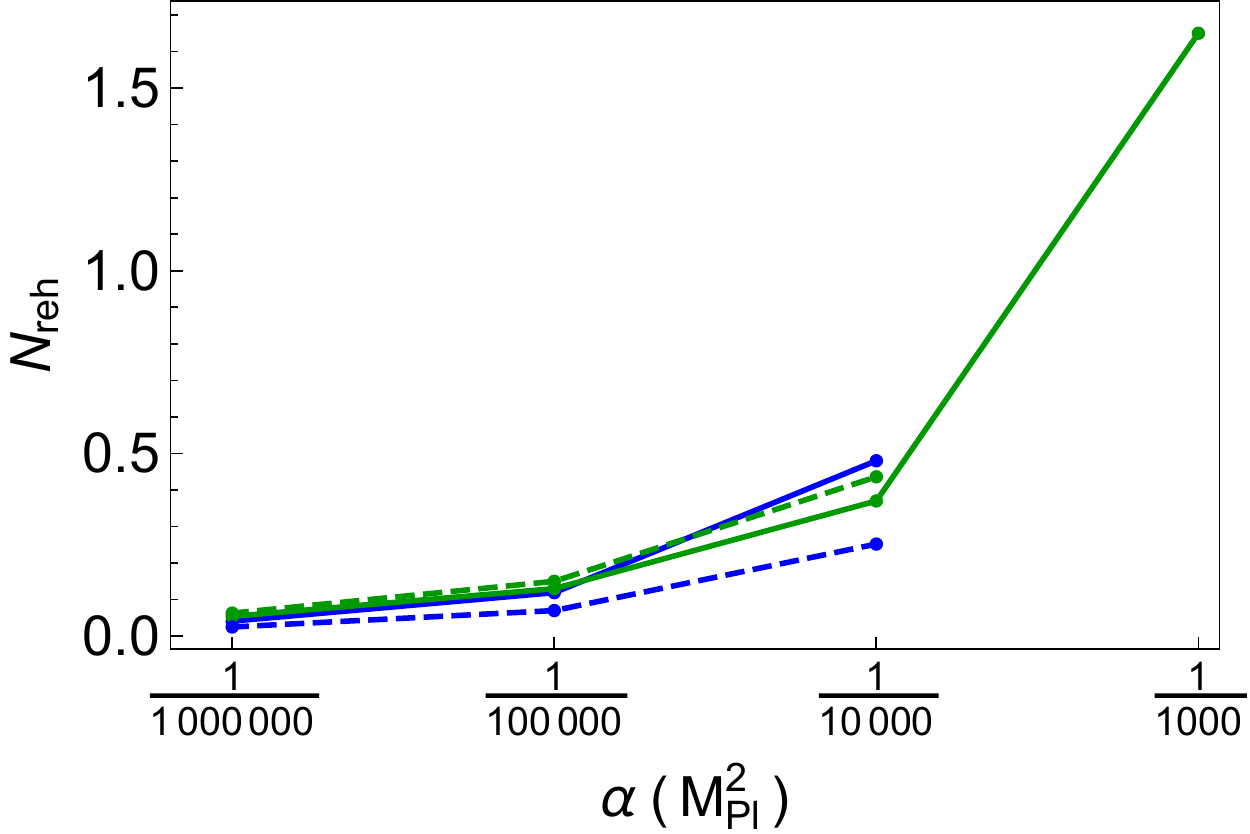}
 \includegraphics[width=.45\textwidth]{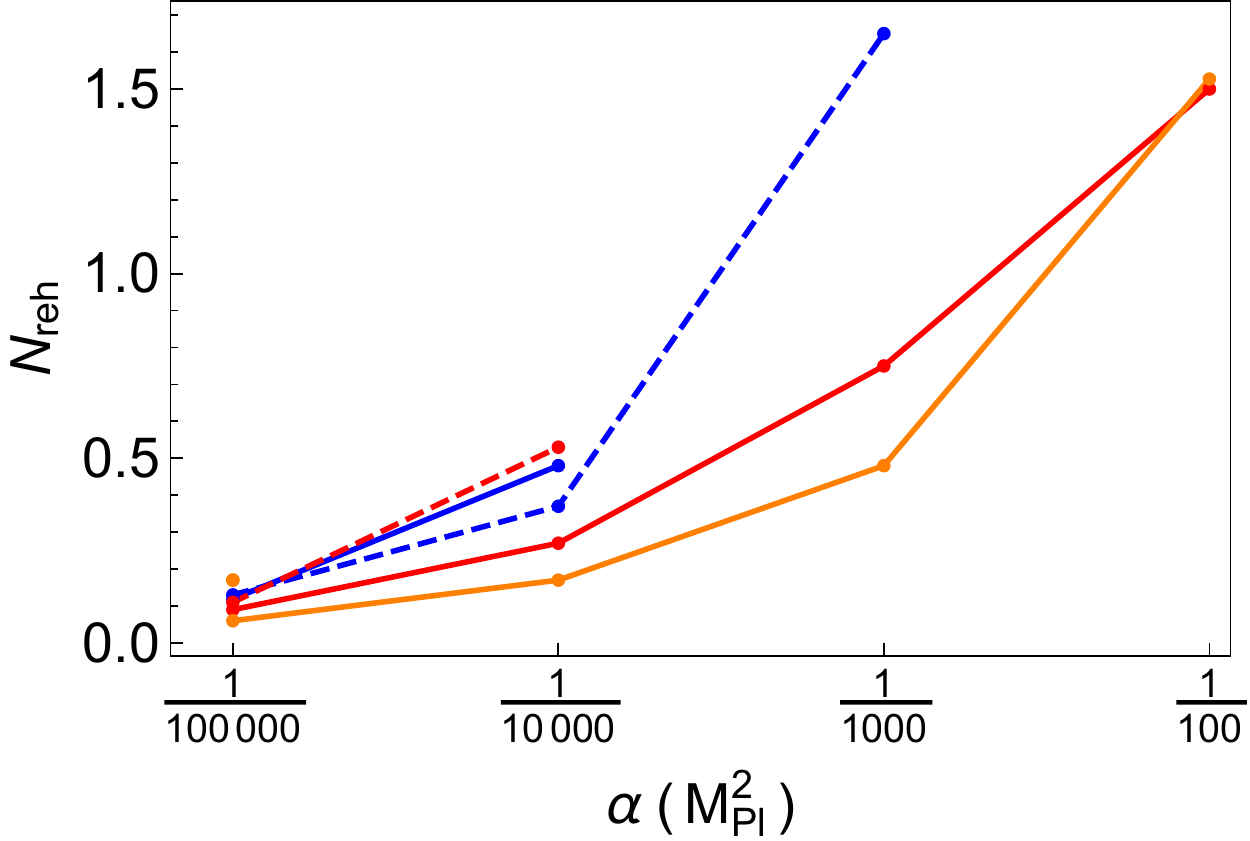}
\caption{{\it Left:}
The time of reheating for $n=1$ through $\phi$ (green) and $\chi$ (blue) fluctuations for the T- and E-models (dashed and solid curves respectively)
{\it Right:} The E-model behavior for $n=1, 3/2, 6$ (blue, red, orange) and resonance through $\phi$ or $\chi$ modes (dashed and solid curves respectively).
For $10^{-4} \lesssim\alpha\lesssim10^{-3}$ the E-model preheats predominately through inflaton self-resonance, while the T-model does not completely preheat. For $n\ge 3/2$ the E-model preheats through amplification of the spectator field for $\alpha \lesssim 0.01$. For small values of $\alpha \lesssim 10^{-4}$ preheating is practically instantaneous (lasting less than one $e$-fold) for any potential parameter $n$. 
}
 \label{fig:Nreh}
\end{figure}

Finally, Fig.~\ref{fig:Nreh} provides a visual summary of the preheating efficiency for different models and parameter values. For the case of massive particles, $n=1$, the T-model exhibits efficient preheating through the $\chi$ field for $\alpha \lesssim 10^{-4}$. On the other hand, parametric resonance in the E-model is more efficient, starting at $\alpha \approx 10^{-3}$, albeit through self-resonance of the $\phi$ field,  since tachyonic production of $\chi$ modes is shut off due to the large positive mass term (see Fig.~\ref{fig:n1meff}). For steeper potentials $n\ge 3/2$, self-resonance of the $\phi$ field becomes progressively more inefficient, while tachyonic resonance of $\chi$ modes becomes efficient already at $\alpha \approx 10^{-3}$ and is able to completely preheat the universe within $1.5$ $e$-folds after the end of inflation, much faster than a naive single-field analysis would suggest.

Overall this means that $\alpha$-attractors with $n=1$ and $\alpha \gtrsim 10^{-3}$, equivalently a tensor to scalar ratio $r\gtrsim 10^{-6}$, can undergo a long matter-dominated expansion after the end of inflation and the decay of the inflaton condensate can proceed only through perturbative decays to other particles. Unfortunately, there is no concrete theoretical motivation for the size of such couplings, hence the transition to radiative degrees of freedom cannot be estimated. For potentials describing massless scalar fields, the decay of the condensate to radiative degrees of freedom can occur very quickly through tachyonic production of the spectator field $\chi$, for both the E-model explored here and the T-model explored in Refs.~\cite{Krajewski:2018moi, Iarygina:2018kee}.

\subsection{Gravitational waves}

It has been shown that efficient preheating leading to a turbulent fluid can lead to the production of gravitational waves. This can occur e.g. through coupling of the inflaton to gauge fields, as well as through inflaton decay through self-resonance. The latter case is similar to the current analysis of parametric resonance in $\alpha$-attractors, where the spectator field is amplified. Using the ``rule of thumb'' estimates of Ref.~\cite{Giblin:2014gra}, the frequency of GW's today is related to the Hubble scale at the time of generation and the dominant wavenumber of the source as
\beq
f\simeq 2.7 \cdot 10^{10} {k_{\rm phys} \over \sqrt{M_{\rm Pl} H}} \, {\rm Hz} \, .
\eeq
In most models, the physical wavenumber is proportional to the Hubble scale, thus reducing the Hubble scale reduces the frequency of the GW signal as $f\propto \sqrt{H}$. As has been extensively shown in the present work and in Refs.~\cite{Iarygina:2018kee, Krajewski:2018moi}, preheating in $\alpha$-attractors occurs at a typical wavenumber $k\sim \mu$, while the Hubble scale scales as $M_{\rm Pl}H\sim \sqrt{\alpha} \mu $. Using these estimates, the peak GW frequency today becomes
\beq
f \sim  {10^{7} \over { \alpha^{1/4} }} \, {\rm Hz}   \, ,
\label{eq:GWfreqvsalpha}
\eeq
where we used the value of $\mu\simeq 6\times 10^{-6}\, M_{\rm Pl}$ required to produce the observed amplitude of density fluctuations. Thus, contrary to the common behavior of GW from preheating, reducing the Hubble scale through reducing $\alpha$ (increasing the field-space curvature) will actually increase the peak frequency of GW's, pushing them further away from the observable range of interferometers. It remains interesting to follow progress in detection strategies for Ultra High Frequency gravitational waves, as many early universe sources operate in this regime.

\section{Summary and Discussion}
\label{sec:Conclusion}

In the present work we revisited the multi-field behavior of the generalized E-model, which consists of two-fields on a hyperbolic manifold. More highly curved manifolds lead to a lower Hubble scale and correspondingly to a smaller tensor to scalar ratio. 
We focus on  the region $10^{-7}\lesssim r \lesssim 10^{-4}$, which is below the direct detection limits of the next generation CMB experiments.
The potential of the inflaton field $\phi$ is asymmetric with respect to the global minimum at the origin. It exhibits a flat plateau where inflation is realized, leading to the usual Starobinsky-like predictions $n_s \sim N_*^{-1}$ and $r\sim N_*^{-2}$
and a sharp potential``wall'', which the field probes after inflation and during preheating.
By contrast,  the potential of the spectator field $\chi$ is symmetric with respect to the minimum at $\chi=0$.
Several studies in the literature have examined the equivalence of the single-field behavior of the E- and T-models during inflation. Going beyond these studies we were able to show that the similarities of the T- and E-model extend beyond the single-field analysis. In fact, their multi-field behavior during inflation is identical up to slow-roll corrections. 
Previous analyses of the E- and T-models have established the existence of a single field attractor along the minimum of the spectator field \cite{Carrasco:2015rva}. In order to assess the possibility of multi-field effects beyond the single-field attractor, we
 examined the basin of attraction by choosing a wide variety of initial conditions along iso-potential surfaces. We showed that the global behavior of this system consists of two straight inflationary trajectories, each keeping one of the fields constant.
 While each of them can be made arbitrarily long by appropriate choice of initial field values, only the final trajectory, the single field attractor along $\chi=0$ gives results that are in agreement with the CMB. 
 It remains to be seen, if similar two-stage behavior appears in other realizations of $\alpha$-attractors and if some well-motivated models exist where both stages can lead to predictions that are consistent with CMB measurements.
Furthermore, the two straight trajectories are joined  by a sharp turn and a brief period of oscillations around $\chi=0$. An assessment of the observability of such a signal \cite{Konieczka:2014zja} at CMB or LSS scales is beyond the scope of this work and is left for future analysis.

Reheating is crucial for connecting inflationary predictions to CMB observables. Especially in the case of inflationary models that follow the predictions of the Starobinsky model, $n_s=1-2/N_*$, the latest {\it Planck} results \cite{Akrami:2018odb} are putting mild pressure on $N_* \simeq 50$, instead preferring a value closer to $N_* = 60$. Such models include the Starobinsky model, Higgs inflation and its generalizations of non-minimally coupled models \cite{KS, nonm1, Bezrukov:2007ep} and of course $\alpha$-attractors. These results and the anticipated improvement from next generation experiments, like LiteBird and CMB-S4, can significantly constrain the existence of a prolonged matter dominated expansion after inflation.

The preheating efficiency depends on the amplification of fluctuations of the inflaton and spectator fields, which is governed by their corresponding effective masses. The coupled metric perturbations component that contributes to the inflaton self-resonance is proportional to $\sqrt{\alpha}$ and  becomes subdominant for  $\alpha \ll 1$.  The term that encodes the space-time curvature is even smaller, being proportional to $\alpha$. The inflaton self-resonance is thus solely determined by the second derivative of the potential, while the resonance structure of the spectator field is determined by the interplay of the potential contribution and the field-space effects, which do not scale  with $\alpha$.
Furthermore, the wavenumbers that are amplified  due to parametric resonance of either the inflaton or spectator fields do not depend on $\alpha$ and scale as $k\lesssim {\cal O}(1) \mu$, where $\mu\simeq 6\times 10^{-6} \,M_{\rm Pl}$ in order to produce  the correct amplitude of density perturbations. By contrast the Hubble scale depends on $\alpha$ as $H\propto \mu\sqrt{\alpha}$. Hence for small $\alpha$ the dominant preheating dynamics is occurring at very sub-horizon (sub-Hubble) scales. This creates the apparent paradox that reducing the inflationary scale will lead to the frequency of GW's from preheating to increase as $f\propto \alpha^{-1/4}\propto H^{-1/2}$ rather than decrease, as in the usual case for low-scale inflation.

The preheating efficiency of the E-model is qualitatively different than that of the T-model \cite{Krajewski:2018moi, Iarygina:2018kee}. The parametric resonance of $\phi$ fluctuations is significantly more enhanced in the $E$-model, as compared to the T-model. This can be traced back to the inherent asymmetry of the E-model potential, which introduced a spike in the effective mass of the fluctuations and higher harmonic content in the background motion.
For massive fields, the tachyonic component of the spectator effective mass in the E-model is canceled by the contribution of the potential term, hence tachyonic amplification is completely shut off. However the spike introduced by the potential term leads to efficient parametric  resonance. However, a similar spike is present in the self-resonance of the inflaton field and is more pronounced than the one in the spectator effective mass. This leads to the E-model preheating predominately through self-resonance for massive fields ($n=1$). Furthermore, preheating in the E-model is efficient for higher values of $\alpha$ than in the T-model ($\alpha \sim 10^{-3}$),  leading to the first distinguishing feature between them.

For massless fields, or equivalently potentials that behave as $V(\phi,\chi) \propto \{ |\phi|^{2n},|\chi|^{2n}\}$ with $n\ge 3/2$ close to the minimum, the spectator field dominates the preheating behavior of the E-model, leading to fast preheating for $\alpha\lesssim 0.01$. For small wavenumbers ($k\lesssim \mu $), the $\chi$ modes grow tachyonically due to the effects of the negative field-space. For larger wavenumbers, the amplification is controlled by the potential spike, leading to parametric resonance and multiple instability bands. For highly curved manifolds $\alpha \lesssim 10^{-4}$, preheating concludes within less than an $e$-fold for any potential choice.
The preheating dynamics of both the E- and T-model reinforce the need for notational clarity regarding $\alpha$-attractors: a flat potential 
of the form $V\propto \left | 1-e^{-\phi/\Lambda}\right |^{2n}$ or $V\propto \left | \tanh(\phi/\Lambda) \right |^{2n}$
 that is usually associated with $\alpha$-attractors should not be regarded as their main characteristic. On the contrary, the hyperbolic manifold, from which the  potential flatness originates, and their multi-field nature, should be taken into account to properly address the dynamics of  $\alpha$-attractor models.
In anticipation of upcoming CMB and LSS data, that hope to further restrict the value of $n_s$ and $r$, theoretical uncertainties must become small enough to allow for an accurate comparison between theory and observation. 
Single-field simulations are unable to capture the most important preheating time-scales, which are controlled by the tachyonic growth of the spectator field in both the E- and T-models of $\alpha$-attractors. We must thus consider the full two-field dynamics in order to put $\alpha$-attractor predictions to the test.

Using the T- and E-models as characteristic examples, we analyzed the various mass-scales that control the tachyonic growth of fluctuations, making a first step towards an Effective Field Theory description of preheating in hyperbolic manifolds \cite{Giblin:2017qjp}.
The necessary presence of a spectator field, as required by the supergravity constructions of $\alpha$-attractors, make it necessary to extend the single-field preheating results found in the literature  \cite{Lozanov:2017hjm} to examine the effects of  efficient tachyonic preheating.
 Having provided a qualitative and quantitative understanding of the relevant time and mass-scales, we leave such two-field lattice simulations for future work.

\acknowledgements{
The work of AA is partially supported by the Netherlands Organization for Fundamental Research in Matter (FOM), by the Basque Government (IT-979-16) and by the Spanish Ministry MINECO (MINECO/FEDER grant  FPA2015-64041-C2-1 and MCIU/AEl/FEDER grant PGC2018-094626-B-C21).
DGW and OI are supported by the Netherlands Organization for Scientific Research (NWO).
EIS gratefully acknowledges support from the Dutch Organisation for Scientific Research (NWO) and the KITP for the generous hospitality while this work was completed.
This research was supported in part by the National Science Foundation under Grant No. NSF PHY-1748958.
 }

\section*{Appendix A: Generalization of the E-model}

For completeness, we describe here the ${\cal N}=1$ Supergravity embedding of the two-field E-model  \cite{Carrasco:2015rva} considered in the main text. We consider the super-potential
\beq
W_H = \sqrt{\alpha}\mu \, S\, F(Z)
\eeq
and K\"ahler potential
\beq
K_H = {-3\alpha \over 2} \log \left [{ (1-Z\bar Z)^2 \over (1-Z^2)(1-\bar Z^2)}\right ]+S \bar S \, .
\eeq
Using the relation between the K\"ahler potential and the superpotential
\beq
Z = {T-1\over T+1}
\eeq
and choosing
\beq
F(Z)=\left (
{2Z\over Z+1}
\right )^n
\eeq
we get
\beq
K_H = {-3\alpha \over 2} \log \left [{      (T+\bar T)^2
 \over 4 T \bar T
 }\right ]+S \bar S
 \label{eq:KH}
\eeq
and
\beq
W_H = \sqrt{\alpha}\mu S \left ( {T-1\over T}  \right )^n \, .
 \label{eq:WH}
\eeq
as in Ref.~\cite{Carrasco:2015rva}.
The potential follows to be of the form
\beq
V=\alpha \mu^2 4^n \left [
(Z \bar Z-1)^2\over (Z^2-1) (\bar Z^2 -1)
\right ]^{-3\alpha/2}
\left [
Z \bar Z \over (1+Z)(1+\bar Z)
\right]^2
\, .
\eeq
One can use multiple field-space bases to describe these models.
The choice
\beq
Z=\tanh\left ( {\phi + i\theta \over \sqrt {6\alpha}}\right )
\label{eq:Tmodeloriginal}
\eeq
was used in Ref.~\cite{Carrasco:2015rva}, leading to the kinetic term
\beq
{\cal L}_{kin} = {1\over 2} {\cal G}_{\phi\phi} \,\partial_\mu \phi\, \partial^\mu \phi
+
{1\over 2} {\cal G}_{\theta\theta}\, \partial_\mu \theta\, \partial^\mu \theta
\,
\eeq
with
\beq
{\cal G}_{\phi\phi} = {\cal G}_{\theta\theta}  ={1\over \cos^2 \left( \beta\theta \right ) }  \, , \quad \beta = \sqrt{2\over 3\alpha} \, .
\eeq
 The corresponding two-field potential is
\beq
V(\phi,\theta) = \alpha\mu^2 \left(
1-2 \cos(\beta\theta)e^{-\beta \phi} + e^{-2\beta\phi}
 \right)^n
\left |
\cos \left ( \beta \theta \right )
\right |^{-3\alpha} \, .
\eeq
 We instead choose the basis used in Ref.~\cite{Krajewski:2018moi},
 which can be derived from Eq.~\eqref{eq:Tmodeloriginal} by performing the transformation
 \beq
 \cos \left ( \beta \theta \right ) = {1\over \cosh \left (\beta \chi \right ) }\, .
 \eeq
 This leads to the kinetic term  ~\eqref{eq:Lphichi}
 \beq
{\cal L}_{kin} = {1\over 2}  \partial_\mu \chi \partial^\mu \chi +{1\over 2} \cosh^2 \left ( \beta \chi \right ) \partial_\mu \phi \partial^\mu \phi   \, ,
\label{eq:kineticterm}
\eeq
 and potential
 \beq
V(\phi,\chi) =  \alpha \mu^2 \left (
1-  {2e^{-\beta \phi} \over \cosh \left (\beta \chi \right ) } + e^{-2\beta\phi}
 \right)^n \left( \cosh(\beta \chi) \right )^{2/\beta^2} \, ,
\eeq
 where again $\beta = \sqrt{2/3\alpha}$.
 It is trivial to see that for $\chi=0$ we recover the usual expression for the E-model
 \beq
 V(\phi,\chi) =  \alpha \mu^2 \left (
1-  e^{-\beta\phi}
 \right)^{2n}  \, .
 \eeq
 where the exponent is $2n$ instead of $2$.

The kinetic term of Eq.~\eqref{eq:kineticterm} is generally written as
\beq
{\cal L} = {1\over 2} {\cal G}_{IJ} \partial_\mu \phi^I \partial^\mu \phi^J \, ,
\eeq
with $\{\phi^1,\phi^2\} \equiv \{\phi,\chi\}$ and
in the basis used the non-zero field-space quantities are
\beqn
{\cal G}_{\phi\phi} =   cosh^2(\beta \chi)
\, , \quad
{\cal G}_{\chi\chi}=1 \, \quad {\cal G}_{\phi\chi}={\cal G}_{\chi\phi}=0 \, .
\eeqn
The corresponding inverse metric is
\beqn
{\cal G}^{\phi\phi} =  \text{sech}^2(\beta  \chi )
\, , \quad
{\cal G}^{\chi\chi}=1 \, , \quad {\cal G}^{\phi\chi}={\cal G}^{\chi\phi}=0
\eeqn
Along the background trajectory $\chi=0$ the field-space is reduced to the unitary matrix ${\cal G}^{IJ}(\phi,\chi=0) = \mathbb{I}$.
 The only non-zero Christoffel symbols are
\beqn
\Gamma^\phi_{\chi\phi} =\beta  \tanh (\beta  \chi )
\, , \quad
\Gamma^\chi_{\phi\phi} =-\frac{1}{2} \beta  \sinh (2 \beta  \chi )
\eeqn
which vanish for $\chi=0$.
The non-zero components of the Riemann tensor are
\begin{equation}
\begin{aligned}
{\cal R}^\phi_{\chi\phi\chi} = -\beta^2 \, ,\quad
{\cal R}^\phi_{\chi\chi\phi} = \beta^2
\, , \quad
{\cal R}^\chi_{\phi\phi\chi} =\beta ^2 \cosh ^2(\beta  \chi )\, , \quad
{\cal R}^\chi_{\phi\chi\phi} =  -\beta ^2 \cosh ^2(\beta  \chi )
\end{aligned}
\end{equation}
which reduce to $\pm \beta^2$ for $\chi=0$.
 The corresponding components of the Ricci tensor can be computed to be
\beqn
{\cal R}_{\phi\phi} =-\beta ^2 \cosh ^2(\beta  \chi )
\, , \quad
{\cal R}_{\chi\chi} =  -\beta^2
\eeqn
leading to  the field-space Ricci scalar
\beqn
{\cal R}=-2\beta^2 =- {4\over 3\alpha} \, ,
\eeqn

This choice of the field-space basis allows an easier comparison between our present work and our related analysis of the two-field 	generalized T-model, as well as the numerical investigations of the T-model found in Ref.~\cite{Krajewski:2018moi}. Furthermore, the equations of motion for both the background as well as for the fluctuations become simple in this basis, since the metric becomes the unit matrix and all Christoffel sumbols vanish at $\chi=0$. This metric, albeit simple, might give the illusion that the two field-space directions are inherently different, one of them even being canonically normalized. However, as evident in the value of the Ricci scaler, this basis describes a field-space with a constant negative curvature at every point.

\end{document}